%% file: PPNP_RAA.tex
\documentclass[review]{elsarticle}

\usepackage[colorlinks,citecolor=blue,linktoc=all,linkcolor=cyan]{hyperref}
\usepackage{graphicx}

\usepackage[T1]{fontenc}
\usepackage{dsfont}               % use mathds instead of mathbb for outline fonts
\usepackage{mathrsfs}             % provides mathscr without overwriting mathcal
\usepackage{slashed}              % For Dirac slash notation.
\usepackage{amsmath}
\usepackage{amssymb}
\usepackage{amsbsy}
\usepackage{amsfonts}
\usepackage{multirow,bigdelim}
\usepackage{lineno}

\numberwithin{equation}{section}
\numberwithin{table}{section}
\numberwithin{figure}{section}

\journal{Progress in Particle and Nuclear Physics}

\usepackage{titlesec}
\usepackage{sectsty}
\titleformat{\section}{\normalfont\Large\bfseries}{\thesection}{1em}{}
\titleformat{\subsection}{\normalfont\large\bfseries}{\thesubsection}{1em}{}
\titleformat{\subsubsection}{\normalfont\normalsize\bfseries}{\thesubsubsection}{1em}{}
\begin{document}
	
	\begin{frontmatter}
		
		\title{Reactor antineutrino flux and anomaly}

		%authors, affiliations, corresponding author mention 
		\author[BNL]{Chao Zhang\corref{correspondingauthor}}
		\ead{czhang@bnl.gov}
		\author[BNL]{Xin Qian}
		\cortext[correspondingauthor]{Corresponding author}
		\author[CNRS]{Muriel Fallot}
		
		\address[BNL]{Physics Department, Brookhaven National Laboratory, Upton, NY, 11973 U.S.A.}
		\address[CNRS]{SUBATECH, CNRS/IN2P3, Universit\'{e} de Nantes, Ecole des Mines de Nantes, F-44307 Nantes, France}
		
		\begin{abstract}
                Reactor antineutrinos have played a significant role in establishing the standard model of particle physics and the theory of neutrino oscillations. In this article, we review the reactor antineutrino
                flux and in particular the reactor antineutrino anomaly (RAA) coined over a 
                decade ago. RAA refers to a deficit of the measured antineutrino inverse beta decay rates 
                at very short-baseline reactor experiments compared to the theoretically improved 
                predictions (i.e.~the Huber-Mueller model). Since the resolution of several previous experimental 
                anomalies have led to the discovery of non-zero neutrino mass and mixing, many efforts have been invested to study the origin of RAA both experimentally and theoretically. The
                progress includes the observation of discrepancies in antineutrino energy spectrum 
                between data and the Huber-Mueller model, the re-evaluation of the Huber-Mueller model uncertainties, the 
                potential isotope-dependent rate deficits, and the better agreement between data and new model predictions using the improved summation method. These developments
                disfavor the hypothesis of a light sterile neutrino as the explanation of RAA and supports the 
                deficiencies of Huber-Mueller model as the origin. Looking forward, more effort from 
                both the theoretical and experimental sides is needed to fully understand the root of RAA and to                 
                make accurate predictions of reactor antineutrino flux and energy spectrum for future discoveries.
		\end{abstract}
		
		
	\end{frontmatter}
	
	\newpage
	
	\thispagestyle{empty}
	\tableofcontents
	
	%to begin the line numbers: 
	% \linenumbers

	%beginning of the core of the manuscript
	\input{Introduction.tex}

%      SNO+ citation ~\cite{SNO:2022qvw}

% Inside the core of a commercial power reactor, a portion of the neutrons are 
%   captured by $^{238}$U because of its much higher concentration, producing new fissile isotopes: 
%   $^{239}$Pu and $^{241}$Pu. Fissions of $^{235}$U, $^{239}$Pu, and $^{241}$Pu are induced by 
%   thermal neutrons ($\sim$0.025-eV kinetic energy). In contrast, fission of $^{238}$U can be 
%   induced only by fast neutrons ($\sim$1-MeV kinetic energy).

\input{Flux_calculation.tex}

    \input{RAA.tex}

    \input{Additional_measurements.tex}

    \input{New_calculation.tex}

    \input{Summary.tex}

	%end of the core of the manuscript
	
    \section*{Acknowledgements}
    We thank Petr Vogel for suggesting writing this review article and David Jaffe for reading the manuscript. The work of C.~Zhang and X.~Qian was supported by the US Department of Energy (DOE) Office of Science, Office of High Energy Physics under Contract No.~DE-SC0012704. The work of M.~Fallot was supported by the CNRS-in2p3 OPALE Master project, the NEEDS/NACRE project and the SANDA European project.

    % \section*{Author's contributions \textit{(optional section)}}
    % Detailing here the contributions of the authors of the review.

    \bibliography{ref}
	%Please use Bib\TeX\ to generate your bibliography and include DOIs whenever available. Example of bib file: 
	
	%%%%%%%%%%%%%%%%%%%%%%%%%%%%%%%%%%%%%%%%%%%%%%%%%%%%%%%%%%%%%%%%%%%
	% Encoding: ISO-8859-1

	%@Article{Eichmann:2016yit,
	%author        = {Eichmann, Gernot and Sanchis-Alepuz, Helios and Williams, Richard and Alkofer, Reinhard and Fischer, Christian S.},
	%title         = {{Baryons as relativistic three-quark bound states}},
	%journal       = {Prog. Part. Nucl. Phys.},
	%year          = {2016},
	%volume        = {91},
	%pages         = {1-100},
	%archiveprefix = {arXiv},
	%doi           = {10.1016/j.ppnp.2016.07.001},
	%eprint        = {1606.09602},
	%owner         = {chfi},
	%primaryclass  = {hep-ph},
	%slaccitation  = {%%CITATION = ARXIV:1606.09602;%%},
	%timestamp     = {2018.08.02},
	%}

	%@Comment{jabref-meta: databaseType:bibtex;}
	%%%%%%%%%%%%%%%%%%%%%%%%%%%%%%%%%%%%%%%%%%%%%%%%%%%%%%%%%%%%%%%%%%%
	
	% \newpage
	% \appendix
	% \renewcommand*{\thesection}{\Alph{section}}
	
	% \section{Appendices}\label{appendix}

\end{document}

%% file: Introduction.tex
\section{Introduction}\label{sec:introduction} 
Neutrinos, first predicted in 1930 by W.~Pauli~\cite{pauli} as a ``desperate remedy'' to rescue the energy conservation law that seemed to have failed in nuclear beta decays, remain a fundamental tool today to study the nature of the universe. 
In the Standard Model of particle physics, there are three flavors of neutrinos ($\nu_e$, $\nu_\mu$, $\nu_\tau$) and they are associated with their corresponding charged leptons ($e^-$, $\mu^-$, $\tau^-$) as weak isospin pairs. Neutrinos only participate in weak interactions and by construction are massless in the Standard Model. 
The discovery of neutrino oscillations in 1998~\cite{Super-Kamiokande:1998kpq} and the subsequent precision measurements of neutrino oscillation properties over the past two decades completely changed the original picture and provided one of the first evidences of physics beyond the Standard Model. 
Neutrino oscillations show that neutrinos transform flavors as they travel in space, and the amplitude of this neutrino flavor mixing is much larger compared to that of the quark mixing~\cite{pdg2022}. This phenomenon implies that neutrinos have a non-zero mass, although the mass is much smaller compared to other fermions. 
The ongoing and future experimental efforts aim to explore the remaining unknowns of neutrinos, such as whether they obey charge-parity conservation, whether they are their own anti-particles, whether they have a fourth or more flavors, and so on. 
Studying the neutrino properties and understanding why they appear to be unique in the family of fundamental particles may hold the key to advance our understanding of the universe.

Since neutrinos interact very weakly with matter, neutrino experiments always involve either an ample neutrino source, or a massive detector, or both. One of the most powerful, cost-effective, controllable, and well-understood sources of neutrinos is the nuclear reactors. 
In fact, using the Hanford and the Savannah River nuclear reactors, F.~Reines and C.~Cowan made the first experimental detection of (anti)neutrinos in 1953--1959~\cite{Reines1,Reines2,Reines3}, nearly 30 years after Pauli's prediction of their existence. 
Since then, reactor antineutrino experiments have played a crucial role in the studies of neutrino oscillations~\cite{Vogel:2015wua}, such as solving the solar neutrino problem on Earth by the KamLAND experiment in 2003~\cite{KamLAND1}, and the discovery of non-zero mixing angle $\theta_{13}$ by the Daya Bay and RENO experiments in 2012~\cite{DayaBay:2012fng,RENO:2012mkc}. 
In a typical reactor antineutrino experiment, antineutrino flux and energy spectrum are measured and compared with theoretical predictions, from which the properties of neutrinos are extracted. 
The last several decades have witnessed a tremendous advancement in both the experimental techniques to improve measurement precision and the theoretical development to improve the reactor antineutrino flux calculations. 
Together, they form the basis of performing successful reactor antineutrino experiments. 

One of the most intriguing events in the past decade involving reactor antineutrinos is the so-called \emph{Reactor Antineutrino Anomaly}, or the RAA. 
The RAA was coined in 2011 as a result of a re-evaluation of reactor antineutrino flux calculations by Mueller \textit{et al.,}~\cite{Mueller:2011nm} and Huber~\cite{Huber:2011wv}.
When the experimental measurements, which agreed in the past with older calculations, are compared with this new evaluation (namely, the Huber-Mueller model), a deficit of about 6\% in the integrated antineutrino flux is observed with a significance of over 2$\sigma$~\cite{Mention:2011rk}. 
The RAA immediately drew substantial attention in the community, partly because one of the possible explanations of the RAA is the existence of four or more flavors of neutrinos, often named the sterile neutrinos. 
Together with hints of sterile neutrinos from earlier experiments such as LSND~\cite{LSND:2001aii} and MiniBooNE~\cite{MiniBooNE:2008yuf}, a new age of sterile neutrino hunting was invigorated. 
It has been over 10 years since the birth of the RAA. During this era, many new experimental results have emerged, which either directly or indirectly shed light onto the origin of the RAA. 
Meanwhile, breakthroughs in the reactor antineutrino flux calculations were also made, and theoretical uncertainties were better understood. It is now a good time to look back and review what we have learned from this saga and summarize our current understanding of the RAA.

This review is organized in an approximately chronological order of historical events. 
In Sec.~\ref{sec:flux_calculation}, the basic mechanisms to produce and detect reactor antineutrinos are introduced. The two principal methods to calculate the reactor antineutrino flux, the summation method and the conversion method, are described. 
In Sec.~\ref{sec:RAA}, the re-evaluation of reactor antineutrino flux in 2011 (i.e.~the Huber-Mueller model) is introduced. The issue of the RAA is explained with the inclusion of new reactor antineutrino flux measurements from modern experiments after 2011, and possible explanations of the RAA are described. 
In Sec.~\ref{sec:additional_mea}, additional new experimental results from the mid 2010s that shed light onto the origin of the RAA are described, including the measurements of the reactor antineutrino energy spectrum, the measurements of the individual reactor antineutrino flux from $^{235}$U and $^{239}$Pu, and the measurement of the $\beta-$spectrum ratio between $^{235}$U and $^{239}$Pu.
Breakthroughs in developing reactor antineutrino flux calculations and understanding their uncertainties are reviewed in Sec.~\ref{sec:new_calculation}. 
Finally, with all the new experimental evidences and theoretical advancements, in Sec.~\ref{sec:summary} we summarize our current understanding of the origin of the RAA, and present an outlook of where to go from here.

%% file: Flux_calculation.tex
\section{Reactor antineutrino basics}\label{sec:flux_calculation} 

Nuclear fission reactors are widely used in the world since the 1950s as an efficient means to generate electricity. 
Since each fission releases approximately 200~MeV of energy, the energy density of a fissile material, such as $^{235}$U, is about 30,000 times higher than that from coal. 
As of 2023, nuclear energy provides about 10\% of the world's electricity from over 430 power reactors. 
In addition, there are about 220 research reactors in over 50 countries for the production of medical and industrial isotopes, and for academia research and training. 
As a by-product of nuclear reactors, pure electron antineutrinos ($\bar\nu_e$) are emitted in the reactor core from $\beta$-decays of the fission products. 
On average, about $6\times10^{20}$ $\bar\nu_e$s are produced every second from a typical 3~GW$_\textrm{th}$ commercial reactor core. 
This intense source of (anti)neutrinos from reactors provides many advantages compared to other natural or artificial neutrino sources. 
The flux is considerably higher than that from accelerators, and it comes nearly free of charge for scientists. 
The flux is composed of pure $\bar\nu_e$s, which significantly simplifies the flux uncertainty determination, one of the main hurdles in neutrino oscillation experiments. 
The operation of nuclear reactors provides a considerable time variation of the antineutrino flux, and in certain circumstances reactors can be turned off completely for refueling or maintenance. This is helpful for experiments to determine their backgrounds. 
Detectors can be placed very close to the core, within a few meters, or very far away from the core at kilometers or even hundreds of kilometers away. 
This is essential for neutrino oscillation studies since the oscillation features vary with distances. In this section, we review the principles of production and detection mechanisms for reactor antineutrinos, and methods to calculate the reactor antineutrino flux and energy spectrum.

\subsection{Production of reactor antineutrinos}\label{sec:nu-production}
\begin{figure}[bhtp]
  \begin{centering}
    \includegraphics[width=0.60\textwidth]{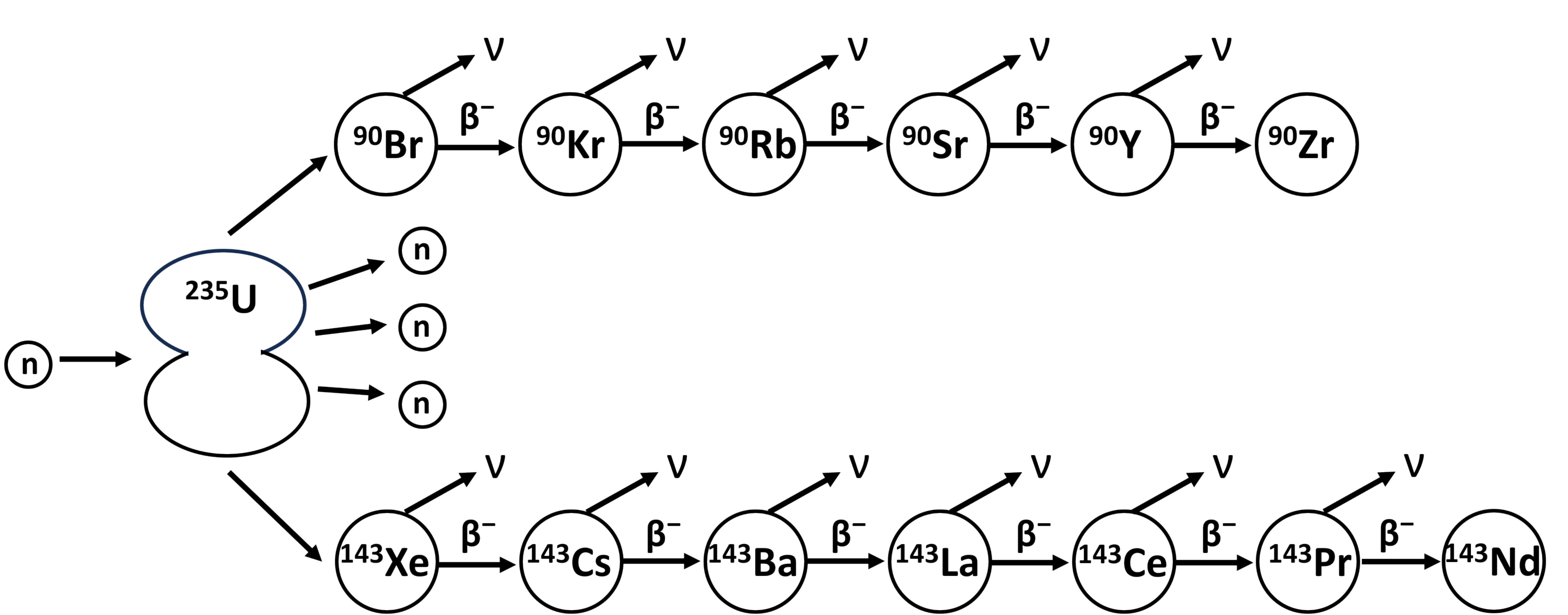}
    \includegraphics[width=0.37\textwidth]{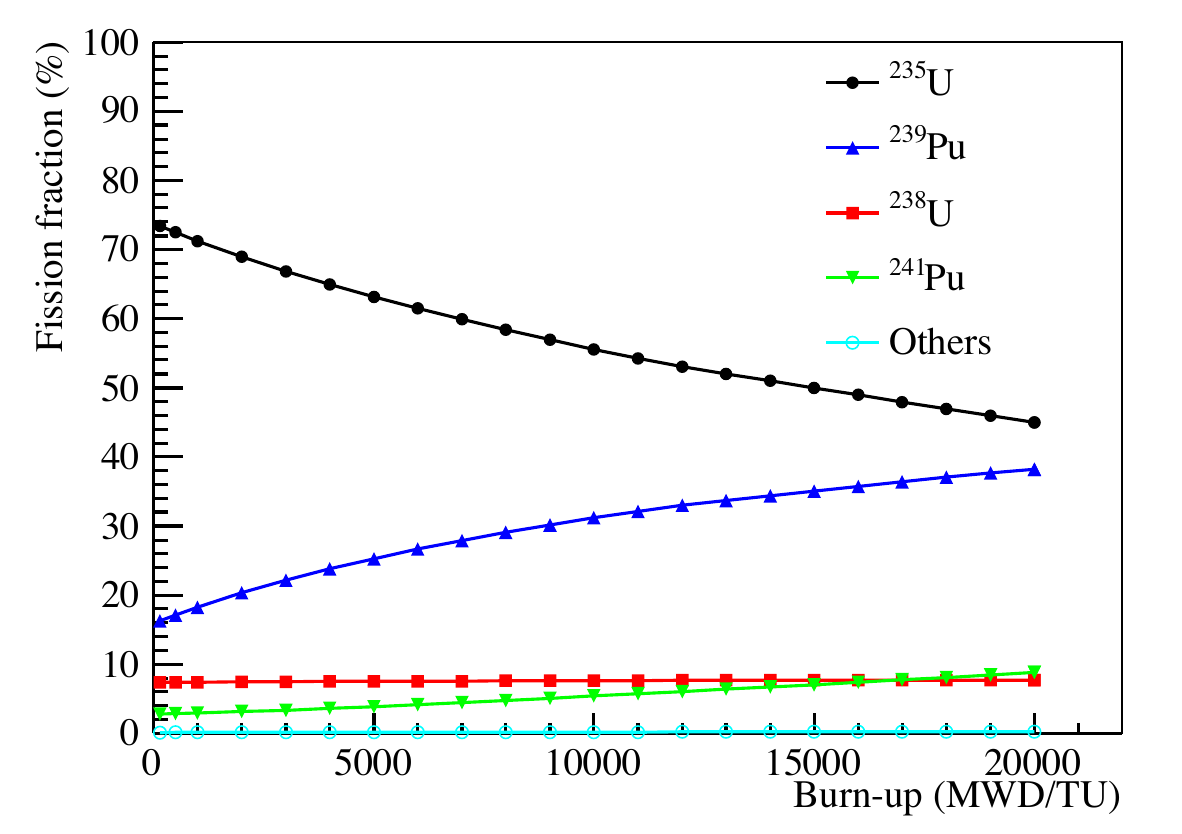} 
    \put(-350,-10){a)}
    \put(-100,-10){b)}
    \par\end{centering}
    \caption{\label{fig:reactor_fission} 
    (Left) An illustration of nuclear fission of $^{235}$U and the subsequent $\beta$-decays from its fission fragments. Electron antineutrinos are emitted from these  $\beta$-decay chains. In this example, 11 $\bar\nu_e$s are emitted after the fission before the fission fragments become stable. On average, 6 $\bar\nu_e$s are emitted per fission in a nuclear reactor.
    (Right) Fission fractions of the four main fission isotopes, $^{235}$U, $^{238}$U, $^{239}$Pu, and $^{241}$Pu, in the Daya Bay reactor core as a function of burn-up (in units of megawatt~$\cdot$~days normalized by the initial uranium mass in the fuel) from a simulation of a complete operation cycle between refueling. Figure taken from Ref.~\cite{DayaBay:2016ssb}.
    }
\end{figure}

Fission is the main process that generates power in a nuclear reactor. In most reactors, $^{235}$U is the primary fissile isotope. Since the natural abundance of $^{235}$U is only 0.7\% in uranium, commercial reactors typically consist of uranium dioxide fuel with about 3--5\% enrichment of $^{235}$U in order to sustain the fission chain reaction in the core. 
Fission happens when $^{235}$U absorbs a thermal neutron and breaks into two smaller-mass fragments, as illustrated in Fig.~\ref{fig:reactor_fission}a. 
Since the binding energy of the daughter nuclei is higher than $^{235}$U, this process releases energy, on average about 202~MeV per $^{235}$U fission. The heat from nuclear fission in the core is transferred through a coolant that is pumped through the reactor and used to drive steam turbines to generate electricity. 
Each $^{235}$U fission also produces about 2.4 neutrons, of which 1.4 neutrons are absorbed or escape the core by engineering design, leaving exactly 1 neutron to initiate the next fission and therefore sustains the fission chain reaction while keeping the stability of the core. 
The neutrons from the $^{235}$U fission have kinetic energy of a few MeV. They are slowed down by neutron moderators in the core to thermal energy ($\sim$0.025~eV kinetic energy), where the cross section of fission is much higher than neutron absorption or other processes.

In a typical commercial reactor, there are three other main fission isotopes, $^{238}$U, $^{239}$Pu, and $^{241}$Pu, besides $^{235}$U. Together these four isotopes contribute more than 99.7\% of all fissions in a commercial reactor. $^{238}$U has 99.3\% natural abundance in uranium. 
Although $^{238}$U does not fission with thermal neutrons, it fissions with fast neutrons when the kinetic energy of the neutron is above $\sim$1~MeV. 
$^{238}$U can also absorb a neutron, and through two subsequent $\beta$-decays turns into a fissile isotope $^{239}$Pu. 
$^{239}$Pu can further absorb two neutrons and turns into another fissile isotope, $^{241}$Pu. 
Both $^{239}$Pu and $^{241}$Pu fission with thermal or epithermal neutrons, mostly the latter, due to the presence of a strong resonance around 0.2--0.3~eV~\cite{Littlejohn:2018hqm}.
Table~\ref{table:fission_isotope} summarizes the processes to generate these four fission isotopes, the average fission cross section of each isotope for thermal neutrons, and the energy released in each fission. 
Figure~\ref{fig:reactor_fission}b shows the change of fission fraction of each isotope as a function of burn-up (in units of megawatt~$\cdot$~days normalized by the initial uranium mass in the fuel) in a typical operation cycle of a reactor core. 
While $^{238}$U contributes to a constant $\sim$8\% of total fission, the $^{235}$U fission fraction decreases with time and $^{239}$Pu and  $^{241}$Pu fission fractions increase with time. 
This cycle resets when the reactor is shutdown to perform a refuel, typically once every 12--18 months. 
The average fission fraction in a typical commercial reactor for $^{235}$U : $^{238}$U : $^{239}$Pu : $^{241}$Pu is about $0.58 : 0.08 : 0.29 : 0.05$.

\begin{table*}[ht!]
\caption{\label{table:fission_isotope} Processes to generate the four main fission isotopes, $^{235}$U, $^{238}$U, $^{239}$Pu, and $^{241}$Pu in a typical commercial reactor, the average fission cross section ($\sigma_{\textrm{fis}}$) of each isotope for thermal neutrons, the average fission fraction of each isotope for a typical reactor, the energy released in each fission that transforms into heat over a finite time interval~\cite{Ma2013PRC}, and the inverse beta decay yield predicted by the Huber-Mueller model (details in Sec.~\ref{sec:huber-mueller}). }
\begin{center}
\begin{tabular}{|c|c|c|c|c|c|}
\hline
\hline
Isotope & Origin  & $\sigma_{\textrm{fis}}$  & Fission fraction & Energy per fission & IBD yield \\ 
 &   & (barn)  &  & (MeV/fission) & ($10^{-43}$~cm$^2$/fission) \\ \hline
$^{235}$U & enrichment & 531 & 0.58 & 202.36 & 6.69 \\
$^{238}$U & natural abundance & --- & 0.08 & 205.99 & 10.10 \\
$^{239}$Pu & $^{238}\textrm{U}(n, \gamma)^{239}\textrm{U}(\beta^-)^{239}\textrm{Np}(\beta^-)^{239}\textrm{Pu}$ & 750 & 0.29 & 211.12 & 4.36 \\
$^{241}$\textrm{Pu} & $^{239}\textrm{Pu}(n, \gamma)^{240}\textrm{Pu}(n, \gamma)^{241}\textrm{Pu} $ & 1010 & 0.05 & 214.26 & 6.05 \\\hline
\hline
\end{tabular}
\end{center}
\end{table*}

Fission breaks a parent nucleus, e.g.~$^{235}$U, into two daughter nuclei of smaller masses, such as $^{90}$Br and $^{143}$Xe. Most of the energy released in a fission is carried away by these fission fragments in the form of their kinetic energy. The mass distribution of the fission fragments, i.e.~fission yields, has a typical double-hump structure and is peaked around mass 90--100 and 135--145. The nuclei of the fission fragments typically have too many neutrons and are therefore unstable. They will go through a series of $\beta^-$ decays before reaching stable isotopes, typically within days. An illustration of these processes is shown in Fig.~\ref{fig:reactor_fission}a. 
It is these $\beta^-$ decays that produce the pure electron antineutrinos inside the reactors. 
On average, about 6 $\bar\nu_e$s are produced in each fission through the $\beta^-$ decay chains. 
Among them, about 30\% of the $\bar\nu_e$s have energy about 1.8~MeV, which is the threshold for detection via the inverse beta decay reaction (see Sec.~\ref{sec:sub:detection} for more discussions). 
Besides the $\bar{\nu}_e$s originated from the fission process, another important source of $\bar{\nu}_e$s comes 
from neutron capture on $^{238}$U: $^{238}\textrm{U}(n,\gamma)^{239}\textrm{U}$. The $\beta^-$ decay of $^{239}$U (Q-value of 1.26~MeV and 
half-life of 23.5 minutes) and the subsequent $\beta^-$ decay of $^{239}$Np (Q-value of 0.72~MeV and half-life of 2.3 days) produce 
a sizeable amount of $\bar{\nu}_e$s below the 1.8~MeV inverse beta decay threshold.
The energy of the reactor $\bar\nu_e$s rarely exceeds $\sim$8 MeV, which is typical for $\beta$-decays.

The energy-dependent reactor antineutrino flux, defined as the total number of electron antineutrinos released from a reactor core per second per neutrino energy interval, can be calculated using the following formula:
\begin{equation}\label{eq:nu_spectra}
\frac{d\Phi}{dE_\nu} = \sum_i 
\left( \frac{W_\textrm{th}}{\sum_j f_j e_j} \right) 
\cdot f_i 
\cdot \frac{d\phi_i}{dE_\nu} ,
\end{equation}
where the index $i$ and $j$ goes over the four fission isotopes $^{235}$U, $^{238}$U, $^{239}$Pu, and $^{241}$Pu. $f_i$ is the fission fraction of each isotope, and $e_j$ is the energy released in each fission that transforms into heat over a finite time interval as listed in Table~\ref{table:fission_isotope}. 
Since $W_\textrm{th}$ is the total thermal power of the reactor core, the term in the parenthesis calculates the total number of fissions in the reactor. 
The remaining unknown, $d\phi_i/dE_\nu$, is the total number of $\bar\nu_e$s per neutrino energy interval per fission from each isotope, and is often called the isotopic reactor antineutrino flux. 
Figure~\ref{fig:spectrum_ibd} shows the calculated $d\phi_i/dE$ above 1.8~MeV for each isotope. 
The two popular methods to calculate $d\phi_i/dE$, the summation method and the conversion method, will be reviewed in detail in Sec.~\ref{sec:sub:flux_calculation}. 
In the following we briefly review how $W_\textrm{th}$, $e_i$, and $f_i$ are obtained and the additional effects to be included in the reactor antineutrino flux calculation.

Reactor thermal power ($W_\textrm{th}$) is typically monitored with multiple online and offline systems by the nuclear power plants, since it is a critical factor for the safe operation of a reactor. Two main methods to measure $W_\textrm{th}$ are measuring the neutron flux near the core, and measuring the heat balance in the primary or secondary loop of the cooling system. The neutron flux monitors usually have larger bias for thermal power because of the change of isotope content in the core. They have to be re-calibrated once the bias exceeds 1--2\%. The heat balance method measures the temperature, pressure, and mass flow rate of the feed water in the cooling system to calculate the thermal power. The largest uncertainty comes from the water flow measurement, which can be improved by the installation of orifice plates in the secondary loop. Overall, the uncertainty of thermal power measurement is about 0.5\% in a modern commercial reactor.

Not all energy from fission is transformed into heat. For instance, the energy carried away by antineutrinos in the $\beta$-decays does not produce heat in the reactor. The heat from $\beta$-decays of fission daughters also has a time constant that can be as long as hundreds of days for long-lived isotopes. On the other hand, non-fission reactions such as neutron captures in the core can generate additional heat in the system. Therefore, the energy released per fission ($e_i$, for the $i$th isotope) is defined as the energy from a fission event that transforms into heat over a finite time interval. It is calculated~\cite{Ma2013PRC, Kopeikin:2004cn} by considering the previously mentioned effects using typical reactor parameters and information from nuclear databases. The uncertainty on the energy released per fission is about 0.2\%.

During a reactor operation cycle, uranium is depleted and plutonium is produced until the next refueling cycle when new fuel rods are introduced and old fuel rods are removed or redistributed in the core. The fission fraction of each fission isotope ($f_i$, for the $i$th isotope) changes as a function of time, which is commonly referred to as fuel evolution. 
Fuel evolution is a dynamic process related to many factors such as power, neutron flux, material composition, type and position of fuel rods, boron concentration, and others. 
Detailed simulations of fuel evolution are typically performed by the nuclear power plants (NPPs) with commercial software for the safe operation of reactors. 
Upon agreement with the NPPs, daily or weekly averaged fission fractions, $f_i(t)$, can be released to neutrino experiments. An example of fuel evolution from a Daya Bay reactor core is shown in Fig.~\ref{fig:reactor_fission}b. 
Open source reactor simulation code, such as DRAGON~\cite{dragon}, is often used by experiments to validate the results from NPPs. The uncertainty of the calculated fission fraction $f_i$ is about 5\% for an individual isotope. However, because of the strong correlation of $f_i$ among the isotopes~\cite{DayaBay:2016ssb}, e.g.~the anti-correlation between $^{235}$U and $^{239}$Pu, the uncertainty on the calculated total number of neutrinos from fission fraction uncertainty is only about 0.6\%.

After burning in the core, the nuclear fuel is removed from the reactor and typically stored in a cooling pool near the reactor core for a long period of time. 
This is often referred to as the spent nuclear fuel (SNF). 
The long-lived isotopes in the SNF will continue to decay and produce electron antineutrinos from their $\beta^-$ decay chains. 
Two examples of such long-lived isotopes are $^{106}$Ru with half-life of 373.6 days and $^{144}$Ce with half-life of 284.9 days. 
The antineutrino flux from SNF needs to be calculated with inputs from the refueling history and SNF inventory information provided by the NPPs. The uncertainty on this calculation can be large if the SNF inventory history information is sparse. Nevertheless, since the contribution to the antineutrino flux from SNF is usually less than 0.3\%, even a conservative 100\% uncertainty can be tolerated. 

Our descriptions so far focused on the most common types of commercial reactors used by reactor antineutrino experiments, which are the light water reactors such as the pressured water reactors (PWR) or the boiling water reactors (BWR). 
There are many other types of reactors depending on how they are cooled with different neutron moderators, such as heavy water, graphite, liquid sodium, molten salt, and others. Different reactor designs have different neutron flux energy distributions, which in turn changes the fraction of fission versus non-fission reactions and results in different fuel evolution in the reactor core. 
Therefore, detailed reactor information has to be obtained by an experiment from all nearby reactors regardless of their types. 
Commercial reactors typically have $\sim$4\% enrichment of $^{235}$U and are categorized as low-enriched uranium (LEU) reactors. On the other hand, many research reactors have more than $\sim$20\% enrichment of $^{235}$U and are categorized as highly-enriched uranium (HEU) reactors. In typical HEU reactors, more than 99\% of fissions come from $^{235}$U, therefore their fuel evolution is nearly constant. HEU reactors typically are less powerful (10--100s of MW$_\textrm{th}$) compared to LEU commercial reactors (a few GW$_\textrm{th}$), but have compact cores (tens of centimeters in diameter) compared to LEU reactor cores (a few meters in diameter). This is often desirable for neutrino experiments that wish to place their detectors as close as possible to the reactor core, and to minimize the smearing effect from neutrino oscillations inside the core itself. 
Finally, we mention that there are proposed R\&Ds of thorium reactors~\cite{thorium-reactor} where the fuel cycle starts by converting the initial fertile $^{232}$Th in the thorium fuel to the fissile $^{233}$U through the $^{232}\textrm{Th}(n, \gamma)^{233}\textrm{Th}(\beta^-)^{233}\textrm{Pa}(\beta^-)^{233}\textrm{U}$ process. Thorium has several advantages over the uranium fuel, such as the abundance and breeding capability. If thorium reactors are realized in the future, they will provide a very different reactor antineutrino flux from the $^{235}$U reactors, and could be very useful for future reactor antineutrino researches. 

\subsection{Detection of reactor antineutrinos}\label{sec:sub:detection}
Since neutrinos do not experience the strong or electromagnetic force, their interaction cross sections with matter are notoriously small, on the order of $10^{-44}$~cm$^2$ for MeV neutrinos, which earned them the nickname ``ghost particles''. 
Therefore, detection of neutrinos requires a massive detector to provide ample amount of target nuclei, and a sensitive method to find the rare signal when an interaction eventually occurs. 
So far, the best method to detect reactor antineutrinos is through the inverse beta decay (IBD) reaction: $\bar{\nu}_e + p \rightarrow e^+ + n$. 
Not only does it have a relatively large cross section ($\sim$6$\times 10^{-43}$~cm$^2$), the presence of both a positron and a neutron in the final state provides a unique coincidence signal to distinguish it from background processes such as radioactive decays from impurities in the detector. 
For these reasons, the IBD reaction is almost exclusively used by experiments to study reactor antineutrino flux and neutrino oscillations, and is the main focus of this section. 
Other channels to detect reactor antineutrinos in the past are summarized in Table~\ref{tab:detection_channels} and briefly reviewed here for completion. $\nu$-electron elastic scattering ($\nu$ES) has been used to measure the weak mixing angle $\theta_{W}$ and to constrain the anomalous neutrino magnetic moments~\cite{TEXONO:2009knm,Reines:1976pv,Vidyakin:1992nf,Derbin:1993wy,MUNU:2005xnz}. 
This channel does not have an energy threshold, so it can probe reactor antineutrinos below the IBD threshold of 1.8~MeV. 
It is also less sensitive to the neutrino flavors or the neutrino--antineutrino differences, making it suitable for certain physics studies. 
However, since the $\nu$ES signal is only a single electron recoil event, the requirement on the radiopurity of detector material and cosmic-ray rejection is stringent. 
Similarly to $\nu$ES, the coherent elastic $\nu$-nucleus scattering (CE$\nu$NS) reaction is a promising threshold-less and flavor-insensitive way to detect reactor antineutrinos. 
It has the largest cross section (enhanced by the square of number of neutrons in the target nucleus) among all methods. The signal, however, is a tiny energy deposition from the recoil nucleus (less than a few hundred eV), therefore its detection above backgrounds is extremely challenging and several groups are working on the first detection~\cite{CONNIE:2021ggh,CONUS:2020skt,CONUS:2021dwh,Ricochet:2021rjo}. In particular, tantalizing evidence of CE$\nu$NS detection using reactor antineutrinos was recently reported in Ref.~\cite{Colaresi:2022obx}.
Finally, $\nu$-deuteron interactions were used in early days to study charged current and neutral current weak interactions with reactors~\cite{Reines:nud1980}. However, this method is hindered by the availability of the large amount of heavy water needed.

\begin{table*}[ht!] 
\caption{Summary of detection channels for reactor antineutrinos. Table adapted from Ref.~\cite{Qian:2018wid}. The cross section is integrated over the reactor antineutrino energy distribution assuming the typical fission fractions from the four main isotopes as listed in Table~\ref{table:fission_isotope}. For the $\nu$ES and the CE$\nu$NS reaction, $Z$ and $N$ are the total number of protons and neutrons in the target nucleus, respectively. }
\begin{center}
\begin{tabular}{|c|c|c|c|}
  \hline\hline
  Channel & Name & Cross Section ($10^{-44}$ cm$^2$) & Threshold (MeV) \\\hline
  $\bar{\nu}_e + p \rightarrow e^+ + n$ & inverse beta decay (IBD) & 63 & 1.8 \\\hline
  $\bar{\nu}_e + e^- \rightarrow \bar{\nu}_e + e^-$ & $\nu$-electron elastic scattering ($\nu$ES) & $0.4 \cdot Z$ & --- \\\hline
  $\bar{\nu}_e + A \rightarrow \bar{\nu}_e + A $   & coherent elastic $\nu$-nucleus scattering (CE$\nu$NS) & $9.2 \cdot N^2$ & ---\\\hline
  $\bar{\nu}_e + d \rightarrow n + n + e^+$ & $\nu$-deuteron charged current (CC) scattering & 1.1 & 4.0\\\hline
  $\bar{\nu}_e + d \rightarrow n + p + \bar{\nu}_e$ & $\nu$-deuteron neutral current (NC) scattering & 3.1 & 2.2\\\hline
  \hline
\end{tabular}\label{tab:detection_channels}
\end{center}
\end{table*}

\begin{figure}[bhtp]
  \begin{centering}
    \includegraphics[width=0.7\textwidth]{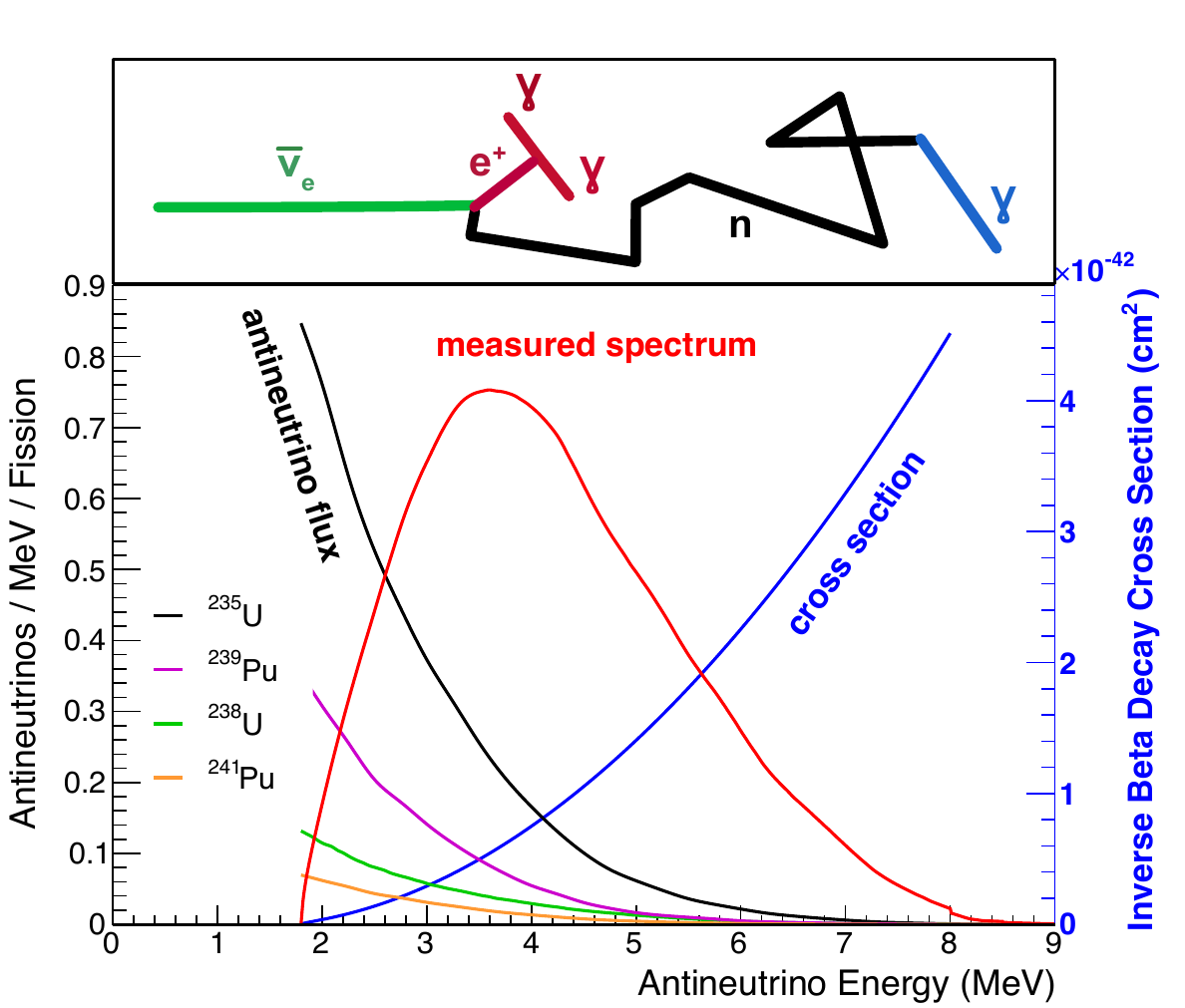}
    \par\end{centering}
    \caption{\label{fig:spectrum_ibd} 
    Reactor antineutrino detection using the inverse beta decay (IBD) reaction. The bottom figure shows the reactor $\bar\nu_e$ flux from the  four main fission isotopes, $^{235}$U, $^{238}$U, $^{239}$Pu, and $^{241}$Pu, weighted by their typical fission fractions in a commercial reactor. The IBD cross section is shown as the blue curve. Their product is the $\bar\nu_e$ energy spectrum measured by the detectors, shown as the red curve. The top figure shows an illustration of the steps involved in the IBD detection. See text for more details. Figure taken from Ref.~\cite{Vogel:2015wua}. }
\end{figure}

In the following, we describe the details of the inverse beta decay (IBD) detection channel. 
The cross section of the IBD reaction, $\bar{\nu}_e + p \rightarrow e^+ + n$, is accurately known since it is directly tied to the well-measured neutron beta decay process: 
\begin{equation}\label{eq:ibd_xs}
    \sigma_{\textrm{IBD}} 
    = \frac{2\pi^2 E_e^{(0)} p_e^{(0)}}{m_e^5 f^R_{p.s.} \tau_n} (1+\textrm{corr.})
    = 9.53 \frac{E_e^{(0)} p_e^{(0)}}{\textrm{MeV}^2} (1+\textrm{corr.}) \times 10^{-44} \textrm{cm}^2,
\end{equation}
where $m_e$ is the mass of the electron, $f^R_{p.s.}=1.7152$ represents the phase space factor of neutron decay, including the Coulomb, weak magnetism, recoil, and outer radiative correction, and $\tau_n = 878.4\pm0.5$~seconds~\cite{pdg2022} is the neutron lifetime. 
$E_e^{(0)}$ and $p_e^{(0)}$ are the zeroth order (in $1/M_n$) energy and momentum of the final state positron after ignoring the neutron recoil energy:  $E_e^{(0)} = E_\nu + M_p - M_n$. It is evident that the neutrino must have at least 1.8~MeV energy to overcome the mass difference between neutron and proton, and to produce a positron.
The correction term in Eq.~\eqref{eq:ibd_xs} is non-negligible, and includes higher order terms in $1/M_n$ after considering the angular distribution of the final-state positron. 
The derivation and numerical forms of the correction term are given in Refs.~\cite{Vogel:1999zy,Strumia:2003zx}. The largest uncertainty in the IBD cross section comes from the experimentally determined neutron lifetime. Although the quoted uncertainty from PDG is quite small, 0.05\%, there is a long-standing disagreement between the in-beam measured neutron lifetime and the more precise measurements using trapped ultracold neutrons. Recent reviews of the history and status of the neutron lifetime anomaly can be found in Refs.~\cite{Wietfeldt:2011suo, Czarnecki:2018okw}. The calculated IBD cross section as a function of $\bar\nu_e$ energy is shown in Fig.~\ref{fig:spectrum_ibd}.

The detection of an IBD reaction consists of a pair of coincident signals from the final state positron and neutron, as illustrated in the top panel of Fig.~\ref{fig:spectrum_ibd}. The positron quickly deposits its energy and then annihilates with a nearby electron to produce a pair of 511~keV $\gamma$-rays. 
This creates a prompt signal with the prompt energy directly related to the $\bar\nu_e$ energy: $E_{\textrm{prompt}} \simeq E_\nu - 0.8$~MeV, where the kinetic energy of the neutron (about tens of keV) can be ignored to the first order. 
The neutron loses most of the kinetic energy by elastically scattering with nearby nuclei until it reaches thermal energy. 
It is then captured by a nearby proton and releases a 2.2~MeV $\gamma$-ray with a capture time of $\sim$200~$\mu$s in a typical liquid scintillator detector. 
This creates a mono-energetic delayed signal. 
This time and spatially correlated prompt-delayed signal pair is unique to the IBD reaction and a powerful tool to reject backgrounds, which are mostly isolated single events from radio-impurity or cosmic-ray spallation. 
A factor of $10^6$ background rejection with respect to all recorded events has been commonly achieved in reactor antineutrino experiments with this technique. 
Typical analysis cuts to select an IBD event include a prompt energy cut to select the $e^+$ signal, a delayed energy cut to select the neutron capture signal, a time correlation cut, a spatial correlation cut, a muon veto cut to reject events after cosmic-ray muons, and a multiplicity cut to remove ambiguous events when two or more IBD-like pairs are found in a group of coincident signals.

Organic liquid scintillator is the most commonly used detector technology in reactor antineutrino detection. Primary scintillation fluors, such as 2,5-Diphenyloxazole (PPO), are dissolved in an organic solvent, such as linear alkylbenzene (LAB) or pseudocumene (PC). 
Secondary wavelength shifters are often added to the solvent to match the sensitive range of the scintillation photon detectors, such as photomultiplier tubes (PMTs). 
There are many advantages of organic liquid scintillator compared to other neutrino detectors.
Organic liquid scintillator is relatively low-cost and can be scaled up to kilotons of mass. 
The scintillator mostly consists of hydrogen and carbon at an approximate ratio of 1:2, which provides many free protons as the targets for the IBD interaction. 
Liquid scintillator can be uniformly contained in a spherical or cylindrical vessel, which is advantageous for event reconstruction. 
The light yield is high, on the order of $10^4$ photons per MeV of the deposited energy.
The time response of the scintillation light is also fast, on the order of a few nanoseconds. These properties make organic liquid scintillator ideal for the detection of low-energy reactor antineutrinos.

Another advantage of liquid scintillator is its capability to uniformly load isotopes with high neutron-capture cross section to further increase the neutron detection efficiency, which is the delayed signal in the IBD reaction. 
For instance, $^{157}$Gd has the highest thermal neutron capture cross section, about 250 kilobarns, among any stable nuclide. 
A 0.1\% gadolinium-loaded liquid scintillator can largely reduce the neutron capture time from $\sim$200~$\mu$s to $\sim$30~$\mu$s. 
After the neutron is captured on Gd, a cascade of $\gamma$-rays is released that adds up to about 8~MeV, which is much higher than the 2.2~MeV $\gamma$-ray from the hydrogen capture. 
Both the shorter capture time and the higher delayed energy help reduce the accidental coincidence background, mostly from low-energy radioactive impurities. Because of these advantages, Gd-loaded liquid scintillator is a popular choice for medium-sized neutrino detectors up to tens of tonnes~\cite{DayaBay:2012fng, RENO:2012mkc, DoubleChooz:2011ymz}. 
For even larger detectors, factors such as cost, optical transparency, and long-term stability become a concern with Gd-loaded liquid scintillators. 
$^6$Li is another popular isotope for loading because of its final states after neutron capture: $n + {}^6\mathrm{Li} \to {}^4\mathrm{He} + {}^3\mathrm{H}$. 
Unlike $\gamma$-rays, which could scatter far away from the neutron capture location, the alpha and triton particles deposit their energies very close to their generation location (within 1~cm). This is desired for segmented neutrino detectors to better localize the neutrino event position~\cite{Declais:1994su,PROSPECT:2020sxr} and has the potential to reconstruct the direction of the neutrino. 
Furthermore, the slower scintillation light from the alpha and triton, being heavily ionizing particles compared to gamma-rays, allows for more efficient pulse shape discrimination (PSD) to suppress backgrounds.

There are two types of remaining backgrounds after the analysis cuts to select IBD candidates. The first type is the accidental coincidence background: two uncorrelated signals accidentally happen close in time and location and mimic an IBD pair. 
The size of the accidental background depends on the isolated single-event rate in the detector, and is dominated by the radioactive decays from impurities inside and around the detector. Typical radioimpurity sources are those from the $^{232}$Th and $^{238}$U decay chains. The accidental background is usually the largest background below $\sim$2.6~MeV, which is the characteristic $\gamma$-ray energy from the $^{208}$Tl decay. On the other hand, the accidental background can be accurately calculated based on the measured single-event rate and the analysis cuts, and cross-checked by changing the coincident time window, since it is expected to have no time correlation. Therefore, only the statistical uncertainty matters for the accidental background. A fiducial volume cut to select the inner regions of the detector can largely reduce this background because most of the impurities concentrate on the surface of the detector sensors, the containment vessels, and other external structures.

The second type of background is the truly correlated events mimicking an IBD pair. They mostly come from cosmic-ray muons interacting in the detector, producing unstable isotopes that $\beta$-decay together with a neutron in the final states~\cite{KamLAND:2009zwo,DoubleChooz:2015jlf}. The $\beta^{-}$ mimics the prompt signal and the neutron capture mimics the delayed signal in the IBD pair. 
The notable isotopes of this kind are $^{9}$Li and $^{8}$He from $\mu$-C spallation in the liquid scintillator. 
Both isotopes have a relatively long lifetime (a few hundred milliseconds) and are difficult to veto efficiently without causing significant dead time in the detector. 
Cosmic ray induced energetic neutrons outside the detector, often referred to as fast neutrons, can also produce correlated backgrounds when they enter the active detector volume. 
The proton recoil from neutron scattering mimics the prompt signal, and the neutron capture after thermalization mimics the delayed signal in the IBD pair. 
For near-surface detectors, the fast neutron background could be significant since the surface cosmic-ray neutron flux is high, and multiple-neutron background becomes another concern~\cite{PROSPECT:2020sxr}.
These cosmic-ray muon and neutron-induced correlated backgrounds can be estimated from the time and energy distribution of post-muon events, but usually have a larger uncertainty of 20--30\%. Experiments where nearby reactors can be turned off for a period of time benefit from being able to measure the backgrounds directly, although they may still be limited in event statistics. Placing the detector deep underground will greatly reduce this background if such a location is available, but it is often unfeasible for detectors very close to a reactor core. 

After subtracting the estimated backgrounds, the measured number of IBD events per energy bin, $dN/dE$, can be compared with the expected antineutrino energy spectrum $S(E)$ calculated using the following equation:
\begin{equation}\label{eq:measured_spectrum}
S(E) = 
N_p \cdot \epsilon_d(E_\nu) \cdot
\frac{P_{\textrm{sur}}(L/E_\nu)}{4\pi L^2} \cdot
\frac{d\Phi}{dE_\nu}(E_\nu) \cdot 
\sigma_{\textrm{IBD}}(E_\nu) \cdot
R(E, E_\nu), 
\end{equation}
where $N_p$ is the total number of target protons, $\epsilon_d$ is the detection efficiency after all analysis cuts, and $P_{\textrm{sur}}$ is the neutrino oscillation probability for a $\bar\nu_e$ to survive as a $\bar\nu_e$ after traveling a distance $L$ from the reactor.
$P_{\textrm{sur}}$ is a function of $L/E_\nu$ from the theory of neutrino oscillations, which will be briefly reviewed in Sec.~\ref{sec:possible_raa_explanation}. 
$d\Phi/dE_\nu$ is the energy-dependent reactor antineutrino flux and $\sigma_{\textrm{IBD}}$ is the inverse beta decay cross section as previously described in Eq.~\eqref{eq:nu_spectra} and Eq.~\eqref{eq:ibd_xs}, respectively. 
It is worth noting that the experimentally measured energy spectrum is typically presented in the experiment's reconstructed energy, and a detector response function $R(E, E_\nu)$, which is usually nonlinear, is needed to convert the true neutrino energy $E_\nu$ to the reconstructed energy $E$.
As shown in Fig.~\ref{fig:spectrum_ibd}, since the reactor antineutrino flux decreases with energy and the IBD cross section increases with energy, the measured spectrum peaks at about 4~MeV in neutrino energy. 
One commonly defined quantity is the IBD yield per fission, $\sigma_f$, which is the product of reactor antineutrino flux and IBD cross section, integrated over energy and normalized by the total number of fissions: $\sigma_f = (\int (d\Phi/dE_\nu) \cdot \sigma_{\textrm{IBD}} \cdot dE_\nu)/N_f$. 
Experimentally extracted IBD yield ($\sigma_f^{\textrm{exp}}$) using Eq.~\eqref{eq:measured_spectrum} can be compared with theoretical predictions ($\sigma_f^{\textrm{th}}$) using Eq.~\eqref{eq:nu_spectra} and Eq.~\eqref{eq:ibd_xs}. 
The associated experimental uncertainty and theoretical uncertainty are conveniently separately in this way. 
The ratio between the two: $R = \sigma_f^{\textrm{exp}} / \sigma_f^{\textrm{th}}$, is often used as a test of reactor antineutrino flux models. This will be discussed in detail in Sec.~\ref{sec:RAA}.

Finally, we outline the common experimental uncertainties in measuring reactor antineutrino flux. Besides the reactor-related uncertainties as discussed in Sec.~\ref{sec:nu-production}, the detector-related uncertainties come from two major sources: the total number of target protons ($N_p$) and the absolute detection efficiency ($\epsilon_d$). 
In many experiments, a fiducial volume has to be defined in the analysis to select the cleanest region of the detector. 
For those experiments, the target protons can be calculated from the size of the fiducial volume ($V_f$), the density of the detector material ($\rho$), and the mass fraction of hydrogen in the detector ($F_H$). 
Density can be measured to high precision ($\sim$0.1\%) with good temperature control of the environment.  
The mass fraction of hydrogen can be determined to about 1\% using  chemical combustion measurements and mass spectroscopy analysis~\cite{DayaBay:2015kir,DoubleChooz:2022ukr}. 
The uncertainty of the fiducial volume is usually the largest in determining $N_p$, since it is dependent on the accuracy of event position reconstruction. 
A comprehensive calibration system that can cover the whole volume is necessary to reduce its uncertainty. As an example, the KamLAND experiment reached a 1.4\% uncertainty on $V_f$ after a dedicated calibration campaign~\cite{KamLAND:2009ply}. Many later reactor antineutrino experiments used gadolinium-doped liquid scintillator (GdLS) as their detector material. Since the neutron capture signal in GdLS is much higher than most of the radioactive backgrounds, fiducial volume cuts are not necessary. In those experiments, $N_p$ is determined by the total mass of GdLS, which can be measured to high precision during detector filling, and the mass fraction of hydrogen to a precision of about 1\%. 

The uncertainty of the detection efficiency ($\epsilon_d$) has to be determined carefully from detector Monte-Carlo simulation and experimental calibrations. Here we use the Daya Bay experiment as an example~\cite{DayaBay:2016ssb}. The three largest uncertainties in determining $\epsilon_d$ in Daya Bay are the Gd capture fraction, delayed energy cut, and the spill-in correction, with each of them being about 1\%. 
The Gd capture fraction is the fraction of neutron capture on gadolinium (nGd) versus other elements, such as hydrogen. It is determined by the gadolinium concentration in the GdLS, but also includes the spill-out effect where the IBD interaction happens in GdLS, but the neutron travels outside GdLS and captures on other elements. 
The delayed energy cut selects the nGd signal, which peaked at about 8~MeV but has a long tail extending to lower energy. 
This tail is caused by partial deposition of the nGd de-excitation $\gamma$-rays in the scintillator. The shape of the tail is dependent on the number and energy distribution of the $\gamma$-rays. 
The spill-in effect describes an extra contribution to the GdLS-volume events, about 5\% in Daya Bay, where the IBD interaction happens outside the GdLS volume (in Daya Bay the volume outside GdLS is the regular liquid scintillator), but the neutron travels inside the GdLS volume and captures on Gd. 
The spill-in and spill-out effects both depend on the neutron transport properties. While all of these factors are modeled in the detector Monte-Carlo simulation, there could be large differences when selecting different simulation models, therefore MC-data comparison is crucial in correctly assessing the detector efficiency $\epsilon_d$. 
Experiments typically use multiple calibration data sets for the MC-data comparison, such as radioactive gamma and neutron sources deployed along the center z-axis and edges of the detector volume, as well as uniformly distributed events from cosmic-ray interactions such as spallation neutrons and $^{12}$B events. 
As an example, the reactor and detector related systematic uncertainties associated with determining reactor antineutrino flux in Daya Bay are summarized in Table~\ref{table:flux_unc}.

\begin{table*}[ht!]
\caption{\label{table:flux_unc}  Systematic uncertainties associated with determining reactor antineutrino flux in the Daya Bay experiment. The uncertainties are divided into either reactor related or detector related. Table compiled from Ref.~\cite{DayaBay:2016ssb}.  }
\begin{center}
\begin{tabular}{|c|c|c|c|}
\hline
\hline
Reactor & uncertainty  & Detector  & uncertainty  \\ \hline
power & 0.5\%  & target protons & 0.9\%  \\
energy/fission & 0.2\%  & Gd capture fraction  & 0.9\%  \\
fission fraction & 0.6\%  & delayed energy cut & 1.0\%  \\
spent fuel & 0.3\%  & spill-in correction & 1.0\%  \\
isotopic spectrum & $>2\%$  &  & \\
\hline
\hline
\end{tabular}
\end{center}
\end{table*}

\subsection{Calculation of isotopic reactor antineutrino flux}\label{sec:sub:flux_calculation}

We now come back to the calculation of isotopic reactor antineutrino flux, $d\phi_i/dE_\nu$ for the $i$th fission isotope in Eq.~\eqref{eq:nu_spectra}, which is the most challenging part in a reactor antineutrino flux model. There are two principal ways to calculate $d\phi_i/dE_\nu$, the summation method and the conversion method, each with its own advantages and challenges.

The summation method (see Refs.~\cite{Davis:1979gg,Vogel:1980bk}) is straightforward conceptually. For each fission isotope, $^{235}$U, $^{238}$U, $^{239}$Pu, or $^{241}$Pu, its time-dependent fission yields $Y_n(Z, A, t)$, that is, the number of fission products per fission for daughter isotope $n$ with atomic number $Z$ and mass number $A$ at time $t$, is known and can be obtained from a nuclear database. 
Combining this with the experimentally determined $\beta$-decay branching ratios $b_{n,m}$ for each fission product $n$ and branch $m$, and a normalized shape function of each $\beta$-decay $P(E_\nu, E_0, Z)$ with Q-value $E_0$ and antineutrino energy $E_\nu$, the isotopic reactor antineutrino flux can be calculated through a summation as follows:
\begin{equation}\label{eq:nuspec_summation}
 \frac{d\phi_i}{dE_\nu} = \sum_n Y_n(Z, A, t) \cdot
\left( \sum_m b_{n,m} \cdot P(E_\nu, E_0, Z) \right),
\end{equation}
where $n$ goes through each fission daughter product and $m$ goes through each $\beta$-decay branch. 
Commonly used nuclear databases for fission yields and $\beta$-decay data are ENDF~\cite{ENDF} from the USA, JEFF~\cite{JEFF} from Europe,  JENDL~\cite{JENDL} from Japan, CENDL~\cite{CENDL} from China, and ROSFOND~\cite{ROSFOND} from Russia. Generally, there are tens of thousands of $\beta$-decay branches involved in the summation, therefore, knowledge about the underlying $\beta$-decay theory is essential. 
The beta decay process was first theorized by Fermi in 1933, more than thirty years after its experimental evidence. 
In Fermi’s model, the beta decay proceeds through the emission of a pair of leptons ($e^-$ and $\bar\nu_e$ in the case of $\beta^-$ decay) with antiparallel spins and a total angular momentum ($L$) equal to zero. 
This type of decays is called Fermi transition and was complemented later by the discovery of transitions where the spins of the electron and the antineutrino are parallel making a total spin equal to one, called Gamow--Teller transitions. 
Both Fermi and Gamow--Teller transitions are called allowed transitions ($L=0$) and they constitute the most probable transitions in a beta decay. The calculation of the shape of the energy spectrum of the electron or the antineutrino for allowed transitions can be performed without the knowledge of nuclear structure of the involved nuclei.
Less probable beta transitions with the exchange of an angular momentum different from zero ($L>0$) were evidenced later and were called forbidden transitions historically. In the case of these forbidden transitions, a shape factor enters the expression of the electron or antineutrino energy spectrum that has to be determined using nuclear structure microscopic models, requiring that the spin and parity of the involved levels are known. 
These forbidden transitions represent about one third of the beta transitions for the fission products, therefore contributing importantly to the reactor antineutrino energy spectrum.

There are several challenges with the summation method. The nuclear databases for fission yields and $\beta$-decay are still incomplete. Some beta decay branching ratios are poorly measured or even unknown, especially for the isotopes with many decay branches. 
Some experimental data are known to have a bias known as the Pandemonium effect~\cite{Hardy:1977suw,Fallot:2012jv}, in particular for $\beta$-decays with large Q-values. 
The $\beta$-decay shape function $P(E_\nu, E_0, Z)$, even for allowed transitions, requires corrections due to the electromagnetic interaction, finite size effects or induced currents, which are not trivial to derive~\cite{Huber:2011wv,HayenSeverijns}. Moreover, about 25\% of the $\beta$-decays are first forbidden decays that change parity, and their individual shape functions are difficult to calculate. 
Because of these, it is difficult to quantitatively evaluate the uncertainties associated with the summation method and generally a 10--20\% total uncertainty is assigned to the prediction, making it undesirable for comparison with experimental measurements.
In the last decade, much progress has been made to overcome the difficulties and improve the summation method. More details about the summation method and its new advancement will be reviewed in Sec.~\ref{sec:new_calculation}.

The other method, the conversion method, uses the experimentally measured cumulative electron spectra from the $\beta$-decays 
of fission products of the main fissile isotopes.  
The electron spectra associated with $^{235}$U, $^{239}$Pu, and $^{241}$Pu were measured with the magnetic beta spectrometer 
BILL~\cite{MAMPE1978127} at the ILL High Flux Reactor in Grenoble for the thermal neutron fission in the 
1980s~\cite{Schreckenbach:1981wlm,VonFeilitzsch:1982jw, Schreckenbach:1985ep, Hahn:1989zr}. 
In these ILL beta spectra measurements, thin actinide foils of
fissile isotope compounds (e.g.~$^{235}$UO$_2$) were evaporated on a small nickel foil and covered by another nickel foil. This target 
was then irradiated in a high flux of thermal neutrons for tens of hours, during when fission and the subsequent beta activities above 2~MeV 
reached an equilibrium. 
The fission products were stopped by the Ni foil while the beta particles emerged and were measured in the spectrometer magnets 13 meters away. Electron spectra were recorded covering the energy range of 1--10~MeV, in steps of 50~keV. 
Backgrounds were measured by replacing the fissile target with a similar layer of non-fissile material. 
Calibration of the spectrometer was performed with conversion electron sources or $(n, e^-)$ reactions on $^{207}$Pb, $^{197}$Au, 
$^{113}$Cd and $^{115}$In targets, providing calibration points up to 7.37~MeV.
The uncertainty of the beta spectrum measurement is dominated by the normalization uncertainty below 8~MeV, while at higher energy the statistical uncertainties dominate. 
The beta spectrum associated with $^{238}$U fission was not measured until 2014 at FRM-II in Garching~\cite{Haag:2013raa} because it requires a fast neutron flux facility.

% taken from the later sections and to be merged here ...
%The converted spectra are based on a unique measurement performed at ILL with the BILL spectrometer~\cite{MAMPE1978127} using thin actinide foils exposed to a well known thermal neutron flux. 
%The irradiation time of the targets was 12 hours to two days.
%Two measurements of the energy spectrum of the electrons coming from the thermal fission of $^{235}$U have been performed 
%with two different normalizations because of the re-evaluation of one of the calibration cross sections between the two 
%measurement periods. It is the second measurement realized for a 12 hours irradiation that is used by the neutrino experiments 
%of the reactors.

These electron spectra are then converted into antineutrino spectra given that the electron and the antineutrino share the total energy of each $\beta$-decay. While this conversion is trivial when individual $\beta$-decay branches involved are known, in reality only the total electron spectrum is measured. Therefore, about 30 virtual branches with equidistant end-point energy spacing are used to fit the total electron spectrum and determine their branching ratios~\cite{VonFeilitzsch:1982jw, Schreckenbach:1985ep, Hahn:1989zr}. Each virtual branch is given an average charge $Z$ based on its end-point energy from a predetermined weighting function. 
The fit assumes the allowed $\beta$-decay spectral shape: $P(E_\nu, E_0, Z)$ as defined in Eq.~\eqref{eq:nuspec_summation}. After the fit, the conversion to antineutrino spectrum is performed on each virtual branch and summed together. 
Since this conversion procedure does not require precise knowledge of fission yields and $\beta$-decay schemes and is constrained by the experimentally determined total electron spectrum, it is believed to be much more precise than the summation method and could reach $\sim$1\% uncertainty on the antineutrino spectra~\cite{Vogel:2007du}. 

However, there are several concerns about the conversion method as well. First, the method relies on the experimental measurements of the cumulative electron spectra from fission isotopes, but these measurements have not been independently repeated or verified at other facilities besides ILL. Any biases in the ILL measurements would be propagated to the converted antineutrino spectra. 
Second, all virtual branches assume an allowed $\beta$-decay shape, but this shape needs corrections for several nuclear effects such as the finite size of the nucleus and the weak magnetism. The presence of many non-unique first forbidden decays introduce additional shape uncertainties that are difficult to evaluate.
Third, the method relies on the average charge $Z$ as a function of the end-point energy, which by itself is a simple fit to the summation calculation to reflect the fact that light fission fragments usually have higher Q values. Therefore, it still carries the uncertainty from the summation method, which could impact the antineutrino spectra~\cite{Vogel:2007du}. 
% These concerns later led to new development of the conversion method, re-calculations of the reactor antineutrino flux in the form of Huber-Mueller model, and re-evaluation of the uncertainties associated with the conversion method. Details will be reviewed in Sec.~\ref{sec:huber-mueller} and \ref{sec:possible_raa_explanation}.

For about 20 years from around 1990 to 2010, the default reactor model to predict isotopic reactor antineutrino flux and compare with experimental measurements was the ILL-Vogel model, which uses the conversion method to predict $^{235}$U, $^{239}$Pu, and $^{241}$Pu flux from the ILL electron spectrum measurements~\cite{VonFeilitzsch:1982jw, Schreckenbach:1985ep, Hahn:1989zr}, and the summation method to predict $^{238}$U flux~\cite{Vogel:1980bk} because the electron spectrum from $^{238}$U fission was not measured until much later~\cite{Haag:2013raa}. 
The larger uncertainty in the summation method is mitigated by the smaller fission fraction of $^{238}$U in a typical commercial reactor. 
The ILL-Vogel model prediction is in a good agreement with a set of about 20 reactor antineutrino experiments performed in the 1980--1990s, which provided confidence in the procedures involved in the calculation of reactor antineutrino flux. Details about these reactor antineutrino experiments are reviewed in Sec.~\ref{sec:model-exp-compare}. The landscape changed drastically in 2011 when two new reactor antineutrino flux calculations, the Mueller~\cite{Mueller:2011nm} and the Huber~\cite{Huber:2011wv} model, appeared to address some of the concerns regarding the ILL-Vogel model. This re-evaluation of reactor antineutrino flux eventually led to the reactor antineutrino anomaly and over 10 years of effort attempting to resolve it both experimentally and theoretically, which will be reviewed in the remaining part of this article.

%% file: RAA.tex
\section{The Reactor Antineutrino Anomaly}\label{sec:RAA} 
The reactor antineutrino anomaly (RAA)~\cite{Mention:2011rk} refers to a deficit of the measured antineutrino
rate in short-baseline reactor experiments ($L < 2$ km) with respect to a new set of 
calculations of the antineutrino flux in 2011~\cite{Mueller:2011nm,Huber:2011wv}, namely the Huber-Mueller model. 
Compared to the previous calculations~\cite{VonFeilitzsch:1982jw,Schreckenbach:1985ep,Hahn:1989zr,Vogel:1980bk},
the new calculations together with the updated neutron lifetime~\cite{pdg2022} 
predicts $\sim$6\% higher antineutrino rate. The inclusion of long-lived fission 
isotopes (i.e.~the non-equilibrium effect) in the prediction 
contributes to 1--2\% higher rate. In comparison, the reduced neutron lifetime leading to a larger IBD
cross section contributes to slightly less than 1\% higher predicted rate. The rest of the rate increase comes from a few nuclear effects re-evaluated in the new models.
The calculated deficit cannot be explained by the quoted uncertainty of the Huber-Mueller model~\cite{Mueller:2011nm,Huber:2011wv}, which is about 2\%. 
The RAA has triggered extensive investigations including new antineutrino flux and energy spectrum measurements from modern reactor antineutrino experiments, searches for a sterile neutrino with 
mass around $\sim$1~eV or higher, a new measurement of $\beta$-spectrum ratio between 
$^{235}$U and $^{239}$Pu fission products, critical evaluations of the uncertainties associated with the conversion method, new measurements of relevant beta spectra with the total absorption gamma-ray spectroscopy (TAGS) technique, and new development in the summation method.

\subsection{Re-evaluation of reactor antineutrino flux: the Huber-Mueller model}\label{sec:huber-mueller}

\begin{figure}[bhtp]
  \begin{centering}
    \includegraphics[width=0.45\textwidth]{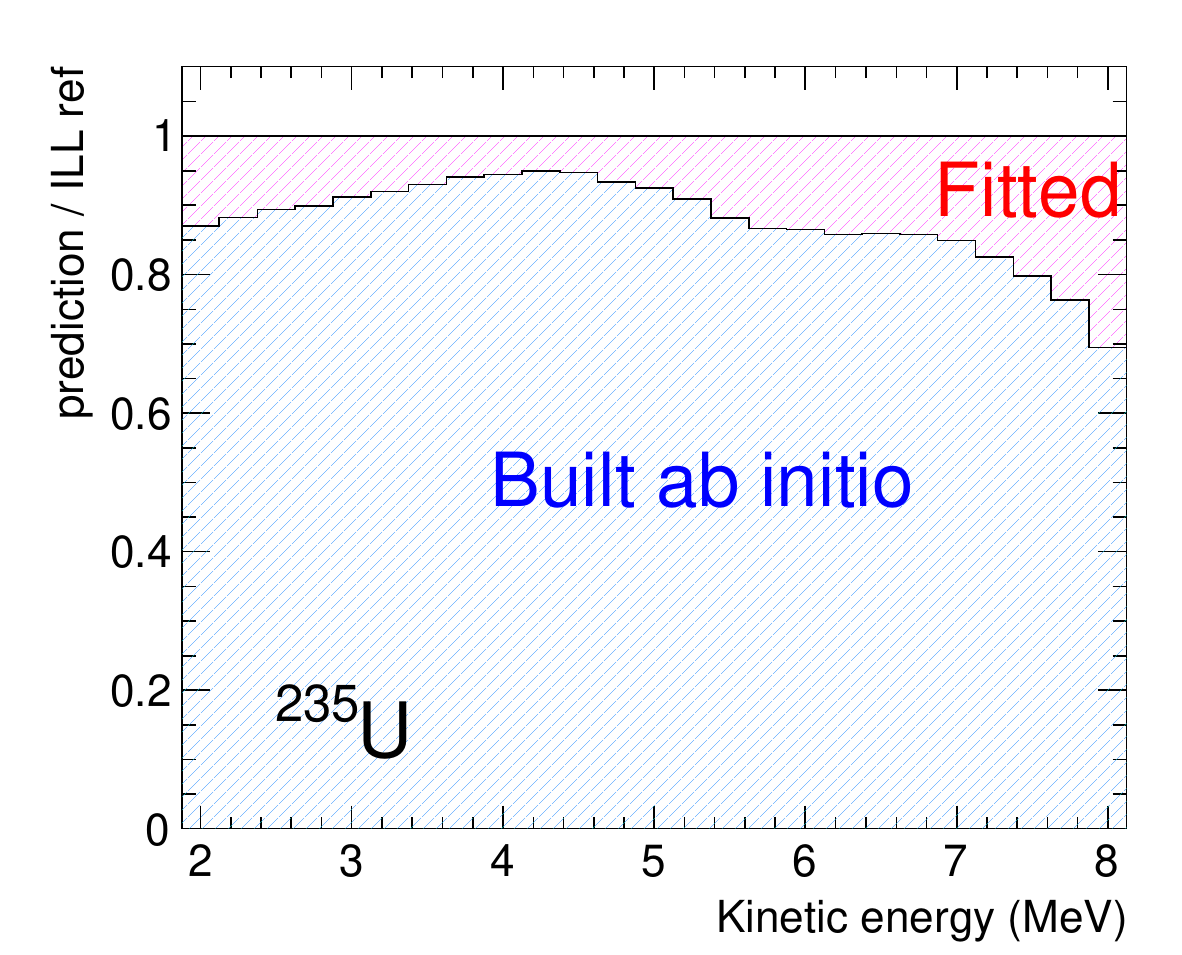} 
    \includegraphics[width=0.48\textwidth]{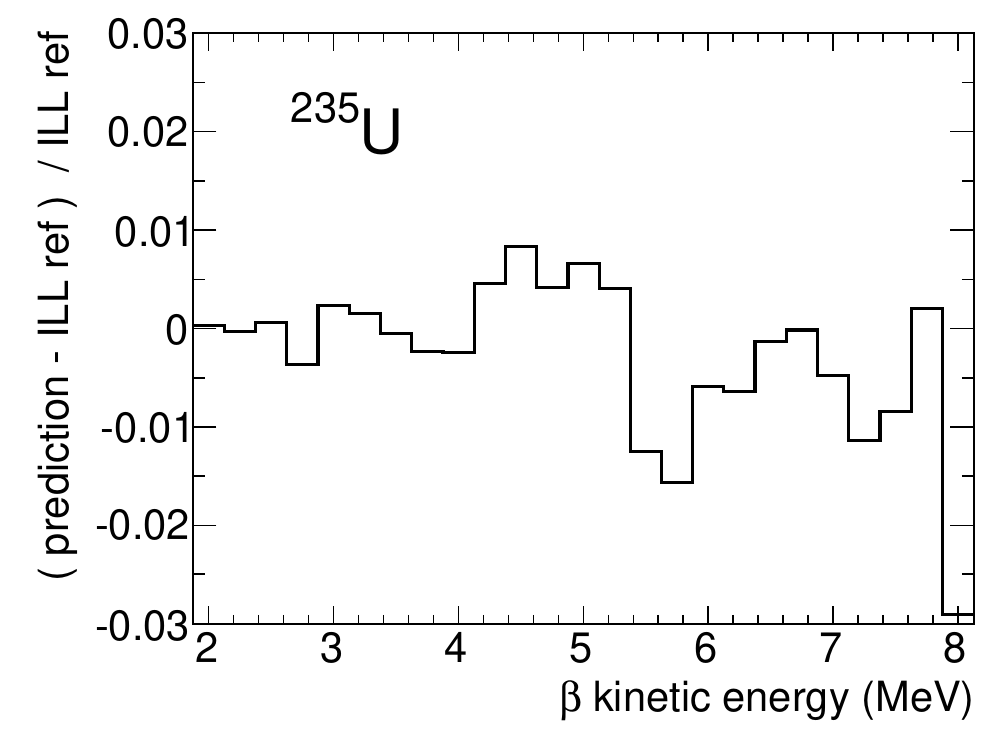}
      \put(-350,-10){a)}
        \put(-100,-10){b)}
    \par\end{centering}
    \caption{\label{fig:Mueller_spectrum} 
    Results from the Mueller calculation~\cite{Mueller:2011nm}. (Left) Comparison of the prediction from the summation method, which 
    is labeled as ``Built ab initio'', with the measured ILL electron spectrum reference. The missing contribution, which is labeled as 
    'Fitted', is fitted using a set of 5 effective virtual $\beta$-branches.
    (Right) Relative difference between the fitted result
    with respect to the ILL $^{235}$U electron spectra. Figures taken from Ref.~\cite{Mueller:2011nm}.}
\end{figure}

In 2011, accompanying a new generation of short-baseline reactor antineutrino experiments including
Double Chooz~\cite{DoubleChooz:2011ymz}, Daya Bay~\cite{DayaBay:2012fng}, and RENO~\cite{RENO:2012mkc} 
aiming at determining the smallest neutrino mixing angle $\theta_{13}$, a set of re-evaluations of the 
reactor antineutrino flux took place. Mueller {\it et al.} proposed a new conversion 
method~\cite{Mueller:2011nm} to include the updated results of the summation method with information of all $\beta$-branches from the ENSDF nuclear 
database~\cite{ENSDF}. As shown in Fig.~\ref{fig:Mueller_spectrum}a, this new 
conversion method restricts the usage of effective virtual branches to  fit only the missing 
$\sim$10\% contribution of the difference between the updated results of the 
summation method and the reference ILL electron data. This approach makes the distributions of 
$\beta$-branches close to the physical ones, so that the corrections related to the Coulomb and 
weak magnetism can be applied at each branch level. For the missing contribution to match the ILL 
electron spectrum, five effective $\beta$-branches with a nuclear charge $\bar{Z}$ of 46 and 
allowed spectral shapes were assumed. For each effective $\beta$-branch, the normalization and the end-point energy
are two free parameters determined by the fit. Figure~\ref{fig:Mueller_spectrum}b
shows the residual difference between the final fitting result and
the measured ILL $^{235}$U electron spectra. The resulting predicted antineutrino spectrum exhibits 
a mean normalization upward shift of about 3\% with respect to that from Ref.~\cite{Schreckenbach:1985ep}.

This net 3\% upward shift in the energy-integrated antineutrino flux was further confirmed by an independent 
calculation by Huber~\cite{Huber:2011wv}. In this work, instead of relying on $\beta$-branches from a nuclear
database, Huber fully employed the virtual effective $\beta$-branches to fit the reference ILL electron 
data and to predict the antineutrino energy spectra. Specifically, 30, 23, and 25 virtual branches are
used to fit the ILL electron spectra for $^{235}$U, $^{239}$Pu, and $^{241}$Pu, respectively. Compared 
to earlier works, Huber improved on the treatment of several nuclear effects in modeling $\beta$-decay spectrum following Ref.~\cite{WILKINSON1989378}:
\begin{equation}\label{eq:beta_spectra}
N_\beta\left(W\right) = Kp^2\left(W-W_0\right)^2 F\left(Z,W\right) L_0\left(Z,W\right) C\left(Z,W\right) S\left(Z,W\right) G_\beta\left(Z,W\right) \left(1+\delta_{\rm WM} W\right),
\end{equation}
where $W=E/\left(m_ec^2\right)+1$ with $E$ and $m_e$ being the energy and mass of 
the electron, $W_0$ is the value of $W$ at the endpoint, and $Z$ is the empirical mean proton number of fission fragments 
calculated from nuclear database for each virtual branch as a function of $W_0$. $K$ is a normalization constant, $p$ is the 
magnitude of electron's momentum. $F\left(Z,W\right)$ is the Fermi function accounting for the fact that 
the outgoing electron is moving in the Coulomb field of the nucleus, and is derived from the solution of the Dirac
equation for a point-like and infinitely heavy nucleus. $L_0\left(Z,W\right)$ and $C\left(Z,W\right)$ are the 
corrections for electromagnetic and weak interactions, respectively, from the finite size of the nucleus.
$S\left(Z,W\right)$ accounts for the screening of the nuclear charge by all the electrons in the atomic bound
state, which effectively reduces the charge seen by the outgoing electron. $G_\beta\left(Z,W\right)$ accounts
for the radiative corrections because of the emission of virtual and real photons by the charged particles present 
in $\beta$-decay. The last term containing $\delta_{\rm WM}$ accounts for the effect of weak magnetism. The exact format
of these terms can be found in Ref.~\cite{Huber:2011wv}.
The neutrino energy spectrum is obtained by replacing $W\rightarrow W_0-W$ and $G_{\beta} \rightarrow G_\nu$ in
Eq.~\eqref{eq:beta_spectra}. 

\begin{figure}[bhtp]
  \begin{centering}
    \includegraphics[width=0.95\textwidth]{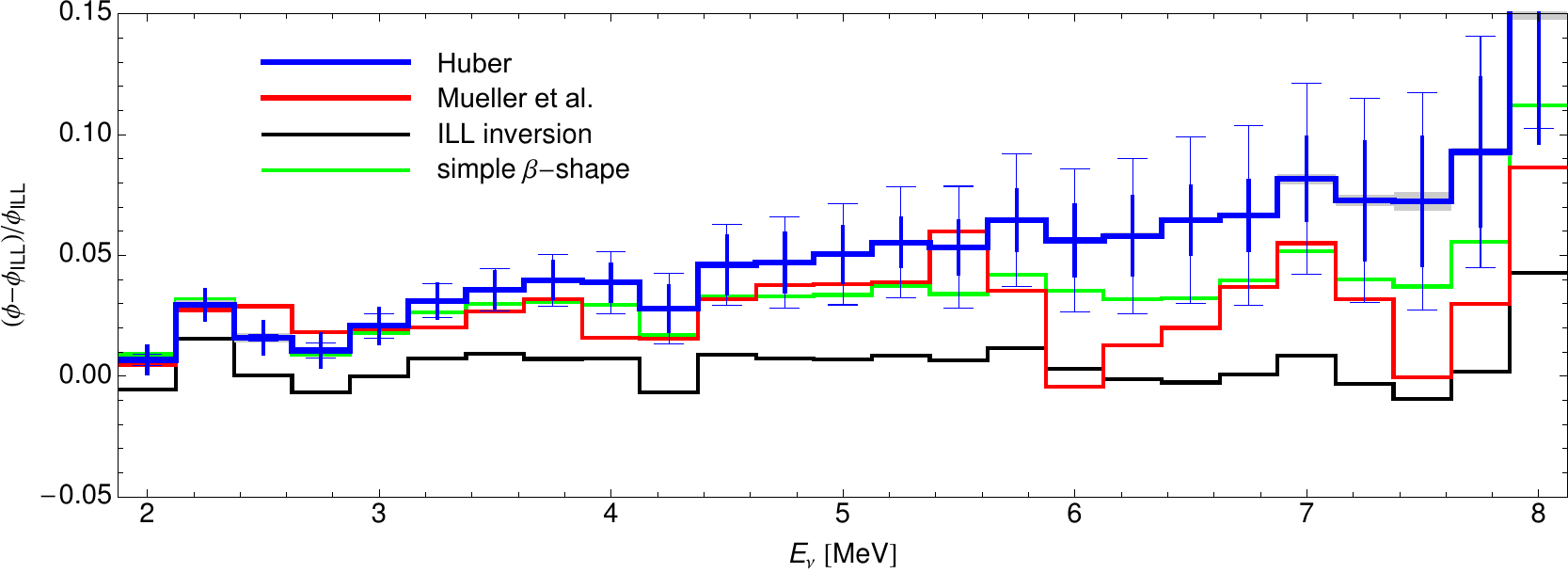}
    \par\end{centering}
    \caption{\label{fig:HM_spectrum} 
     Comparison of $^{235}$U $\nu$ energy spectra from different conversion methods: red line labeled as ``Mueller et al.'' from Ref.~\cite{Mueller:2011nm}, black line labeled as ``ILL inversion'' from Ref.~\cite{Schreckenbach:1985ep}, and the 
     rest from Ref.~\cite{Huber:2011wv}. 
     For the blue line labeled as ``Huber'', the thin error bars show the theoretical uncertainties from the effective nuclear charge $\bar{Z}$ and weak magnetism. 
     The thick error bars are the statistical errors, whereas the light gray boxes show the error from the applied 
     bias correction. The green line, referred to as ``simple $\beta$-shape'', shows the result of Ref.~\cite{Huber:2011wv} if the 
     same description of $\beta$-decay as in Ref.~\cite{Mueller:2011nm} is used. 
     % The black line, referred to as ILL inversion, shows the result of Ref.~\cite{Huber:2011wv}, if the procedure outlined in Ref.~\cite{Schreckenbach:1985ep},is completely followed. 
     Figure modified and taken from Ref.~\cite{Huber:2011wv}.  }
\end{figure}

Figure~\ref{fig:HM_spectrum} compares the predicted $^{235}$U $\nu$ energy spectra from different conversion 
strategies. 
It turns out the inclusion of $\beta$-branches from nuclear databases as in Ref~\cite{Mueller:2011nm} does not make a large difference compared to the case where only virtual branches are used. 
The net 3\% upward shift with respect to the original ILL inversion is traced back to the form of $\bar Z(W_0)$ in Ref.~\cite{Schreckenbach:1985ep} and the fact that all corrections 
to an allowed $\beta$-spectrum shape were applied in an average fashion (i.e. independent of $\bar{Z}$ of each virtual branch). 
This is consistent with the findings in Ref.~\cite{Mueller:2011nm} where the predicted 3\% shift in the low energy region is the result of the branch-level corrections for Coulomb and weak magnetism and the dominate source of the shift in the higher energy region comes from the parameterization of the charge distribution associated with the virtual $\beta$-branches. 
The improved treatment of $\beta$-decay spectrum as in Eq.~\eqref{eq:beta_spectra} compared to that in Ref.~\cite{Mueller:2011nm}
is responsible for a sizeable fraction of difference at energies above 5 MeV.

%% discussion of model uncertainties ... 
For $^{235}$U, $^{239}$Pu, and $^{241}$Pu, the net effect from Huber's conversion method on the integrated detected antineutrino rate is an increase by 3.7\%, 4.2\%, 4.7\%, respectively~\cite{Huber:2011wv}. 
The uncertainties of this conversion method are shown in Fig.~\ref{fig:HM_spectrum}, which include
i) statistical uncertainties, ii) theoretical uncertainties associated with various corrections, and iii) unfolding (or bias)
uncertainties. The total uncertainties from Ref.~\cite{Huber:2011wv} associated with the integrated detected antineutrino rate is estimated to be 1.5\%, 1.5\%, and 1.6\% for $^{235}$U, $^{239}$Pu, and $^{241}$Pu, respectively. 

For the $^{238}$U results obtained from the summation method in Ref.~\cite{Mueller:2011nm}, 
the net effect on the integrated detected antineutrino rate is an increase by 9.8\% compared to that in 
Ref.~\cite{Vogel:1980bk}. The uncertainties associated with the $^{238}$U include 
i) those associated with the nuclear database, ii) the assumption of treating all forbidden
transitions as the lowest possible unique type, which is estimated by comparing the results assuming that all transitions 
are allowed, iii) corrections associated with Coulomb and weak magnetism, and iv) missing information. The uncertainties 
are dominated by the missing information, which is 10--20\%.  

In addition to the improvements in the conversion method, the inclusion of long-lived fission isotopes, which is commonly referred to as the non-equilibrium or off-equilibrium correction, also makes a sizeable change to the predicted reactor antineutrino flux~\cite{Mueller:2011nm}. 
This correction is due to the fact that the ILL
spectra were measured after a short irradiation time in a quasi-pure thermal neutron flux between 12 and 43 hours,
depending on the isotopes. For reactors, the irradiation time is typically one reactor refueling cycle ($\sim$1 year). 
There are about 10\% fission products that have a $\beta$-decay lifetime longer than a few days, and they would enhance the measured rates compared to those in ILL. 
Furthermore, in a standard light-water power reactor, the neutron energy spectrum
contains more epithermal and fast neutrons than those in the ILL measurements. 
Authors of Ref.~\cite{Mueller:2011nm}
studied this effect with the MURE simulation~\cite{mure} of a PWR reactor assembly and their results
are shown in Table~\ref{tab:off_eq2}, which were further validated by the results from the Chooz 
experiment~\cite{CHOOZ:1999hei,CHOOZ:2002qts}. These non-equilibrium corrections need to be evaluated for each 
experiment and typically leads to 1--2\% increase in the predicted antineutrino rate with a relative uncertainty 
of about 30\%. 

\begin{table}[htp]
\caption{
Relative off-equilibrium correction (in \%) to be applied to the 
reference antineutrino spectra for several energy bins and several irradiation-time periods much longer than the ILL
references (12~h for U and 36~h for Pu). Effects of neutron captures on fission products are included and computed 
using the simulation of a PWR fuel assembly with the MURE code. Table taken from Ref.~\cite{Mueller:2011nm}. 
\label{tab:off_eq2}}
\begin{center}
\begin{tabular}{|c|c|c|c|c|c|}
\hline \hline
\multicolumn{6}{|c|}{$^{235}$U} \\   
\hline
  & 2.0 MeV & 2.5 MeV & 3.0 MeV & 3.5 MeV & 4.0 MeV \\ 
\hline
36 h    & 3.1 & 2.2 & 0.8 & 0.6 & 0.1 \\
100 d   & 4.5 & 3.2 & 1.1 & 0.7 & 0.1 \\
1E7 s   & 4.6 & 3.3 & 1.1 & 0.7 & 0.1 \\
300 d   & 5.3 & 4.0 & 1.3 & 0.7 & 0.1 \\
450 d   & 5.7 & 4.4 & 1.5 & 0.7 & 0.1 \\
\hline
\hline
\multicolumn{6}{|c|}{$^{239}$Pu} \\   
\hline
  & 2.0 MeV & 2.5 MeV & 3.0 MeV & 3.5 MeV & 4.0 MeV \\ 
\hline
100 d & 1.2 & 0.7  & 0.2 & $<0.1$  & $<0.1$ \\
1E7 s & 1.3 & 0.7  & 0.2 & $<0.1$  & $<0.1$ \\
300 d & 1.8 & 1.4  & 0.4 & $<0.1$ & $<0.1$ \\
450 d & 2.1 & 1.7  & 0.5 & $<0.1$  & $<0.1$ \\
\hline
\hline
\multicolumn{6}{|c|}{$^{241}$Pu} \\   
\hline
  & 2.0 MeV & 2.5 MeV & 3.0 MeV & 3.5 MeV & 4.0 MeV \\ 
\hline
100 d & 1.0 & 0.5 & 0.2 & $<0.1$ & $<0.1$ \\
1E7 s & 1.0 & 0.6 & 0.3 & $<0.1$ & $<0.1$ \\
300 d & 1.6 & 1.1 & 0.4 & $<0.1$ & $<0.1$ \\
450 d & 1.9 & 1.5 & 0.5 & $<0.1$ & $<0.1$ \\
\hline \hline
\end{tabular}
\end{center}
\end{table}

With this round of re-evaluation of reactor antineutrino flux, the combined result of $^{235}$U, $^{239}$Pu, and $^{241}$Pu from the 
conversion method in Ref.~\cite{Huber:2011wv} and $^{238}$U from the summation method in Ref.~\cite{Mueller:2011nm}, together with the updated neutron lifetime~\cite{pdg2022} and the inclusion of non-equilibrium effect from long-lived fission isotopes,
is commonly referred to as the Huber-Mueller model. Because of the improvements as described in this section, the Huber-Mueller model has replaced the previous ILL-Vogel model and  become the default reactor antineutrino flux model used by many reactor antineutrino experiments since 2011.

\subsection{Comparison with global data from reactor antineutrino experiments}\label{sec:model-exp-compare}

With this new set of calculations of the antineutrino flux in 2011, Mention 
{\it et al.} first re-evaluated the comparison between the measured antineutrino rate in 
existing short-baseline reactor experiments ($L < 2$ km) and the predicted rate from Ref.~\cite{Mueller:2011nm}, and found a
a global deficit of about 6\% in the measured rate at a significance of $2.5~\sigma$~\cite{Mention:2011rk}.  This finding was coined by the authors as the Reactor Antineutrino Anomaly (RAA). 
The initial calculation of this deficit in Ref.~\cite{Mention:2011rk} was biased toward a larger 
value by about 1.5\% because of an improper treatment of flux uncertainties 
in the covariance matrix, as demonstrated in Ref.~\cite{DAgostini:1993arp,Zhang:2013ela}. 
Later analyses~\cite{Gariazzo:2017fdh,Giunti:2021kab} improved upon this treatment, 
updated the reactor antineutrino flux prediction to the Huber-Mueller model~\cite{Mueller:2011nm,Huber:2011wv}, 
and included more recent results from modern reactor antineutrino experiments. 
The latest evaluation of the measured reactor antineutrino rate to prediction ratio from Ref.~\cite{Giunti:2021kab}
is: 
\begin{equation}\label{eq:RAA_R}
\bar{R} = 0.936 ^{+0.024} _{-0.023} \approx 0.936 \pm 0.005~({\rm exp.}) \pm 0.023~({\rm model}),
\end{equation}
where the RAA significance becomes $2.6~\sigma$ and the uncertainty is dominated by the model uncertainty in the prediction.

\begin{table*}[htb!]
\caption{\label{table:results} Tabulated results of 27 experiments with their detection technology, fission fractions, ratio of measured antineutrino rate to the prediction from the Huber-Mueller model, the experimental uncertainties as well as the estimated correlated uncertainties among experiments, the baseline to reactors, and the year of the publications. Experiments are categorized into different groups with horizontal lines. Table modified from Ref.~\cite{Giunti:2021kab}. }
\begin{center}
\begin{tabular}{|c|c|c|c|c|c|c|c|c|c|c|c|c|}
\hline
\hline
Ref. & result     & Det. type         & $^{235}$U & $^{239}$Pu & $^{238}$U & $^{241}$Pu & ratio & $\sigma_{exp}$(\%) & $\sigma_{corr}$ (\%)&  L(m)    & Year \\\hline
\cite{Declais:1994ma} & Bugey-4    & $^3$He+H$_2$O   & 0.538     & 0.328      & 0.078     & 0.056      & ~0.927~ & 1.4     & \rdelim\}{2}{20pt}[1.4]      & 15       &  ~1994\\
\cite{Kuvshinnikov:1990ry} & ROVNO91    & $^3$He+H$_2$O   & 0.614     & 0.274      & 0.074     & 0.038      & ~0.940~ & 2.8     &       & 18        &  ~1991\\\hline
\cite{Afonin:1988gx}& ROVNO88-1I & $^3$He+PE       & 0.607     & 0.277      & 0.074     & 0.042      & ~0.902~ & 6.4     & \rdelim\}{2}{20pt}[3.1] \rdelim\}{5}{20pt}[2.2]     & 18      &  ~1988\\
\cite{Afonin:1988gx}& ROVNO88-2I & $^3$He+PE       & 0.603     & 0.276      & 0.076     & 0.045      & ~0.931~ & 6.4     &        & 18       &  ~1988\\
\cite{Afonin:1988gx}& ROVNO88-1S & Gd-LS             & 0.606     & 0.277      & 0.074     & 0.043      & ~0.956~ & 7.3     & \rdelim\}{3}{45pt}[3.1]       & 18       &  ~1988\\
\cite{Afonin:1988gx}& ROVNO88-2S & Gd-LS             & 0.557     & 0.313      & 0.076     & 0.054      & ~0.956~ & 7.3     &        & 25       &  ~1988\\
\cite{Afonin:1988gx}& ROVNO88-3S & Gd-LS             & 0.606     & 0.274      & 0.074     & 0.046      & ~0.922~ & 6.8     &        & 18        &  ~1988\\\hline
\cite{Declais:1994su} & Bugey-3-I  & $^6$Li-LS      & 0.538     & 0.328      & 0.078     & 0.056      & ~0.930~ & 4.2     & \rdelim\}{3}{20pt}[4.0]      & 15       &  ~1995\\
\cite{Declais:1994su} & Bugey-3-II & $^6$Li-LS       & 0.538     & 0.328      & 0.078     & 0.056      & ~0.936~ & 4.3     &       & 40        &  ~1995\\
\cite{Declais:1994su} & Bugey-3-III& $^6$Li-LS       & 0.538     & 0.328      & 0.078     & 0.056      & ~0.861~ & 15.2    &       & 95        &  ~1995\\\hline
\cite{CALTECH-SIN-TUM:1986xvg} & Goesgen-I  & $^3$He+LS       & 0.619     & 0.272      & 0.067     & 0.042      & ~0.949~ & 5.4     & \rdelim\}{3}{20pt}[2.0] \rdelim\}{4}{20pt}[3.8]      & 38        &  ~1986\\
\cite{CALTECH-SIN-TUM:1986xvg} & Goesgen-II & $^3$He+LS       & 0.584     & 0.298      & 0.068     & 0.050      & ~0.975~ & 5.4     &     & 46       &  ~1986\\
\cite{CALTECH-SIN-TUM:1986xvg} & Goesgen-III& $^3$He+LS       & 0.543     & 0.329      & 0.070     & 0.058      & ~0.909~ & 6.7     &      & 65       &  ~1986\\
\cite{Kwon:1981ua,HOUMMADA1995449} & ILL        & $^3$He+LS       & $\approx$1& -          & -         & -          & ~0.787~ & 9.1     &      & 9        &  ~1981\\\hline
\cite{Vidyakin:1987ue}& Krasn. I   & $^3$He+PE       & $\approx$1& -          & -         & -          & ~0.920~ & 5.2     & \rdelim\}{2}{20pt}[4.1]      & 33        &  ~1987\\
\cite{Vidyakin:1987ue}& Krasn. II  & $^3$He+PE       & $\approx$1& -          & -         & -          & ~0.935~ & 20.5    &       & 92        &  ~1987\\ 
\cite{Vidyakin:1990iz,Vidyakin:1994ut}& Krasn. III & $^3$He+PE       & $\approx$1& -          & -         & -          & ~0.929~ & 4.2     & 0      & 57        &  ~1994\\
\cite{Kozlov:1999ct}& Krasn. IV &  $^3$He+PE       & $\approx$1& -          & -         & -          & ~0.948~ & 3.0     & 0      & 34        &  ~1999\\\hline
\cite{Greenwood:1996pb}& SRP-I      & Gd-LS             & $\approx$1& -          & -         & -          & ~0.934~ & 2.8     & 0 & 18       &  ~1996\\
\cite{Greenwood:1996pb}& SRP-II     & Gd-LS             & $\approx$1& -          & -         & -          & ~0.998~ & 2.9     & 0 & 24       &  ~1996\\\hline
\cite{NUCIFER:2015hdd}&Nucifer     &  Gd-LS              & 0.926    &0.061        & 0.008      & 0.005      & ~1.007~& 10.8     & 0  & 7       & 2016 \\
\cite{CHOOZ:2002qts}& Chooz      & Gd-LS             & 0.496     & 0.351      & 0.087     & 0.066      & ~0.990~ & 3.2     & 0      & $\approx$1000  &  ~1999\\
\cite{Boehm:2001ik}& Palo Verde & Gd-LS             & 0.60      & 0.27       & 0.07      & 0.06       & ~0.991~ & 5.4     & 0      & $\approx$800   &  ~2001\\
\cite{DayaBay:2016ssb}&Daya Bay    & Gd-LS            & 0.564       & 0.304      & 0.076    & 0.056       & ~0.950  & 1.5     & 0     & $\approx$550   & 2016 \\
\cite{RENO:2020dxd}& RENO       & Gd-LS               & 0.571       & 0.300      & 0.073    & 0.056       &~0.941~&  1.6      & 0     & $\approx$411    & 2020 \\
\cite{DoubleChooz:2019qbj}& Double Chooz&  Gd-LS/LS             & 0.520        & 0.333     &0.087      & 0.060      &~0.918~&   1.1     & 0     & $\approx$415     & 2020 \\
\cite{STEREO:2020fvd}& STEREO      &  Gd-LS             & $\approx$1   & -           & -       & -           & 0.948 & 2.5      & 0     & 9-11             & 2020 \\\hline
\hline
\end{tabular}
\end{center}
\end{table*}

\begin{figure}[th!]
  \begin{centering}
    \includegraphics[width=0.98\textwidth]{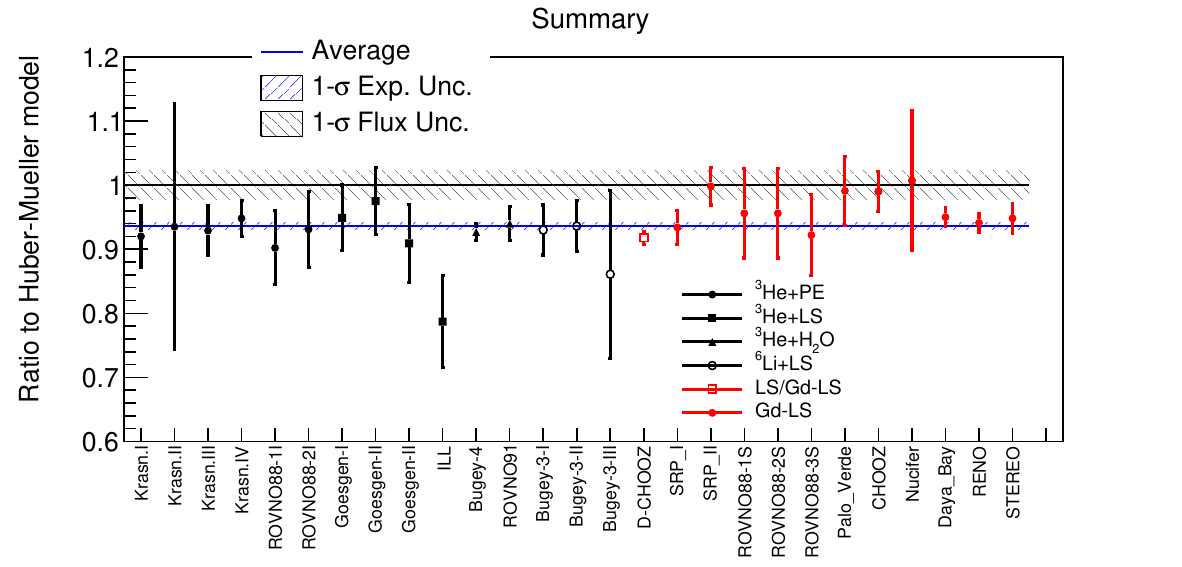}
    \includegraphics[width=0.48\textwidth]{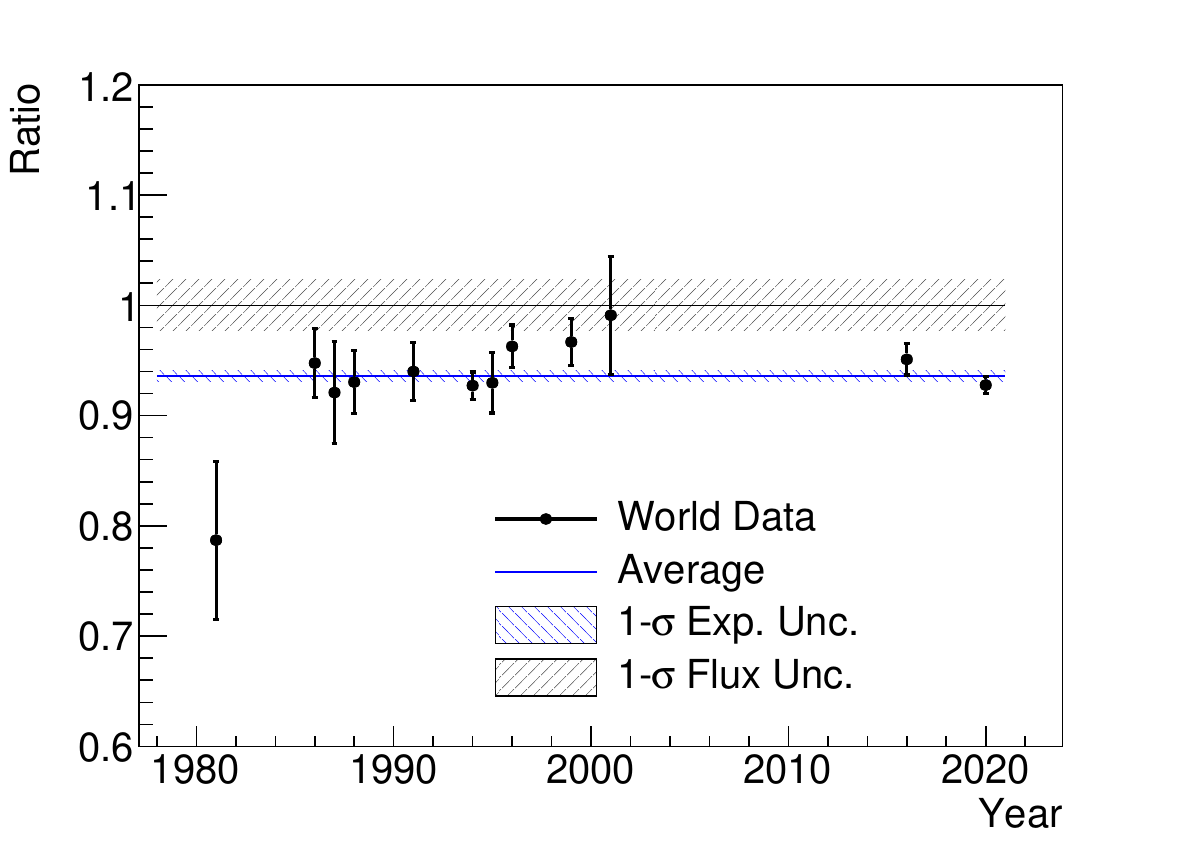}
    \includegraphics[width=0.48\textwidth]{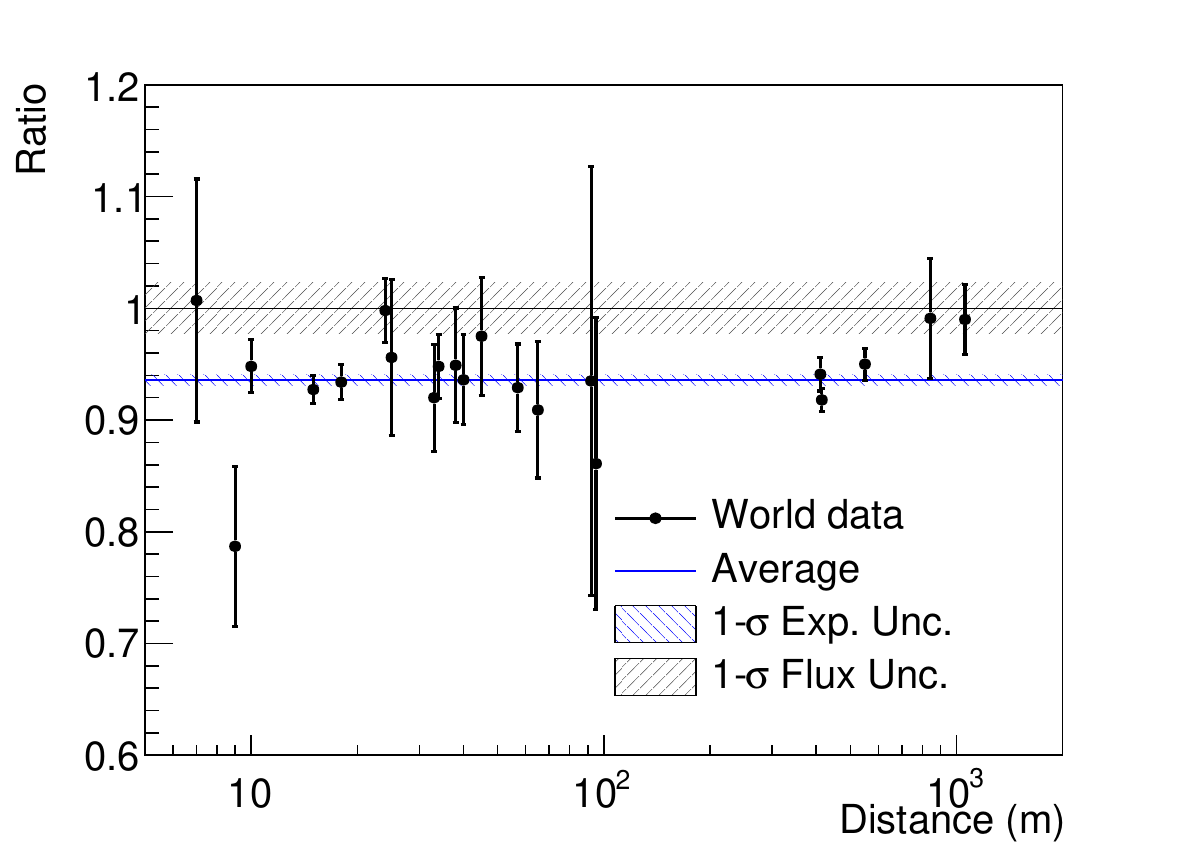}
     \put(-270,190){a)}
      \put(-380,-10){b)}
        \put(-120,-10){c)}
    \par\end{centering}
    \caption{\label{fig:dyb_global_deficit}
      (Top)  The measured reactor $\bar{\nu}_e$ rate from 27 experiments, normalized to the 
       prediction of the Huber--Mueller model~\cite{Huber:2011wv,Mueller:2011nm}.
      The rate is corrected for three-flavor neutrino oscillation at each baseline. The blue shaded
      region represents the global average and its 1$\sigma$ uncertainty. The 2.3\% model
      uncertainty is shown as a band around unity. Measurements with different detection 
      technologies are separately labeled. 
        (Bottom left) The ratio of measured reactor $\bar{\nu}_e$ rate to the prediction 
       of the Huber--Mueller model~\cite{Huber:2011wv,Mueller:2011nm} is shown as a function of the year
       of the corresponding publication. Measurements at the same year are combined for clarity.
       (Bottom right) The ratio of measured reactor $\bar{\nu}_e$ rate to the prediction 
       of the Huber--Mueller model~\cite{Huber:2011wv,Mueller:2011nm} is shown as a function
       of the baseline.  Measurements at the same baseline are combined for clarity. 
      }
\end{figure}

%new average
Table~\ref{table:results}, which is updated from  Ref.~\cite{Giunti:2021kab},
summarizes the key information from 27 experimental measurements of reactor antineutrino rate using the inverse 
beta decay (IBD) channel. Among them, 20 measurements were made before year 2000. The experiments used different detector technologies. 
In the ILL~\cite{Kwon:1981ua,HOUMMADA1995449} and Goesgen~\cite{CALTECH-SIN-TUM:1986xvg} 
experiments, the liquid scintillator target cells were interspaced with $^3$He neutron counters. 
In the Bugey-3 experiment~\cite{Declais:1994su}, the liquid scintillator was loaded with $^6$Li to improve the detection of
neutron captures. In the Bugey-4~\cite{Declais:1994ma} and Rovno91~\cite{Kuvshinnikov:1990ry} experiments, 
only neutron captures were detected. The detector in these two experiments consisted of a water target with 
embedded $^3$He detectors. In the Krasnoyarsk~\cite{Vidyakin:1987ue,Vidyakin:1990iz,Vidyakin:1994ut,Kozlov:1999ct} and Rovno88~\cite{Afonin:1988gx} experiments, again, 
only the total neutron capture rates were measured. The detectors consisted of polyethylene neutron moderators with $^3$He 
neutron counters embedded in them. Finally, the Savannah River experiment (SRP)~\cite{Greenwood:1996pb}, 
and the Rovno88~\cite{Afonin:1988gx} experiment used the Gd-loaded liquid scintillators. After year 2000, two generations 
of experiments aiming at determination of the third neutrino mixing parameter $\theta_{13}$ with baselines
around 1--2~km provided five new measurements using Gd-loaded liquid scintillators. The first generation includes 
Chooz~\cite{CHOOZ:2002qts} and Palo Verde~\cite{Boehm:2001ik} with a single-detector configuration setting 
limits on the $\theta_{13}$ value. The second generation includes Daya Bay~\cite{DayaBay:2016ssb}, 
RENO~\cite{RENO:2020dxd}, and Double Chooz~\cite{DoubleChooz:2019qbj}, which eventually established the non-zero value of $\theta_{13}$~\cite{DayaBay:2012fng,RENO:2012mkc} with the near- and far-detector configurations. The small oscillation-effect (a few \%) driven by $\theta_{13}$ needs to be corrected for these km-baseline experiments~\cite{Zhang:2013ela}. In addition, motivated by the RAA, 
two recent experiments: Nucifer~\cite{NUCIFER:2015hdd} and STEREO~\cite{STEREO:2020fvd} performed new measurements of antineutrino rate
at very-short-baselines of $\sim$10 m. Figure~\ref{fig:dyb_global_deficit} compares the ratio of measured IBD rate to the Huber-Mueller model
prediction from all 27 measurements, categorized by the detector technology (panel a), publication year (panel b),
and baseline (panel c). The global fit of the world average from Eq.~\eqref{eq:RAA_R} and its 1$\sigma$ uncertainty are shown as the blue shaded region. The 2.3\% model uncertainty is shown as the gray band around unity.   

%% higgly enriched vs. low-enrished 
Among these 27 measurements, 9 of them were made using highly enriched uranium (HEU) fuel
in research reactors, where the $^{235}$U fission is the dominating process. 
% The average number of emitted neutrons is 2.44 per $^{235}$U~\cite{IAEA-CRP-STD}. 
The other 18 measurements were made
using low-enriched uranium (LEU) fuel in commercial power reactors, 
where fissions of $^{235}$U, $^{239}$Pu, $^{241}$Pu, and $^{238}$U are involved.
% The average number of emitted neutrons are 2.88~\cite{IAEA-CRP-STD}, 2.95~\cite{IAEA-CRP-STD}, and 2.82~\cite{ENDF} per $^{239}$Pu, $^{241}$Pu, and $^{238}$U fission, respectively. 
The average fission fractions for these 27 measurements 
are included in Table~\ref{table:results}. 
Further investigation regarding RAA in terms of different fission isotopes will be discussed in Sec.~\ref{sec:evolution}.
 
\subsection{Possible explanations of RAA}~\label{sec:possible_raa_explanation}

The observed deficit in Eq.~\eqref{eq:RAA_R} (i.e.~the RAA) cannot be explained by the combined experimental 
uncertainty or the quoted uncertainty of the reactor antineutrino flux model (i.e.~the Huber-Mueller model). Additional mechanisms are required
to resolve the RAA. One popular hypothesis is the existence of a sterile neutrino with its corresponding
mass around 1~eV or higher~\cite{Mention:2011rk}, which can lead to a disappearance
of $\bar{\nu}_e$s at short baselines (a few meters). 

Before discussing the sterile neutrinos, let us briefly review the standard three-neutrino oscillation
framework in which the three neutrino flavors are superpositions of three light-mass states $\nu_1$, $\nu_2$,
$\nu_3$ with different masses, $m_1,~m_2,~m_3$:
\begin{equation}\label{eq:pmns_matrix}
\left( \begin{array}{c} \nu_{e} \\ \nu_{\mu} \\ \nu_{\tau} \end{array} \right) = 
\left ( \begin{array}{ccc} U_{e1} & U_{e2} & U_{e3} \\
	U_{\mu1} & U_{\mu2} & U_{\mu3} \\
	U_{\tau1} & U_{\tau2} & U_{\tau3} 
\end{array} \right) \cdot \left( \begin{array}{c} \nu_{1} \\ 
\nu_{2} \\ \nu_{3} \end{array} \right).
\end{equation}
The unitary $3\times 3$ mixing matrix, $U$,  called the Pontecorvo--Maki--Nakagawa--Sakata (PMNS) matrix~\cite{ponte1,Maki,ponte2}, 
can be parameterized by three Euler angles, $\theta_{12}$, $\theta_{13}$, and $\theta_{23}$, plus one or three phases 
(depending on whether neutrinos are Dirac or Majorana particles), potentially leading to CP violation in the leptonic sector. The mixing matrix $U$ is
conventionally expressed as the following product of matrices:
\begin{equation}~\label{eq:3x3_rot}
  U =R_{23}(c_{23},s_{23},0) \cdot R_{13}(c_{13},s_{13},\delta_{CP}) \cdot R_{12}(c_{12},s_{12},0)  \cdot  R_{M}
\end{equation}
with $R_{ij}$ being the $3\times3$ rotation matrices, e.g.,
\begin{equation}\label{eq:rot_13}
R_{13} = \left( \begin{array}{ccc}
	 c_{13} & 0 & s_{13} \cdot e^{-i\delta_{CP}} \\
	0 & 1 & 0 \\
	-s_{13} \cdot e^{i\delta_{CP}} & 0 & c_{13} 
\end{array} \right), \nonumber
\end{equation}
and $R_M$ being a diagonal matrix:
\begin{equation}
  R_{M} = \left(\begin{array}{ccc}
          e^{i \alpha}& 0& 0\\
          0& e^{i \beta}& 0\\
          0& 0& 1\end{array}
          \right). \nonumber 
\end{equation}
Here $c_{ij} = \cos\theta_{ij}$, $s_{ij} = \sin\theta_{ij}$. $\delta_{CP}$ is the Dirac phase that could lead to CP violation. Majorana phases are denoted by $\alpha$ and $\beta$ if neutrinos are their own antiparticles.

The phenomenon of neutrino flavor oscillations arises because neutrinos are produced and detected in their flavor
eigenstates, but propagate as a mixture of mass eigenstates. For example, in vacuum, the neutrino mass eigenstates 
with energy $E$ propagate as:
%{\setlength{\mathindent}{0cm}
  \begin{eqnarray}\label{eq:osc_prop}
\frac{d}{dL} \left( \begin{array}{c} \nu_{1}(L) \\ \nu_{2}(L) \\ \nu_{3}(L) \end{array} \right) 
= -i \cdot V \cdot  \left( \begin{array}{c} \nu_{1}(L) \\ 
\nu_{2}(L) \\ \nu_{3}(L) \end{array} \right) 
= -i 
\left ( \begin{array}{ccc} \frac{m_1^2}{2E} & 0 & 0 \\
	0 & \frac{m_2^2}{2E} & 0 \\
	0 & 0 & \frac{m_3^2}{2E}
\end{array} \right) \cdot \left( \begin{array}{c} \nu_{1}(L) \\ 
\nu_{2}(L) \\ \nu_{3}(L) \end{array} \right)
\end{eqnarray}
after traveling a distance $L$. The above equation leads to the solution
$\nu_i(L) = e^{-i \frac{m_i^2}{2E}\cdot L}\nu_i(0)$. Therefore, for a neutrino produced
with flavor $l$, the probability of its transformation to flavor $l'$ is 
expressed as:
\begin{equation}\label{eq:osc_dis}
P_{ll^\prime} \equiv P(\nu_{l}\rightarrow \nu_{l^\prime}) = |<\nu_{l^\prime}(L)|\nu_{l}(0)>|^2  
  =   \left |\sum_{j} U_{lj}U^{*}_{l'j}e^{-i(V_{jj})L} \right | ^2  
 =  \sum_{j}|U_{lj}U^*_{l'j}|^2 +  \sum_{j} \sum_{k \neq j} U_{lj} U^{*}_{l'j} U^{*}_{lk} U_{l'k} e^{i\frac{\Delta m^2_{jk} L}{2E}},  
\end{equation}
where $\Delta m^2_{jk} = m^2_j - m^2_k$. From Eq.~(\ref{eq:osc_dis}), it is obvious
that the two Majorana phases are not involved in neutrino flavor
oscillations.   More specifically, we have 
\begin{eqnarray}\label{eq:PueFullnoCP}  
  P_{\nu_\mu\rightarrow\nu_e}(L/E) &=&  \left |\sum_{i=1}^3 U_{\mu i}U^{*}_{ei}e^{-i(m_{i}^2/2E)L} \right | ^2,  \\
  P_{\nu_\mu\rightarrow \nu_\mu}(L/E) &\equiv &  P_{\bar{\nu}_\mu\rightarrow \bar{\nu}_\mu}(L/E) 
   = \,1-4\sum_{k>j}|U_{\mu k}|^2|U_{\mu j}|^2\sin^2\left(\frac{\Delta m^2_{kj}L}{4E}\right), \nonumber \\ 
  P_{\overline\nu_e\rightarrow\overline\nu_e}(L/E) &\equiv & P_{\nu_e\rightarrow \nu_e}(L/E)
   =  \,1-4\sum_{k>j}|U_{ek}|^2|U_{ej}|^2\sin^2\left(\frac{\Delta m^2_{kj}L}{4E}\right).  \nonumber
\end{eqnarray}

While the majority of neutrino oscillation experimental data can be successfully explained by this three-neutrino
framework~\cite{pdg2022}, the 
exact mechanism by which neutrinos acquire their mass remains unknown. 
The fact that the mass of electron neutrino 
is at least 5 orders of magnitude smaller than that of electron~\cite{Otten:2008zz} is particularly puzzling. 
The possible existence of additional, possibly very heavy, neutrinos beyond the known three 
may provide a natural explanation of the origin of neutrino mass~\cite{King:2003jb}.
Given the experimental constraints from the precision electroweak measurements~\cite{ALEPH:2005ab},
these additional neutrinos cannot participate in any fundamental interaction of the Standard Model,
which gives rise to the name of sterile neutrinos~\cite{ponte2}. 
While there is no known mechanism to detect them directly, their existence could be detected indirectly through neutrino oscillations if their eigenstates mix with the
three active neutrino flavors. 
The inclusion of one sterile neutrino flavor $\nu_s$ with mass state $\nu_4$ into the mixture of three active neutrinos would 
expand the 3$\times$3 unitary matrix $U$ into a $4\times4$ unitary matrix:
\begin{equation}\label{eq:pmns_matrix_sterile}
  \left( \begin{array}{c} \nu_{e} \\ \nu_{\mu} \\ \nu_{\tau} \\ \nu_{s}\end{array} \right) = 
\left ( \begin{array}{cccc} U_{e1} & U_{e2} & U_{e3} & U_{e4}\\
	U_{\mu1} & U_{\mu2} & U_{\mu3} & U_{\mu4}\\
	U_{\tau1} & U_{\tau2} & U_{\tau3} & U_{\tau4} \\
        U_{s1}   & U_{s2}    & U_{s3}    & U _{s4}
\end{array} \right) \cdot \left( \begin{array}{c} \nu_{1} \\ 
\nu_{2} \\ \nu_{3} \\ \nu_4 \end{array} \right),
\end{equation}
where subscript $s$ stands for the added sterile neutrino flavor. In addition to the 
three existing mixing angles and one phase, this expansion would introduce three mixing angles $\theta_{14}$, $\theta_{24}$,
$\theta_{34}$ and two additional phases $\delta_{24}$, $\delta_{34}$. Therefore, 
the matrix $U$ can be parameterized~\cite{Harari:1986xf} as:
\begin{eqnarray}\label{eq:4x4_rot}
  U = R_{34} \left(c_{34},s_{34},\delta_{34} \right) \cdot R_{24}\left( c_{24}, s_{24}, \delta_{24} \right) \cdot R_{14}\left(c_{14},s_{14},0 \right) 
  \cdot R_{23}\left(c_{23},s_{23},0 \right) \cdot R_{13} \left( c_{13}, s_{13}, \delta_{CP} \right) \cdot
  R_{12}\left( c_{12}, s_{12}, 0 \right),
\end{eqnarray}
where $R$s are $4\times4$ rotation matrices with the following convention:
\begin{equation}
R_{14} = 
\left ( \begin{array}{cccc} c_{14} & 0 & 0 &s_{14}  \\
  0 & 1 & 0 & 0 \\
  0 & 0 & 1 & 0 \\
  -s_{14}   & 0    & 0    & c_{14}
\end{array} \right).
\end{equation}
Note that the definition of mixing angles in Eq.~\eqref{eq:4x4_rot}
depends on the specific ordering of the matrix
multiplication. With this definition, we have
\begin{eqnarray}\label{eq:DisapToApp}
|U_{e4}|^2 &=& \,\,s^2_{14}, \nonumber \\
|U_{\mu4}|^2 &=& \,\,s^2_{24}c^2_{14}, \nonumber \\
4|U_{e4}|^2|U_{\mu4}|^2 &=&\,\, 4 s^2_{14}c^2_{14}s^2_{24}\equiv \sin^22\theta_{\mu e}. 
\end{eqnarray}
The last line in Eq.~\eqref{eq:DisapToApp} is crucial in the region
where $\Delta m^2_{41}$ $\gg$ $|\Delta m^2_{32}|$ and for short baselines
($\Delta_{32} \equiv \frac{\Delta m^2_{32}L}{4E} \sim 0$). 

With Eq.~\eqref{eq:pmns_matrix_sterile}, often referred to as the $3+1$ model, the neutrino oscillation probabilities in Eq.~\eqref{eq:osc_dis} can be
updated to:
\begin{equation}
  P_{\nu_l\rightarrow \nu_{l'}}(L/E) =  \left |\sum_{i=1}^4 U_{li}U^{*}_{l'i}e^{-i(m_{i}^2/2E)L} \right | ^2,
\end{equation}
and Eq.~(\ref{eq:PueFullnoCP}) can then be updated and simplified to:
%\begin{strip}
  \begin{eqnarray}\label{eq:PueApproxMtrx}
    P_{\nu_\mu\rightarrow\nu_e}(L/E) &\approx& P_{\bar{\nu}_\mu\rightarrow \bar{\nu}_e}(L/E) 
    \approx  \sin^22\theta_{\mu e} \sin^2 \Delta_{41} , \nonumber \\
    P_{\nu_\mu\rightarrow \nu_\mu}(L/E) &\equiv& P_{\bar{\nu}_\mu\rightarrow \bar{\nu}_\mu}(L/E)  
    \approx  1 - \sin^2 2\theta_{24} \sin^2\Delta_{41} 
     -\sin^22\theta_{23} \cos2\theta_{24} \sin^2 \Delta_{31} , \nonumber \\
    P_{\overline\nu_e\rightarrow\overline\nu_e}(L/E) & \equiv & P_{\nu_e\rightarrow\nu_e}(L/E) 
    \approx  1 - \sin^2 2\theta_{14} \sin^2 \Delta_{41}
     - \sin^22\theta_{13} \sin^2 \Delta_{31},
  \end{eqnarray}
  in which the values of additional phases are irrelevant~\footnote{This statement is no longer true 
  if there are two or more sterile neutrino flavors.}.  The $\sin^2\Delta_{31}$ terms were kept in the 
  disappearance formulas, since they are important in some disappearance experiments. 
  Given $\Delta_{41}$, the three oscillations in Eq.~\eqref{eq:PueApproxMtrx} depend only on 
  two unknown parameters: $\theta_{14}$ and $\theta_{24}$. 
  Hence, from measurements of any two oscillations, the third one can be deduced. 
  At short-baselines, a large $\Delta m^2_{41}$ could introduce fast oscillations well beyond 
  the experimental detection resolution, which leads to an average constant oscillation probability:
    \begin{eqnarray}\label{eq:PueApproxMtrx_const}
    P_{\nu_\mu\rightarrow\nu_e}(L/E) &\approx& P_{\bar{\nu}_\mu\rightarrow \bar{\nu}_e}(L/E) 
    \approx  \frac{1}{2} \sin^22\theta_{\mu e}  , \nonumber \\
    P_{\nu_\mu\rightarrow \nu_\mu}(L/E) &\equiv& P_{\bar{\nu}_\mu\rightarrow \bar{\nu}_\mu}(L/E)  
    \approx  1 - \frac{1}{2} \sin^2 2\theta_{24} 
      , \nonumber \\
    P_{\overline\nu_e\rightarrow\overline\nu_e}(L/E) & \equiv & P_{\nu_e\rightarrow\nu_e}(L/E) 
    \approx  1 - \frac{1}{2}\sin^2 2\theta_{14}.
  \end{eqnarray}
Note that the last line in Eq.~\eqref{eq:PueApproxMtrx_const} could explain the observed deficit in RAA with $\sin^22\theta_{14} \sim 0.1$.

\begin{figure}[btph]
  \begin{centering}
    \includegraphics[width=0.50\textwidth]{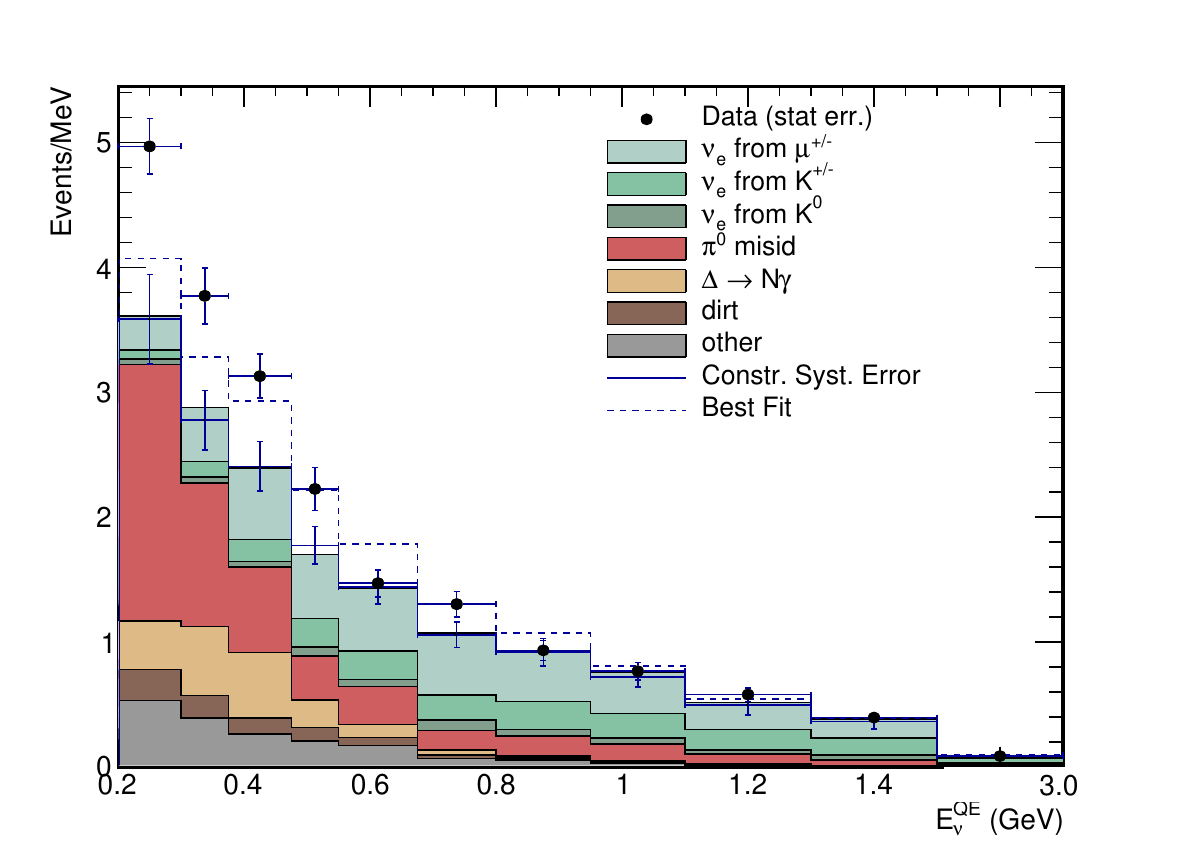}
      \includegraphics[width=0.47\textwidth]{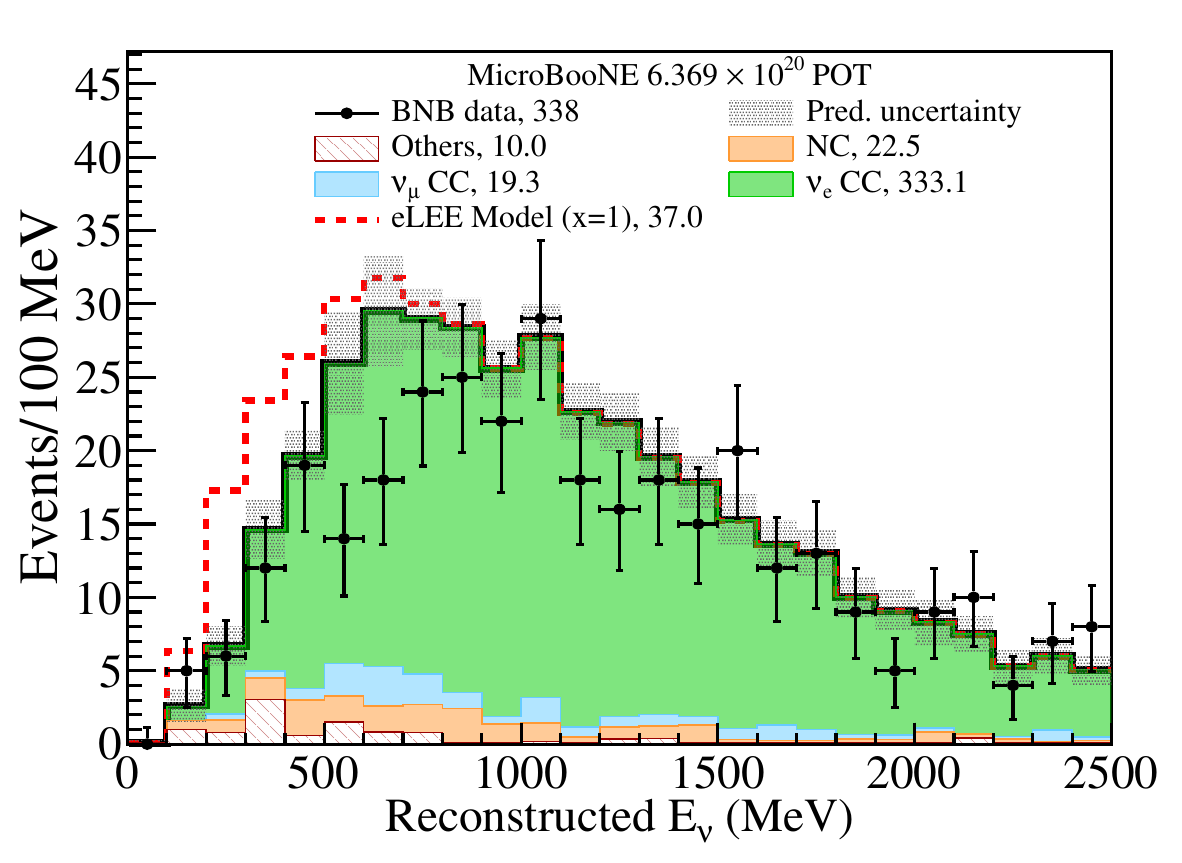}
     \put(-370,-10){a)}
      \put(-120,-10){b)}
    \par\end{centering}
    \caption{\label{fig:miniboone_microboone}
 (Left) Observed low-energy excess of $\nu_e$-like events in the MiniBooNE experiment~\cite{MiniBooNE:2018esg}. 
 The significance of the excess is about 4.8~$\sigma$. 
 (Right) The search for a low-energy excess in the inclusive $\nu_e$ charged-current interactions
 in the MicroBooNE experiment~\cite{MicroBooNE:2021tya} using the same Booster Neutrino Beam at Fermilab as in MiniBooNE. The numbers in the legend show the total number of selected candidates and predicted events in each category. Since no low-energy excess was observed in MicroBooNE, the hypothesis that the low-energy excess observed in MiniBooNE are all from $\nu_e$s is disfavored at over 2.6~$\sigma$. 
     }
\end{figure}

\begin{figure}[th!]
  \begin{centering}
    \includegraphics[width=0.8\textwidth]{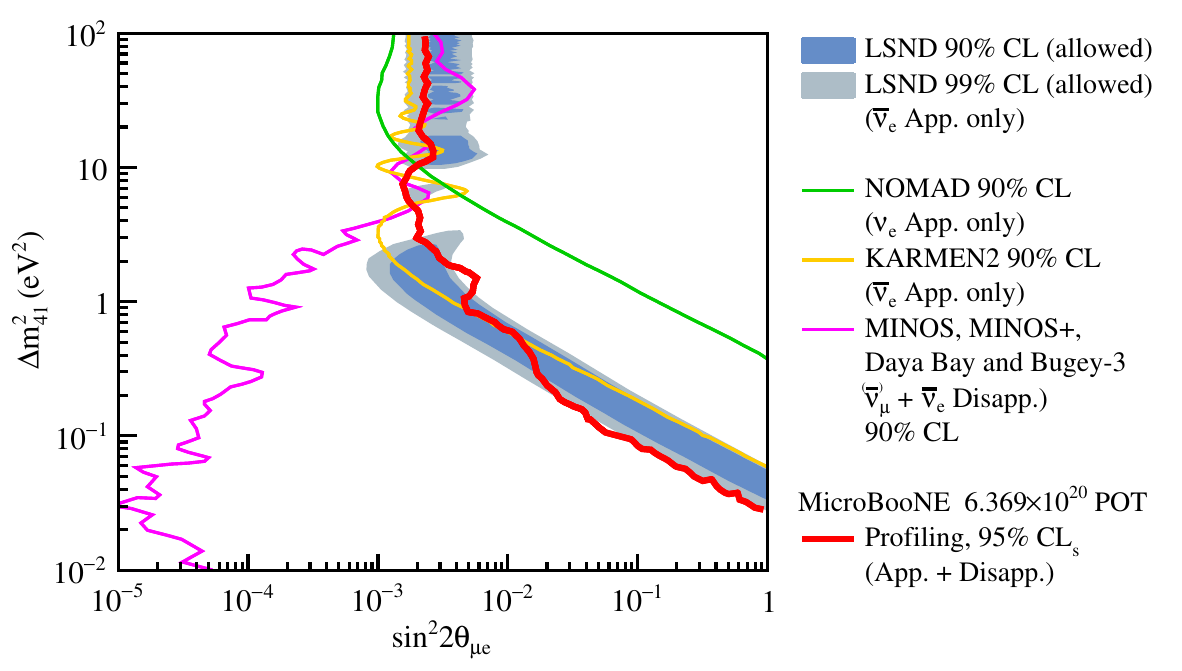}
    \par\end{centering}
    \caption{\label{fig:global_ue}
    The LSND 90\% and 99\% CL allowed regions~\cite{LSND:2001aii} 
    in the parameter space of $\Delta m^2_{41}$ and $\sin^22\theta_{\mu e}$
    using the $\nu_e$ appearance-only approximation are compared with the exclusion contours 
    at the 90\% CL from NOMAD~\cite{NOMAD:2003mqg}, KARMEN2~\cite{KARMEN:2002zcm}, the combined 
    analysis of MINOS, MINOS+, Daya Bay and Bugey-3~\cite{MINOS:2020iqj,MINOS:2017cae,ACHKAR1995503,DayaBay:2016ggj}, 
    and the 95\% CLs exclusion contour from MicroBooNE~\cite{MicroBooNE:2022wdf}. Figure taken from Ref.~\cite{MicroBooNE:2022wdf}.
     }
\end{figure}

Besides the theoretical motivations in searching for sterile neutrinos and the RAA, there were several other 
experimental anomalies that could be explained by an additional light sterile neutrino at the 
$\sim$eV mass scale. In 2001, an excess of $87.9\pm22.4\pm6.0$ electron antineutrino events was reported by the LSND experiment in searching for $\bar\nu_\mu \to \bar\nu_e$ oscillations 
using muon antineutrinos from $\mu^+$ decay at rest~\cite{LSND:2001aii}. In order to examine
the results from LSND, the MiniBooNE experiment performed the searches for (anti-)$\nu_{\mu}\rightarrow$ (anti-)$\nu_e$ oscillations using Booster Neutrino Beam (BNB) with a mean 
energy around 0.8 GeV. A low-energy excess of $\nu_e$-like events  
observed in MiniBooNE~\cite{MiniBooNE:2008yuf,MiniBooNE:2018esg} for both the neutrino (shown in 
Fig.~\ref{fig:miniboone_microboone}a) and the antineutrino data taking. If the low-energy excess in MiniBooNE are 
originated from a sterile-neutrino-induced (anti-)$\nu_{\mu}\rightarrow$ (anti-)$\nu_e$ oscillation, it would be quantitatively compatible with the previous LSND observation~\cite{MiniBooNE:2018esg}.
However, with the Cerenkov detector
technology, MiniBooNE was not able to differentiate the electrons (from $\nu_e$ charged-current interactions)
from the photons (from non-$\nu_e$CC processes).  In order to understand the 
nature of the low-energy excess events observed in MiniBooNE, the MicroBooNE experiment
performed a search using the Liquid Argon Time Projection Chamber (LArTPC) technology, a tracking 
calorimeter that can provide excellent differentiation between electrons and photons utilizing its high
position resolution and low energy threshold~\cite{MicroBooNE:2016pwy}. 
In 2022, MicroBooNE reported their search results, which did not observe the low-energy excess in the $\nu_e$ charged-current interactions as shown in Fig.~\ref{fig:miniboone_microboone}b, and excluded the hypothesis that MiniBooNE's low-energy excess was solely
originated from (anti-)$\nu_{\mu}\rightarrow$ (anti-)$\nu_e$ oscillations at a high significance~\cite{MicroBooNE:2021tya,MicroBooNE:2021nxr,TheMicroBooNECollaboration:2021cjf,MicroBooNE:2021pvo}. This means that other mechanisms to explain the MiniBooNE low-energy excess anomaly are required. 
Figure~\ref{fig:global_ue} summarizes the current status in searching for a sterile-neutrino
induced (anti-)$\nu_{\mu}\rightarrow$ (anti-)$\nu_e$ oscillation. The LSND allowed regions
are compared with the exclusion limits of direct appearance searches from NOMAD~\cite{NOMAD:2003mqg}, 
KARMEN2~\cite{KARMEN:2002zcm}, and MicroBooNE~\cite{MicroBooNE:2022wdf} as well as 
indirect constraints from a combined analysis of disappearance searches from 
MINOS, MINOS+, Daya Bay and Bugey-3~\cite{MINOS:2020iqj,MINOS:2017cae,ACHKAR1995503,DayaBay:2016ggj}.
A strong tension between the appearance and disappearance measurements was 
apparent~\cite{Giunti:2019aiy}. In the future, direct searches of 
(anti-)$\nu_{\mu}\rightarrow$ (anti-)$\nu_e$ oscillation in the SBN~\cite{MicroBooNE:2015bmn} and 
JSNS2~\cite{JSNS2:2013jdh} will provide more clarity. 

\begin{figure}[th!]
  \begin{centering}
    \includegraphics[width=0.8\textwidth]{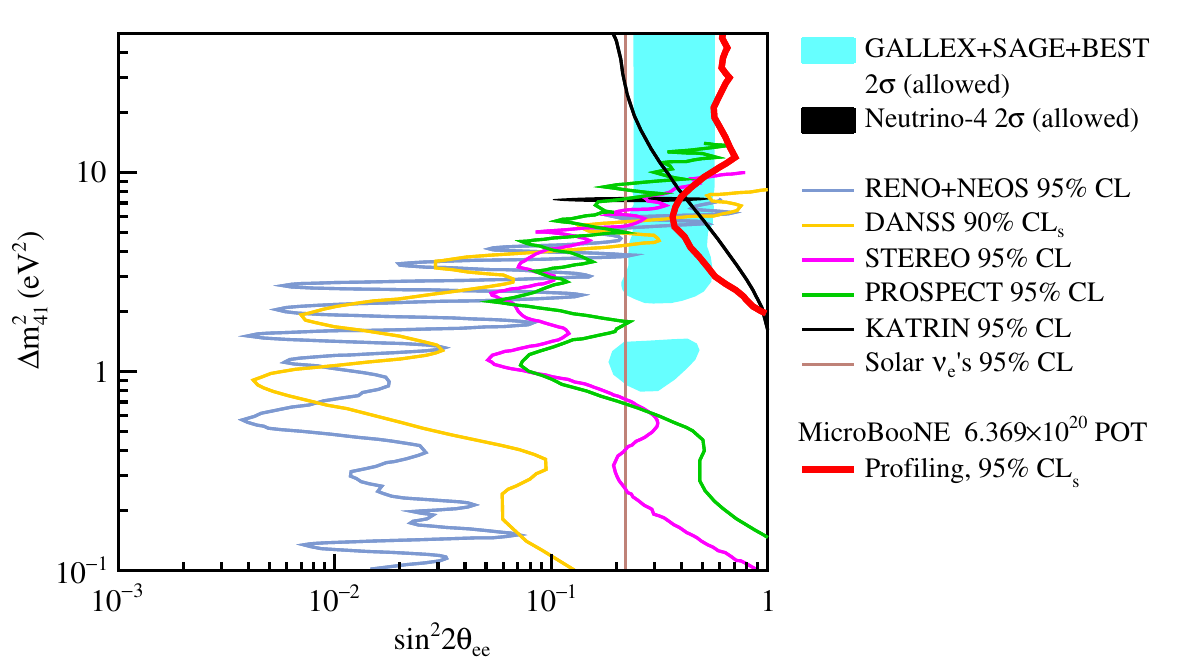}
    \par\end{centering}
    \caption{\label{fig:global_ee}
    The 2$\sigma$ allowed region in the parameter space of $\Delta m^2_{41}$
    and $\sin^22\theta_{ee}$ from combined results of GALLEX, SAGE, and BEST~\cite{Barinov:2021asz}, and the 2$\sigma$ allowed region of the Neutrino-4~\cite{Serebrov:2020kmd} experiment are compared with 
    the exclusion contours from KATRIN~\cite{KATRIN:2022ith}, PROSPECT~\cite{PROSPECT:2020sxr}, 
    STEREO~\cite{STEREO:2022nzk}, DANSS~\cite{Danilov:2022bss}, the combined analysis of RENO and NEOS~\cite{RENO:2020hva}, solar $\nu_e$ experiments~\cite{Giunti:2021iti}, and MicroBooNE~\cite{MicroBooNE:2022wdf}. Figure taken from Ref.~\cite{MicroBooNE:2022wdf}.  
      }
\end{figure}

In the disappearance channel, GALLEX~\cite{Kaether:2010ag} and SAGE~\cite{SAGE:2009eeu}, both being solar neutrino experiments using Gallium as the detector target, observed lower rates of detected $\nu_e$ than expected for their calibration $\nu_e$ sources ($^{51}$Cr for GALLEX, $^{51}$Cr and $^{37}$Ar for SAGE). 
Their observations were recently confirmed at high significance by the BEST experiment using a $^{51}$Cr source~\cite{Barinov:2021asz}. Together, they are often referred to as the Gallium anomaly. Similar to the RAA,  the Gallium anomaly could be explained by $\nu_e$ disappearance induced by a eV-mass-scale sterile neutrino. 
Furthermore, the Neutrino-4 experiment in 2020 claimed an observation of  reactor $\bar{\nu}_e$ oscillations in the range of 6--12 meters with the expected $L/E$ dependence driven by a eV-mass-scale sterile neutrino~\cite{Serebrov:2020kmd}, but it has not yet been independently confirmed by other reactor antineutrino experiments searching for similar spectral shape distortion at varying baselines.
Figure~\ref{fig:global_ee} summarizes the current status in searching for a sterile neutrino
induced (anti-)$\nu_e$ disappearance oscillation. For $\Delta m^2_{41}$ below 10 eV$^2$, 
a large portion of the allowed regions from the Gallium anomaly (GALLEX+SAGE+BEST) and the Neutrino-4 
anomaly has been excluded by direct searches from short-baseline reactor antineutrino experiments: PROSPECT~\cite{PROSPECT:2020sxr}, 
STEREO~\cite{STEREO:2022nzk}, DANSS~\cite{Danilov:2022bss}, the combined analysis of RENO and
NEOS~\cite{RENO:2020hva}, and the MicroBooNE~\cite{MicroBooNE:2022wdf} experiment. 
The high $\Delta m^2_{41}$ region has been significantly constrained by 
KATRIN's limit on the neutrino mass through the measurement of electron 
energy spectrum near tritium $\beta$-decay end-point~\cite{KATRIN:2022ith}. At the same time, the allowed regions are also in 
strong tension with the solar neutrino measurements~\cite{Giunti:2021iti} and the first row of PMNS matrix being 
consistent with unitary~\cite{Qian:2013ora, Parke:2015goa}. 
While the observed experimental anomalies have not yet been
completely resolved, the simple explanation with sterile neutrino oscillations (i.e. the 3 active $+$ 1 sterile 
neutrino model) is no longer
appealing. 

An alternative solution to the RAA resorts to nuclear physics. 
Additional uncertainties could affect the original cumulative beta spectra that the conversion method is based on. For instance, some (n, e$^-$) reaction cross sections and internal conversion coefficients used in the ILL measurements for the calibration and normalization of the beta spectra to the total number of fissions have been revised since the experiments~\cite{onillon2018}. The recent measurement of the 
$\beta$-spectrum ratio between $^{235}$U and $^{239}$Pu at the Kurchatov Institute (KI)~\cite{Kopeikin:2021rnb} also
suggests that there could be an issue in the original ILL measurements (see details in Sec.~\ref{sec:beta_spectrum_ratio}).
% It is possible that systematic uncertainties of the Huber-Mueller model have been underestimated or misunderstood. In fact, the normalization of the spectra of the total electrons themselves are based on reaction cross sections that may have been updated since then.
Another avenue is the procedure for adjusting the conversion method itself, whose systematic uncertainties 
are questionable.  
Depending on the adopted average effective Z distributions used in the fit of the ILL spectra, converted spectra could vary by 1\%~\cite{Hayes:2013wra}.
% 5\%~\cite{HayesSolvay2017,IAEA2019}, which is significantly larger than the model uncertainty quoted in Eq.~\eqref{eq:RAA_R}.
Moreover, known nuclear structure elements, 
in particular the contributions of non-unique forbidden transitions, could significantly increase the published systematic 
uncertainty and distort the converted $\bar\nu_e$ spectrum. 
Indeed, forbidden transitions contribute to 30\% of the total spectrum, and they dominate the spectrum beyond 4~MeV. Several theoretical 
works have tried to estimate the potential uncertainties introduced by this lack of knowledge~\cite{Hayes:2013wra,Fang:2015cma,Hayen:2019ieh}. 

%Authors of 
%Ref.~\cite{Hayes:2013wra} carefully examined the neutrino flux spectrum prediction 
%and concluded that the uncertainties of the flux calculation should be larger than 5\%

\begin{figure}[th!]
  \begin{centering}
    \includegraphics[width=0.8\textwidth, trim=0cm 1.5cm 0cm 3cm,clip]{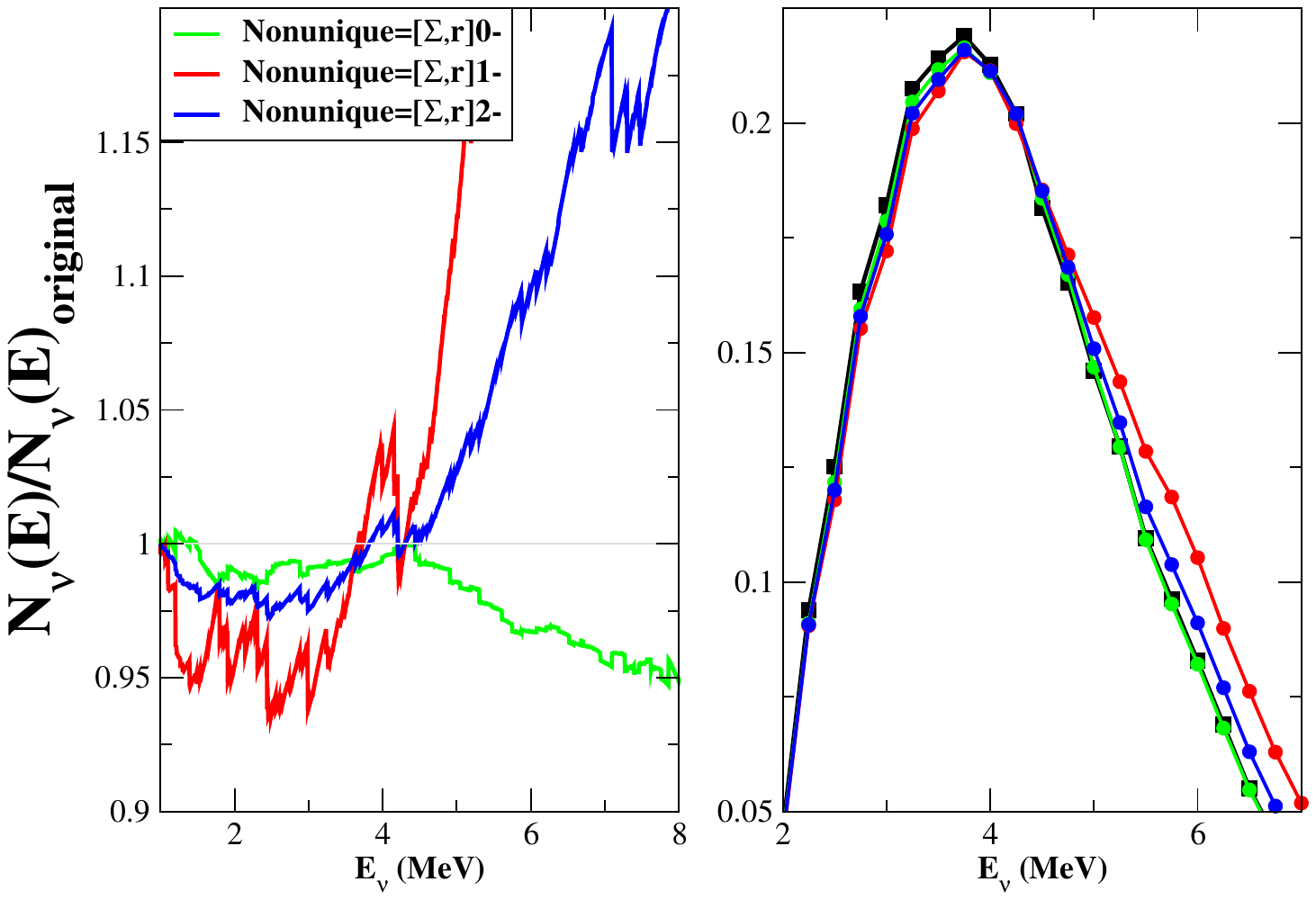}
    \par\end{centering}
    \caption{\label{fig:different_forbidden}
       Different treatments of the
      forbidden Gamow--Teller transitions contributing to the antineutrino spectrum 
      of a typical power reactor. 
      The left panel shows the ratio of these antineutrino spectra relative to that assuming all forbidden decays having the shape of allowed decays~\cite{Schreckenbach:1985ep}. 
      The right panel shows the spectra weighted by the inverse beta decay cross section.
      The corresponding change in the number of detectable antineutrinos relative to the reference is -0.75\%, 5.8\% and 1.85\% for the 0$^-$; 1$^-$, and 2$^-$ forbidden operators, respectively. 
      Figure taken from Ref.~\cite{Hayes:2013wra}.}
\end{figure}

Authors of Ref.~\cite{Hayes:2013wra} examined the uncertainties from the treatment of the forbidden non-unique decays in different ways. 
They show that when the forbidden transitions are treated in various approximations, the shape and magnitude of the spectra change
significantly. 
Figure~\ref{fig:different_forbidden} shows the comparison of the ratios between the antineutrino 
spectra computed when all unique forbidden transitions are treated as unique first-forbidden Gamow--Teller (GT) transitions, and treat non-unique 
forbidden transitions in one of the following ways: (1) as allowed GT (original); (2) as unique first-forbidden GT with the operator 
$\left[ \Sigma, r\right]^{2-}$; (3) with the operator $\left[ \Sigma, r\right]^{0-}$; (4) with the operator $\left[ \Sigma, r\right]^{1-}$, where the transition is noted by its change in spin and parity $\Delta J^\pi$.
The resulting change in the number of detectable antineutrinos can be as large as 6\%.
An exact treatment of the non-unique forbidden transitions would require the knowledge of the spin-parities of all the involved transitions, 
which is not yet accessible. These tests thus provide estimates that the uncertainty on the total $\bar\nu_e$ spectrum induced by 
forbidden transitions is at least 
at the level of 5\%.

%In particular, the form of the corrections are found to be 
%very uncertain for the 30\% of the flux that arises from forbidden decays~\cite{Hayes:2013wra}. 
%Figure~\ref{fig:different_forbidden} shows the predicted antineutrino spectra of a 
%typical power reactor derived from the measured electron spectra with different treatments 
%of the forbidden Gamow-Teller transitions. 

In Ref.~\cite{Fang:2015cma}, the authors use both Shell Model (SM, NuShellX code from Ref.~\cite{Brown:2014bhl}) and 
Quasi-Particle Random Phase (QRPA from Ref.~\cite{Fang:2013uwa}) with two quasi-particle excitations (i.e.~pn-QRPA 
for $\beta$-decay calculations) in the mass region of the heavy fission fragments to investigate the effect of first 
forbidden (FF) beta decay on the shape of the antineutrino spectra. 
They compare the change in the antineutrino spectrum relative to the allowed shape for different channels with the SM and QRPA methods for two even--even nuclei 
($^{136}$Te and $^{140}$Xe) and find that the change  could be as large as 4.5\% due to 
$1^-$ FF transitions. 
They also compare their full microscopic calculations to the approximations made in 
Ref.~\cite{Hayes:2013wra} where 4 out of 9 matrix elements are used (affecting the $0^-$ and $1^-$ decays) and find different 
behaviors in the relative changes with respect to allowed transitions. However, considering the nuclear chart, they find 
that $0^-$ transitions constitute a large fraction of the branching ratios and that the average change over all types of FF 
transitions, end-point energies and log(ft) values would result in a smaller value of 1--2\%. We should note 
as the number of valence nucleons increase, the shell model calculations become impossible and only alternative models 
can be used. 
These studies indicate that the uncertainty associated with the forbidden decays is large and not well-understood, and is unlikely to be reduced without accurate direct measurements of the antineutrino flux.
%The Shell Model can be applied to the decays of nuclei with both even and odd numbers of protons or neutrons, which is for the moment at variance with the pnQRPA calculations.

Besides forbidden non-unique decays, another potential source of uncertainty arises from the weak magnetism correction 
that enters the spectral calculation~\cite{Huber:2011wv,Wang:2017htp, Hayes:2016qnu} and is not well constrained experimentally in the fission product mass region. 
In Ref.~\cite{Huber:2011wv}, an average value of the weak 
magnetism (WM) correction, which acts as a slope on the energy spectrum, is derived from the analysis of 13 nuclei. 
These nuclei are light nuclei, not in the fission product range, because of lack of nuclear data. 
Among them, 3 nuclei exhibit much larger 
weak magnetism corrections because of unusually large log(ft) values. 
These three cases are excluded from the computation of the average correction, but a discussion is made about the potential uncertainty affecting this WM term. 
Indeed, a shift in the mean value of the slope by +0.5\%~MeV$^{-1}$ causes a shift of the antineutrino rate of about -1\%. It 
is not clear how large weak magnetism would be in forbidden decays. In Ref.~\cite{Huber:2011wv}, Huber concludes 
that the size of weak magnetism, or more generally induced currents, is a major source of theoretical uncertainties.
In Ref.~\cite{Wang:2017htp}, the orbital term of the WM correction, which is often approximated
to be proportional to the spin contribution, is explored in the case of allowed transitions in fission products. 
The authors find that this approximation is generally not valid and that the one-body WM corrections require detailed nuclear structure calculations. Nevertheless, the overall 
impact on the total antineutrino spectra is found to be small, and this approximation introduces less than a 1\% 
uncertainty in the fission antineutrino spectra. 

In parallel to these theoretical efforts to re-evaluate the uncertainties associated with the conversion method, there are several additional experimental results from reactor measurements that support the needs to increase the model uncertainties, including the measurement of antineutrino energy spectrum, the measurement of the individual $^{235}$U and $^{239}$Pu antineutrino flux, and the measurement of the $\beta$-spectrum ratio between $^{235}$U and $^{239}$Pu. Those measurements will be discussed in detail in Sec.~\ref{sec:additional_mea}.
% Daya Bay, RENO, and Double Chooz highlighted the 
% shape anomaly of the reactor antineutrinos (details in Sec.~\ref{sec:nu_energy_spectra}), which 
% points to a notable difference between the shapes of spectra predicted by the Huber-Mueller model and 
% measurements made with their near detectors, and later confirmed by recent short-baseline experiments 
% at research reactors. The observed discrepancies support the needs to increase the model uncertainties.
% In addition to the treatment of the forbidden Gamow-Teller transitions, it is also 
% possible that there are unintentional errors in 
% measuring the underlying electron spectrum to explain why the calculated antineutrino 
% flux doesn’t agree with the measured one, which will be further discussed in 
% Sec.~\ref{sec:beta_spectrum_ratio}. 
To understand what could explain these anomalies, the understanding 
of the underlying nuclear physics is essential. Indeed, only the decomposition of the reactor antineutrino 
spectrum into its individual contributions and the study of the missing nuclear 
physics information will provide reliable predictions.   
Further microscopic theoretical 
developments for the calculation of the forbidden shape factors are also needed.
In Sec.~\ref{sec:new_calculation}, we will discuss the progress toward predicting 
the antineutrino spectrum directly using the summation method.

%To fully address the reactor anomaly, 
%more accurate approach describing how the beta decay of each fission 
%product contributes to the antineutrino spectrum is needed. 

%% file: Additional_measurements.tex
\section{Additional reactor measurements}\label{sec:additional_mea}

While the reactor antineutrino anomaly (RAA) as described by Eq.~\eqref{eq:RAA_R} only focuses on the deficit in the measured antineutrino rate, the source of the RAA can be explored by other properties of reactor antineutrinos, such as their energy spectrum, isotopic dependence, and their corresponding electron spectra from the $\beta$-decay of the fission products. In this section, we review the experimental efforts to investigate the origin of the RAA with these indirect measurements.

\input{Energy_spectrum.tex}

    \input{Fuel_evolution.tex}

    \input{Beta_spectrum_ratio.tex}

%% file: Energy_spectrum.tex
\subsection{Measurement of reactor antineutrino energy spectrum}\label{sec:nu_energy_spectra}

%	Daya Bay, RENO, Double Chooz, NEOS, PROSPECT, STEREO, etc. (5 MeV bump)
%% Measurement Principle ...
Besides predicting the overall antineutrino rate, the Huber-Mueller model~\cite{Mueller:2011nm,Huber:2011wv} 
also makes a prediction on the energy spectrum of reactor antineutrinos. 
A systematic comparison between the predicted reactor antineutrino energy spectrum and the experimental measurements can thus shed light on the origin of RAA. 
With the overall antineutrino detection rate normalized, 
an observation of an excess in the 4--6~MeV region in the prompt energy spectrum was first reported by the RENO experiment~\cite{RENO:2015ksa} in the Neutrino 2014 
conference.
% ~\cite{neutrino2014} 
This finding was soon confirmed by other experiments including Daya Bay~\cite{DayaBay:2015lja}, 
Double Chooz~\cite{DoubleChooz:2014kuw}, and NEOS~\cite{NEOS:2016wee}. 
In the following, we discuss the measurement of reactor antineutrino energy spectrum
in detail. 

As introduced in Sec.~\ref{sec:sub:detection}, the primary method to detect the reactor $\bar{\nu}_e$
is the IBD reaction: $\bar{\nu}_e + p \rightarrow e^+ + n$.
%, since the cross section is accurately known~\cite{Vogel:1999zy,Strumia:2003zx}. 
The energy threshold of this process is 
about 1.8 MeV, and the energy of the prompt positron signal is related to the neutrino energy via 
$E_{\bar{\nu}_e} = E_{\rm prompt} + 0.78~{\rm MeV} + T_n$, where $T_n$ is the kinetic energy of the recoil neutron.
Since $T_n$ is only of the order of tens of keV and can be ignored to the first order, the neutrino energy can be accurately determined by the prompt energy.

%% Examine Daya Bay's measurement prediction of background, nonlinearity and detector resolution ... ???
While the relation between the neutrino energy and the prompt energy is straightforward, the reconstruction
of the prompt energy is not as simple. 
In liquid scintillator, the light yield is nonlinear with respect 
to the deposited energy of the charged particles because of the quenching effect described by the Birks' law~\cite{Birks:1964zz}
and Cerenkov radiation~\cite{Cherenkov:1934ilx}. 
The quenching effect between the excited molecules and the surrounding substrate reduces the scintillation light yield 
at high ionization density, i.e.~when the energy deposition per unit length, $dE/dx$, is high. 
This is the case for heavily ionizing particles such as $\alpha$ particles and protons, and for low-energy electrons and positrons.
The Cerenkov radiation contributes additional optical photons when a charged particle travels faster than the phase velocity of light in the detector medium. 
In addition, a nonlinear response can be introduced by the electronic readout system when estimating the number of photoelectrons detected, and by the position-dependent detector response in the event reconstruction.  
Therefore, a comprehensive calibration program is essential to properly 
reconstruct the prompt energy. 

 \begin{figure}[ht!]
  \begin{centering}
    \includegraphics[width=0.48\textwidth]{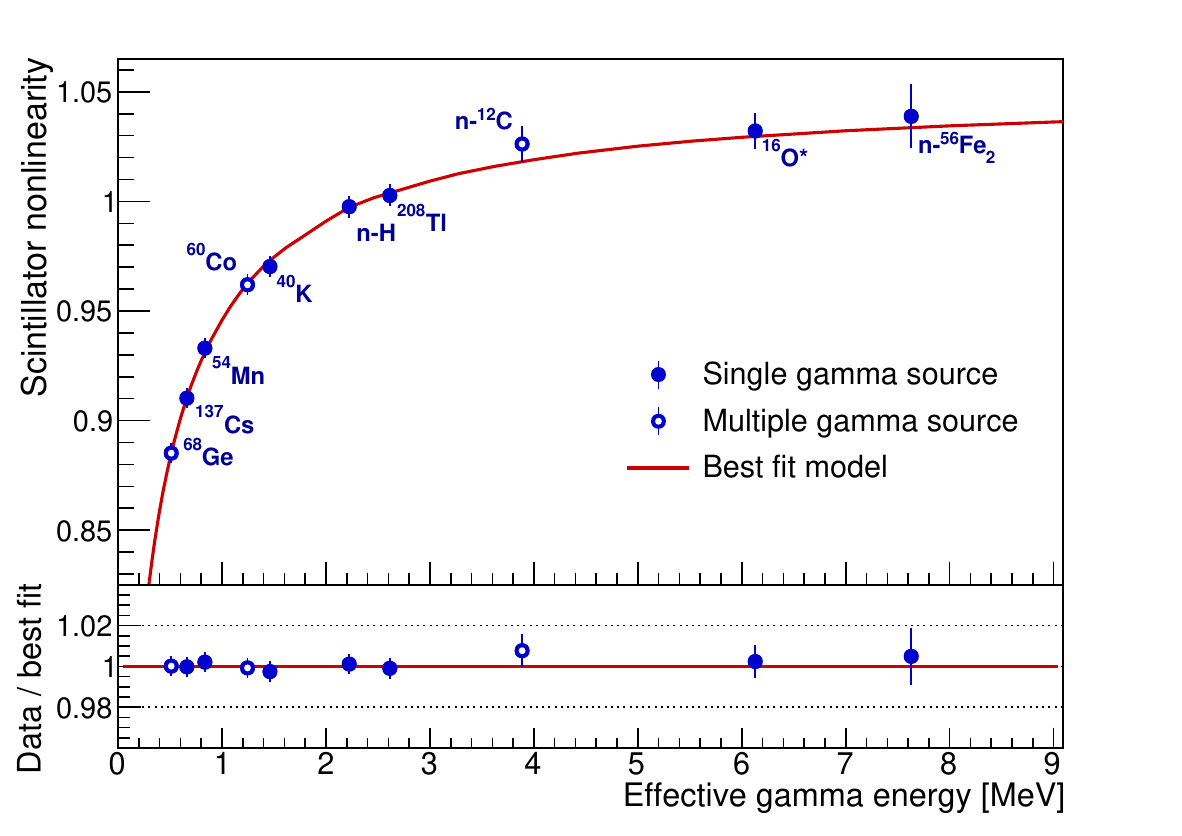}
        \includegraphics[width=0.48\textwidth]{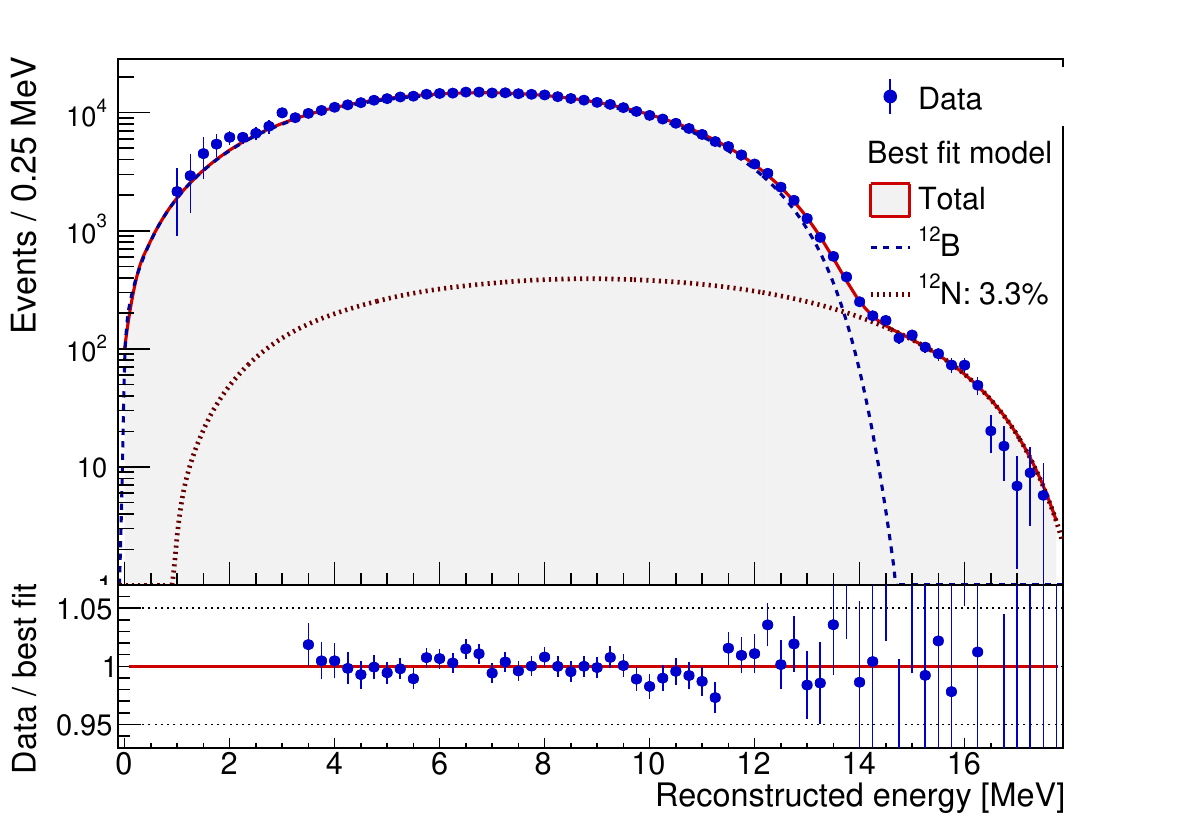}
    \par\end{centering}
    \caption{\label{fig:dyb_calibration}
    (Left) The best-fit non-linearity (red line) from the liquid scintillator is 
    compared with the measured ones from calibration $\gamma$-rays in the Daya Bay experiment. 
    (Right) The reconstructed $^{12}$B energy spectrum is compared with the predicted one in the Daya Bay experiment. Figures taken from Ref.~\cite{DayaBay:2019fje}. }
\end{figure}

As an example, Fig.~\ref{fig:dyb_calibration} illustrates the understanding of the overall detector nonlinear response in Daya Bay after a dedicated calibration campaign. 
The electronics-induced non-linearity was separately measured with a dedicated flash analogy-to-digital converter readout system. 
To address the nonlinear energy response of the liquid scintillator, both $\gamma$-ray calibration points and $\beta$-decay spectrum from $^{12}$B are used. 
The $\gamma$-ray energy calibration includes 
the deployed $^{68}$Ge, $^{137}$Cs, $^{54}$Mn, $^{60}$Co, $^{40}$K
sources, and the naturally occurring $^{208}$Tl from the surrounding materials.
Additional $\gamma$ calibration points are from  the three neutron calibration sources based on 
the~($\alpha,~n$)~reaction of Pu-$^{13}$C, $^{241}$Am-$^{13}$C, and $^{241}$Am-$^{9}$Be. 
The continuous $\beta$ energy spectrum from the $^{12}$B, which is produced inside the liquid scintillator by cosmic-ray muons, is also used. 
The primary decay channel of $^{12}$B is to the ground state of $^{12}$C with a branching ratio of 98.3\%
% ~\cite{KELLEY201771} 
and Q-value of 13.37 MeV. 
There are two more decay channels to the excited states of $^{12}$C (4.44 MeV and 7.65 MeV), each with about 1\% level branching ratio.
All three decay branches are considered in predicting the energy spectrum of $^{12}$B, an allowed transition of the Gamow--Teller type. 
After this dedicated calibration campaign,
less than 0.5\% uncertainty in the energy nonlinearity calibration is achieved for positrons of kinetic energy greater than 1 MeV~\cite{DayaBay:2019fje}.

 \begin{figure}[ht!]
  \begin{centering}
    \includegraphics[width=0.6\textwidth]{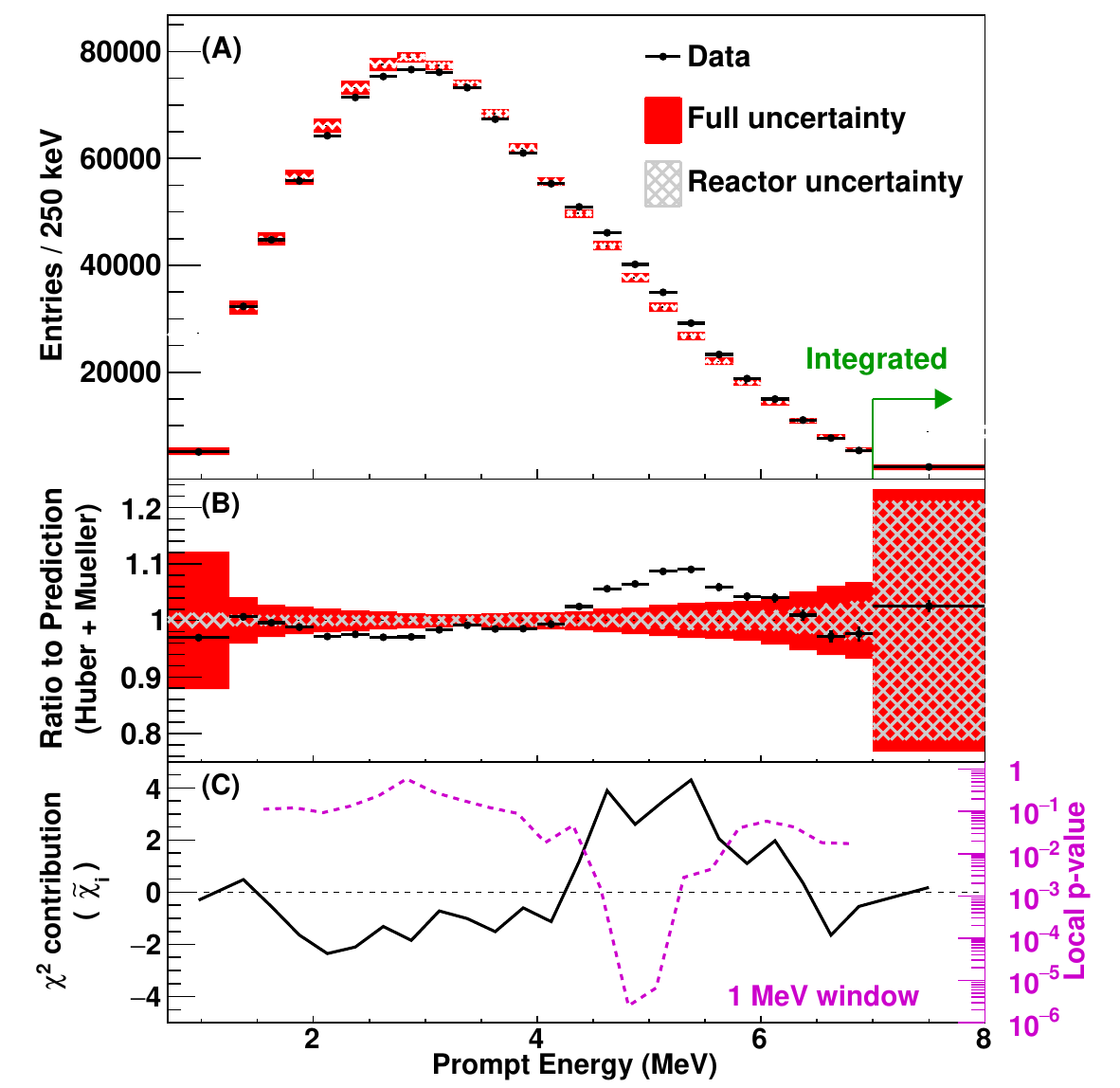}
    \par\end{centering}
    \caption{\label{fig:dyb_bump}
      Predicted and measured prompt-energy spectra, taken from
      Ref.~\cite{DayaBay:2016ssb}. The prediction is based on the Huber--Mueller
      model~\cite{Mueller:2011nm,Huber:2011wv} and normalized
      to the number of measured events. The highest energy bin contains
      all events above 7~MeV. The gray hatched and red filled bands represent
      the square-root of diagonal elements of the covariance matrix
      for the reactor-related and the full (reactor, detector, and background)
      systematic uncertainties, respectively. The error bars on the
      data points represent the statistical uncertainty.
      The ratio of the measured prompt-energy spectrum to the predicted spectrum is shown in the middle panel. 
      The defined $\chi^2$ distribution of each bin (black solid curve) and local p-values for 1-MeV energy
      windows (magenta dashed curve) are shown in the bottom panel. }
\end{figure}

Figure~\ref{fig:dyb_bump} shows the measured prompt energy spectrum from Daya Bay~\cite{DayaBay:2016ssb,DayaBay:2015lja} 
in comparison with the model prediction and its associated uncertainties. The prediction of the total detection rate
has been normalized to that of the data. 
An excess between the 4 and 6~MeV prompt energy beyond the model uncertainties can be clearly seen, which is often referred to as the ``5 MeV bump'' and indicates an underestimation of the model uncertainties. 
Taking into account the entire energy range, this result disfavors the Huber-Mueller model prediction~\cite{Mueller:2011nm,Huber:2011wv} at about 2.9~$\sigma$. 
The local significance for a discrepancy is greater than 4$\sigma$ 
at the highest point around 5 MeV. In addition, the local significance for the 2 MeV window between 4 and 6 MeV corresponds to the p-value of 9.7$\times$10$^{-6}$ (4.4 $\sigma$).
It should also be noted that the observation of 4--6 MeV excess is after the normalization of total antineutrino detection rate. 
Prior to such normalization, the agreement between data and Huber-Mueller model prediction is good around 5 MeV, but a systematic deficit in data is observed below 5 MeV~\cite{DayaBay:2021owf}. 
The reason for this common choice of normalization treatment is to separate the effects between rate deficit and shape distortion, which could have different origins either in the experiments or in the model building.
%After the normalization, the 4 MeV - 6 MeV excess
%is not limited to the Huber-Mueller model~\cite{Huber:2011wv,Mueller:2011nm}, 
%and is having a similar degree of deviation was also observed when compared with the 
%ILL+Vogel~\cite{Vogel:1980bk,VonFeilitzsch:1982jw,Schreckenbach:1985ep,Hahn:1989zr}
%model calculation.

\begin{figure}[ht!]
  \begin{centering}
    \includegraphics[width=1.0\textwidth, trim=0cm 8.5cm 0cm 1cm,clip]{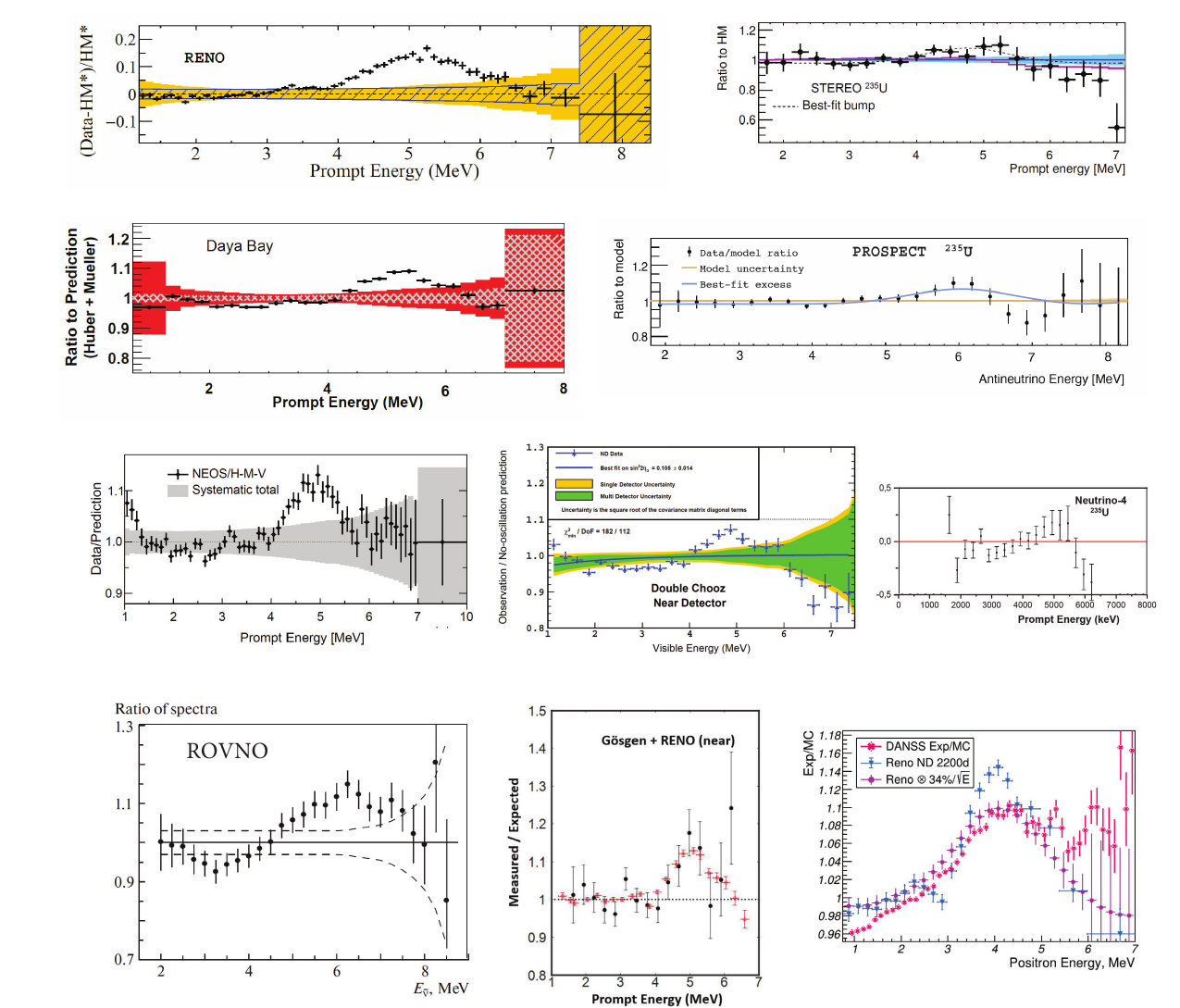}
    \par\end{centering}
    \caption{\label{fig:other_bumps} Summary of observations of the 5-MeV excess in prompt energy (or 6-MeV in antineutrino energy) with respect to the model prediction~\cite{Mueller:2011nm,Huber:2011wv}, 
      including RENO~\cite{RENO:2020dxd}, Daya Bay~\cite{DayaBay:2016ssb}, 
      Double Chooz~\cite{DoubleChooz:2019qbj}, NEOS~\cite{NEOS:2016wee}, 
      (preliminary) DANSS~\cite{Danilov:2022bss}, Neutrino-4~\cite{Serebrov:2020kmd},
       STEREO~\cite{STEREO:2020hup}, PROSPECT~\cite{Adriamirado:2022kmz},
       G\"{o}sgen~\cite{CALTECH-SIN-TUM:1986xvg,Zacek:2018bij} and ROVNO~\cite{Kopeikin:1997ve,Kopeikin:2017cfw}. See text for more discussions.
      % While most of the results are presented in terms of prompt energy (visible energy or positron energy), results from the ROVNO and PROSPECT are presented in terms of antineutrino energy. 
      % Among these, the Neutrino-4, STEREO, and PROSPECT results are from research reactors with highly enriched uranium (HEU) fuel (>20\% $^{235}$U concentration), while the rest are from commercial power reactors with low enriched uranium (LEU) fuel (3-5\% $^{235}$U concentration). The significance of the bump observations is higher from the LEU than that from HEU. 
}
\end{figure}

Figure~\ref{fig:other_bumps} compiles a collection of observations of this 5-MeV excess in prompt energy (or 6-MeV in antineutrino energy) with respect to the model prediction~\cite{Mueller:2011nm,Huber:2011wv}, 
including RENO~\cite{RENO:2020dxd}, Daya Bay~\cite{DayaBay:2016ssb}, Double Chooz~\cite{DoubleChooz:2019qbj}, NEOS~\cite{NEOS:2016wee}, Neutrino-4~\cite{Serebrov:2020kmd}, STEREO~\cite{STEREO:2020hup}, and PROSPECT~\cite{Adriamirado:2022kmz} at various baselines from O(10) m to O(1) km. 
The preliminary result from the DANSS~\cite{Danilov:2022bss} experiment is also included. 
In addition, re-analysis of positron spectra from the G\"{o}sgen experiment, which was performed with a nuclear power plant in Switzerland in the 1980s~\cite{CALTECH-SIN-TUM:1986xvg}, and the ROVNO experiment, which was performed at the Rovno nuclear power station in the Soviet Union in the 1980s~\cite{Kopeikin:1997ve} also revealed a similar excess~\cite{Zacek:2018bij,Kopeikin:2017cfw}. 
Most of these measurements are from commercial power reactors with low enriched uranium (LEU) fuel (3-5\% $^{235}$U concentration) producing electron antineutrinos from $^{235}$U, $^{238}$U, $^{239}$Pu and $^{241}$Pu. 
The Neutrino-4, STEREO, and PROSPECT results are from research reactors with 
highly enriched uranium (HEU) fuel (>20\% $^{235}$U concentration) producing electron antineutrinos mostly from $^{235}$U. 
  %%equal excess peak ??? 
A quantitative comparison of the significance of the 5-MeV excess between Daya Bay and PROSPECT~\cite{Adriamirado:2022kmz} favors the hypothesis that all fission isotopes at the commercial reactor contribute equally, 
and disfavors the hypothesis that only (no) $^{235}$U contributes at 2$\sigma$ (3.7$\sigma$) level. We will revisit this point in Sec.~\ref{sec:evolution}. 
%%Unfolding results ???
Most of the results in Fig.~\ref{fig:other_bumps} are presented in terms of prompt energy (also noted as visible energy or positron energy), results from  ROVNO and PROSPECT are presented in terms of antineutrino energy after unfolding. 
Besides this collection, the unfolded antineutrino energy spectrum are also reported by Daya Bay~\cite{DayaBay:2021dqj}, STEREO~\cite{STEREO:2020hup}, a joint analysis between PROSPECT and STEREO~\cite{Stereo:2021wfd}, and a joint analysis between PROSPECT and Daya Bay~\cite{DayaBay:2021owf}. 
Although the antineutrino energy spectra are reported from different experiments, one should compare them with caution. 
As pointed out in Ref.~\cite{Tang:2017rob}, the unfolding procedure typically introduces an additional smearing matrix, which leads to a bias in the unfolded results. Such a bias can be evaluated and removed by applying an additional smearing matrix to the expected spectrum from model predictions.  
Ref.~\cite{DayaBay:2021dqj} provides a detailed description of how to properly utilize the unfolded results. 

%% Expand on different explanations ??? or refer to a later section
The observation of this 5-MeV prompt energy excess has motivated many studies attempting 
to explain its origin (See Refs~\cite{Hayen:2019ieh,Huber:2016xis,Dwyer:2014eka,Hayes:2015yka,Sonzogni:2016yac,Berryman:2018jxt,Hayen:2019eop,Yoshida:2018zga,Letourneau:2022kfs,STEREO:2022nzk}, among others). 
For example, authors of Ref.~\cite{Hayen:2019ieh} study the dominant forbidden transitions contributing to the antineutrino spectra above 4 MeV using the nuclear shell model. 
Using a complete theoretical formalism~\cite{HayenSeverijns} they find that the shape factors strongly deviate from the usual allowed approximation. 
Using a Monte-Carlo method, they generalize the shape factors for all the fission products and by combining them with fission yields, they find the obtained correction on the total spectrum to be of similar shape and magnitude to the observed spectral discrepancies. 
However, the calculated form factors are larger than those obtained by other models, and the study is not performed with a full summation model free of systematic effects such as the Pandemonium effect, which will be discussed in Sec.~\ref{sec:new_calculation}. 
Measurements of the shape factors for the most important forbidden decays are thus needed to disentangle models, and understand the spectral anomaly. 
At the moment, the exact origin of the 5-MeV prompt energy excess is still under debate. Nevertheless, the observation of this 5-MeV prompt energy excess 
has indicated that the original 2--3\% quoted
Huber-Mueller model uncertainty~\cite{Mueller:2011nm,Huber:2011wv} was underestimated. Furthermore, since an eV-mass-scale sterile neutrino induced oscillation cannot explain this shape distortion at multiple baselines from a few meters to a few kilometers, this observation supports a nuclear-physics origin of the RAA as discussed in 
Sec.~\ref{sec:possible_raa_explanation}.

 \begin{figure}[ht!]
  \begin{centering}
    \includegraphics[width=0.65\textwidth]{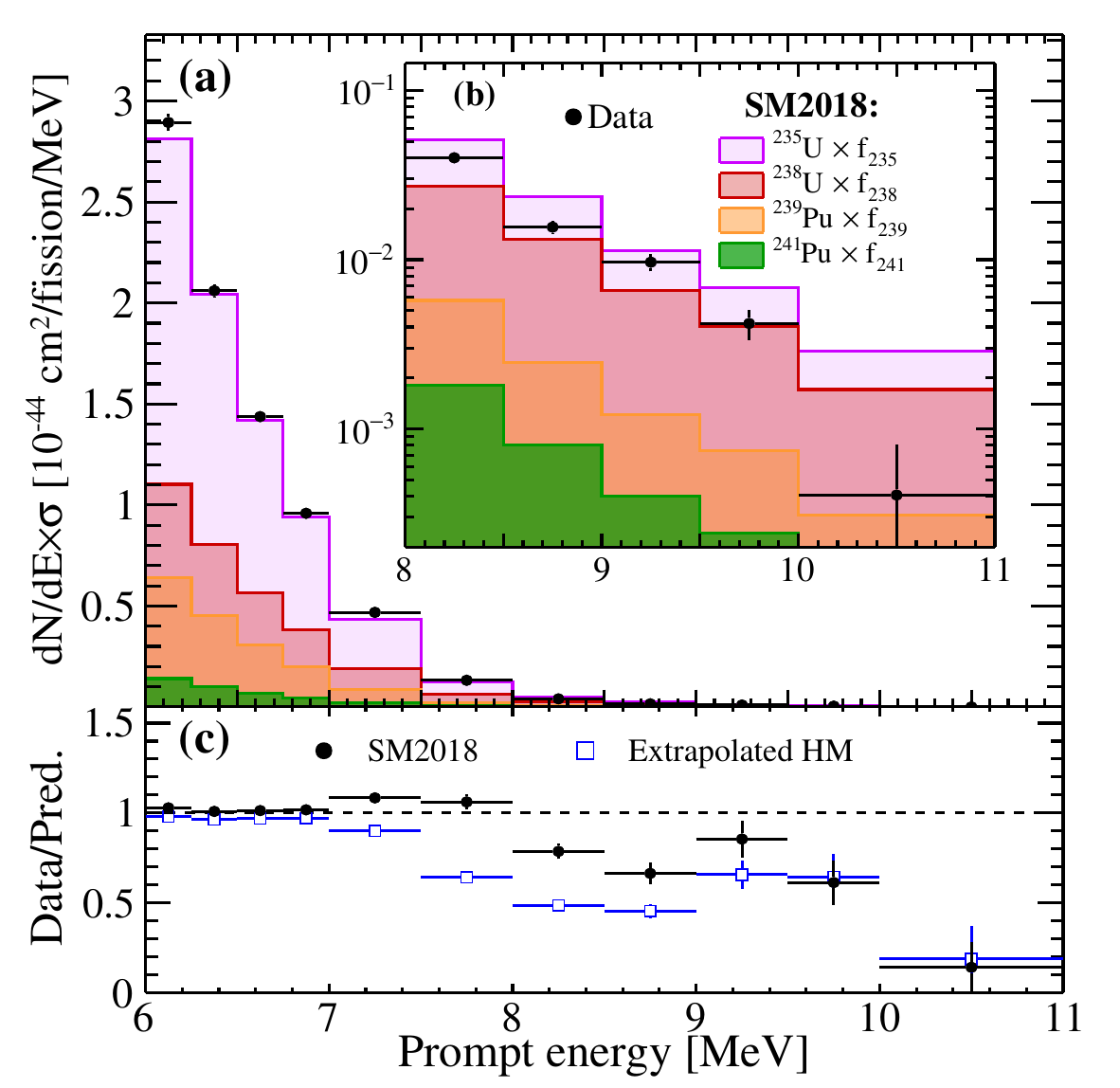}
    \par\end{centering}
    \caption{\label{fig:dyb_herav}  
    (a) The Daya Bay measured prompt energy spectrum in the 6--11~MeV range compared with the prediction from the 
    SM2018 model~\cite{Estienne:2019ujo}, a new prediction based on the summation method (see Sec.~\ref{sec:new_calculation}).  (b) A enlarged plot of (a) above 8 MeV with logarithmic vertical scales. (c) Ratio of the measurement over the prediction from SM2018 and the extrapolated Huber-Mueller (HM) model~\cite{Mueller:2011nm,Huber:2011wv}. Figure taken from Ref.~\cite{DayaBay:2022eyy}. }
\end{figure}

Before concluding this section, we show the recent first measurement of high-energy antineutrinos with a prompt energy above 8 MeV from the Daya Bay~\cite{DayaBay:2022eyy} experiment in Fig.~\ref{fig:dyb_herav}. 
With nearly 9000 high-energy IBD candidates observed over 1958 days of data taking, the hypothesis of no reactor antineutrinos with neutrino energy above 10 MeV is rejected with a significance of 6.2$\sigma$~\cite{DayaBay:2022eyy}. 
%A 29\% antineutrino flux deficit in the prompt energy region of 8-11 MeV is observed
%compared to a recent model prediction.
The measured high-energy prompt energy spectrum was compared to the prediction 
from Huber-Mueller model~\cite{,Mueller:2011nm,Huber:2011wv}, which 
was extrapolated to $E_{\nu} >8$ MeV using a polynomial function~\cite{Huber:2011wv}
and the SM2018 model~\cite{Estienne:2019ujo} (details in Sec.~\ref{sec:sub:new-dev-summation}). 
The extrapolated prediction is larger than the measurement by 30\% or more for prompt energy higher than 7.5 MeV. 
In this energy range, the uncertainties of the SM2018 models are much larger than those in the low energy range because even though the number of nuclei involved is much smaller than in the lower energy range, these fission products are very short-lived and often beta-delayed neutron emitters, and exhibit one or several isomers. 
The fission yields of these nuclei exhibit very large uncertainties, and in the case of isomers they are mostly unmeasured~\cite{Sears2021APS}. 
The beta decay properties of the isomeric states are often unmeasured as well because of the difficulty to separate them from the ground state nucleus. 
More nuclear experimental constraints on both fission yields and beta decay properties are thus necessary to improve the knowledge of the high energy range of the reactor antineutrino spectrum. 
While the high-energy antineutrino spectrum contributes little to the RAA, this observation exposes the shortcoming of the models in the high-energy region.

%A first step towards the use of antineutrino spectral measurements as integral data for nuclear data testing, is the 
%measured antineutrino energy spectrum between 8 and 11 MeV~\cite{DayaBay:2022eyy} mentioned in
%Sec.~\ref{sec:nu_energy_spectra} by Daya Bay in 2022. They have compared their measurement with two models, the SM2018 from~\cite{Estienne:2019ujo} and a polynomial extrapolation of the Huber-Mueller model~\cite{Huber:2011wv}. Figure~\ref{fig:dyb_herav} reproduced from~\cite{DayaBay:2022eyy} shows a deficit of the measured antineutrino by 29\% with respect to SM2018 in this energy range. 

%% file: Fuel_evolution.tex
\subsection{Measurement of the individual $^{235}$U and $^{239}$Pu reactor antineutrino flux} \label{sec:evolution}
	
While the reactor antineutrino anomaly refers to the $\sim$6\% deficit in the total flux when compared to the Huber-Mueller model prediction, it is intriguing to ask whether this deficit is the same for different fission isotopes. 
If the RAA is indeed caused by sterile neutrino oscillations, then the deficit would not depend on fission isotopes. 
On the other hand, if the RAA is caused by issues in the reactor modeling, one may observe different deficits for different fission isotopes.

\begin{figure}[bthp]
  \begin{centering}
    \includegraphics[width=0.7\textwidth]{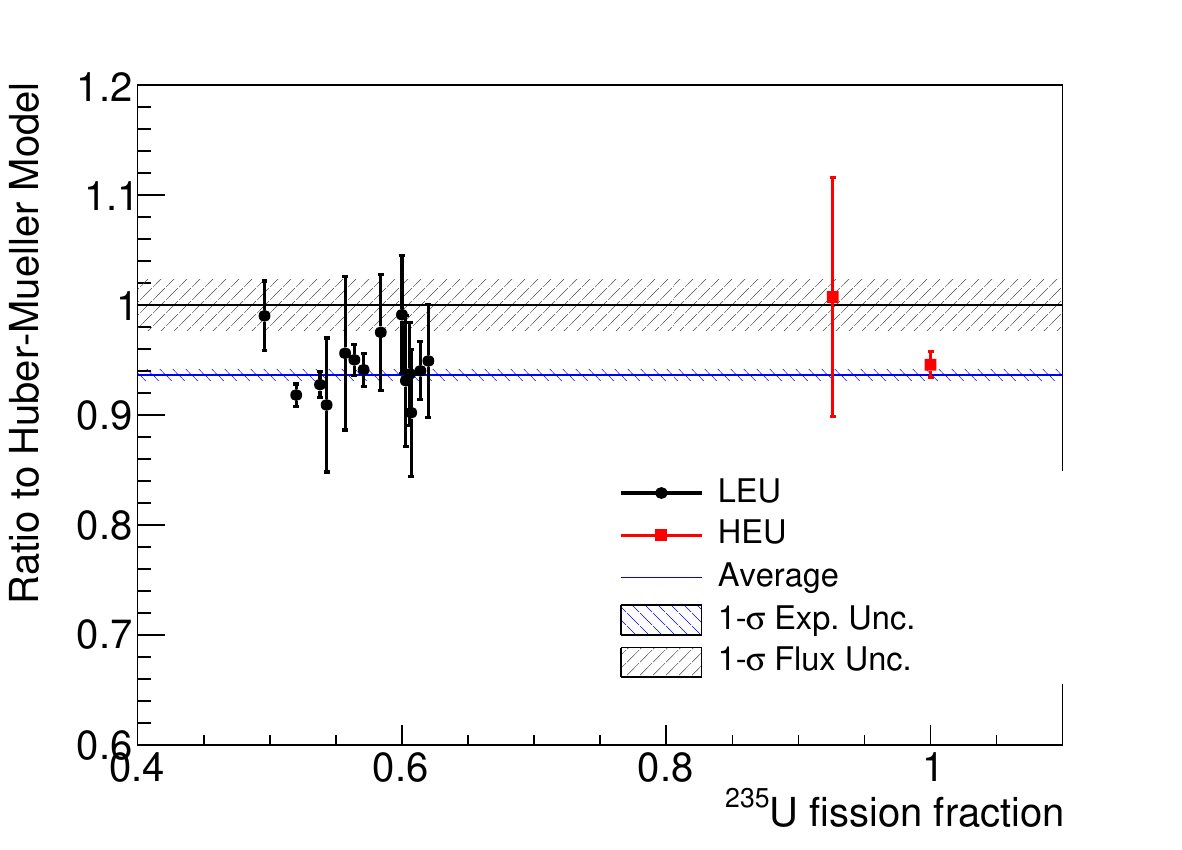}
    \par\end{centering}
    \caption{\label{fig:dyb_global_deficit_fission}
       The measured reactor $\bar{\nu}_e$ rate from 27 experiments as summarized in Table~\ref{table:results}, 
       normalized to the prediction of the Huber--Mueller model~\cite{Mueller:2011nm,Huber:2011wv}, is plotted
       as a function of the $^{235}$U fission fraction. 
       The blue shaded region represents the global average and its 1$\sigma$ uncertainty. The 2.3\%-model
      uncertainty is shown as a band around unity. Measurements at the same $^{235}$U fission fraction
      are combined for clarity.
      }
\end{figure}

Figure~\ref{fig:dyb_global_deficit_fission} provides a first look at this question using data from Table~\ref{table:results} by binning the results as a function of $^{235}$U fission fraction ($f_{235}$). 
The data around $f_{235}$ of 0.5--0.6 come from low-enriched uranium (LEU) reactors, and the data around $f_{235}$ of 0.9--1 come from highly-enriched uranium (HEU) reactors. 
The former are mostly commercial reactors with thermal power of a few gigawatts, and the latter are mostly research reactors with thermal power of 10--100 megawatts. 
There is no obvious difference in the ratio measurements between the LEU and HEU experiments, hinting that the deficit for different isotopes may be similar. 
It should be noted that since the IBD yield of $^{235}$U is about 50\% larger than $^{239}$Pu (Table~\ref{table:fission_isotope}), even for LEU reactors about 70\% of the neutrinos still come from the $^{235}$U fission. 
It would be desirable to have data points at even lower $^{235}$U fission fractions, or higher $^{239}$Pu fission fractions. One possibility, as studied in Ref.~\cite{Fujikake:2023luo}, is to perform new measurements at reactors burning mixed oxide (MOX) fuels.

\begin{figure}[btph]
  \begin{centering}
    \includegraphics[width=0.45\textwidth]{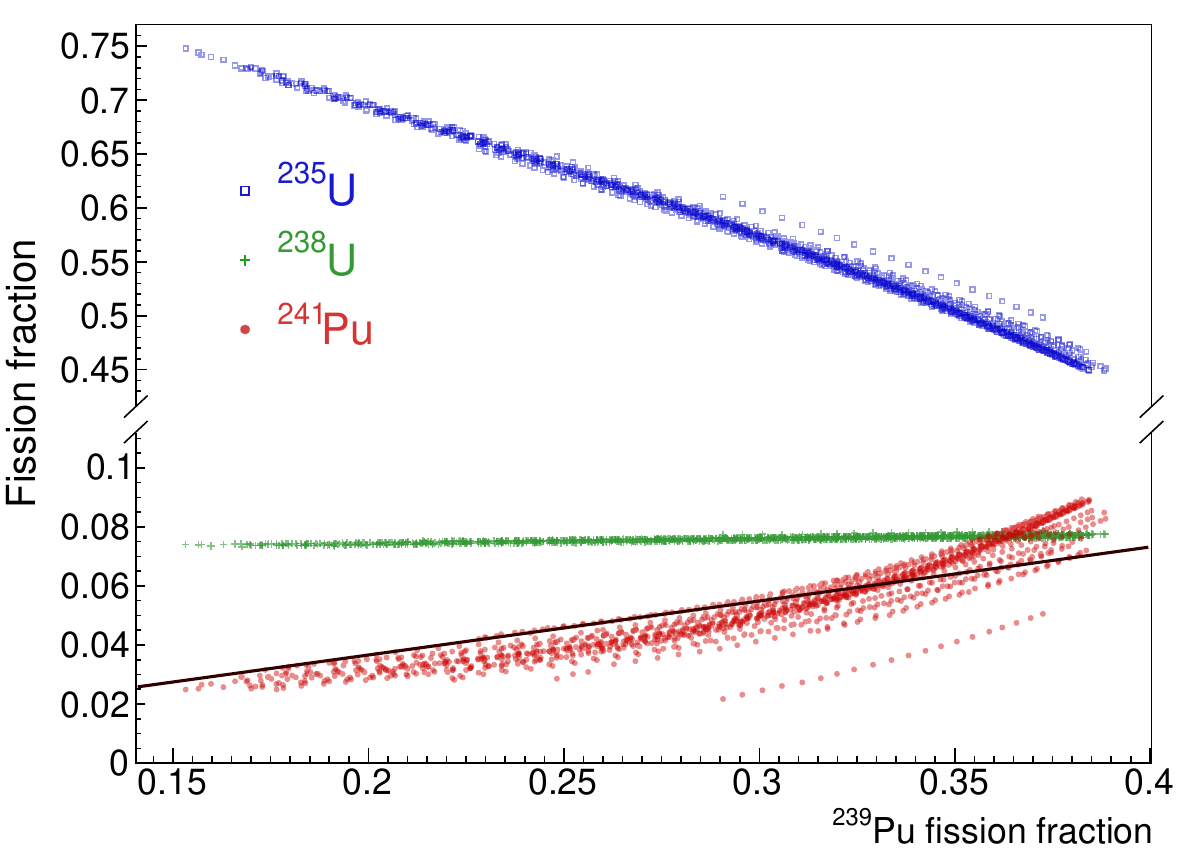}
      \includegraphics[width=0.52\textwidth]{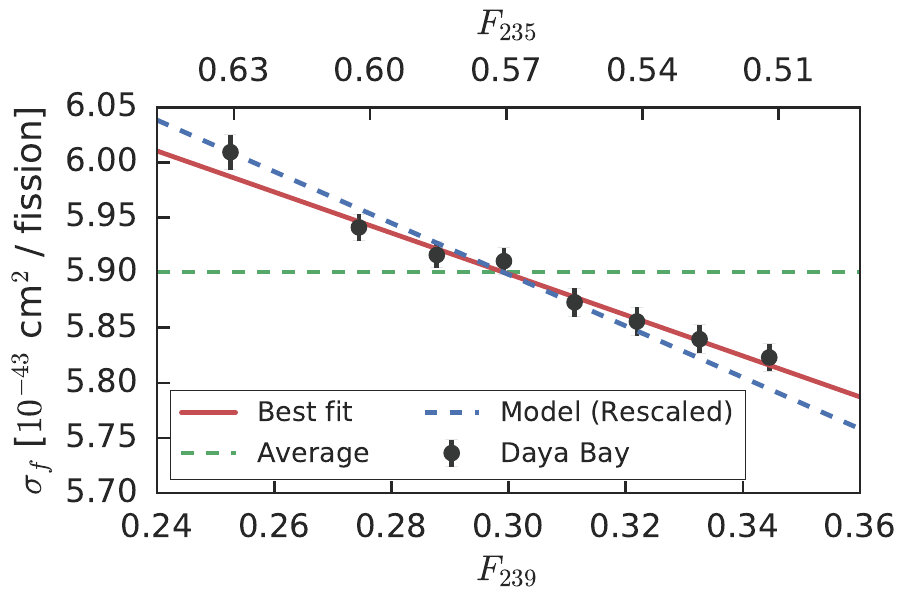}
     \put(-370,-10){a)}
      \put(-120,-10){b)}
    \par\end{centering}
    \caption{\label{fig:ff_correlation_evolution}
 (Left) The weekly fission fractions for the four major isotopes in the six reactors from the Daya Bay experiment. The data include 1958 days covering four to six refueling cycles for each reactor core. The solid line represents an approximately linear relation between fission fractions of $^{239}$Pu and $^{241}$Pu. Figure taken from Ref.~\cite{DayaBay:2019yxq}.
 (Right) IBD yield per fission as a function of effective fission fraction of $^{239}$Pu (lower axis) or $^{235}$U (upper axis). The three lines are average yield (green line), variable yield (red line) fitted to a linear function, and predicted yield from the Huber-Mueller model (blue line), scaled to account for the difference in total yield between the data and prediction. Figure taken from Ref.~\cite{DayaBay:2017jkb}.
     }
\end{figure}

More insights can be gained by studying fuel evolution in a commercial reactor, where the fission fractions of the main fission isotopes gradually change during a refueling cycle, as shown in Fig.~\ref{fig:reactor_fission}b. 
Figure~\ref{fig:ff_correlation_evolution}a further shows the weekly averaged fission fraction scatter plot for the four major isotopes from the Daya Bay experiment after accumulating about 2000 days of data covering four to six refueling cycles in their six reactor cores. 
One can see there are enough statistics to cover many bins of $^{239}$Pu fission fraction ($f_{239}$) ranging from 0.15 to 0.38. 
As a result, a measurable continuous change in the reactor antineutrino flux can be observed during this fuel evolution. 
Furthermore, unlike those shown in Fig.~\ref{fig:dyb_global_deficit_fission}, the data points from one experiment are mostly correlated, making the fuel evolution measurement potentially more sensitive to test the isotopic dependence of the RAA.
This effect was first demonstrated in Refs.~\cite{Klimov:1994aa,Bowden:2008gu} and later exploited to extract the individual $^{235}$U and $^{239}$Pu reactor antineutrino flux by Daya Bay~\cite{DayaBay:2017jkb}, RENO~\cite{RENO:2018pwo}, and NEOS-II~\cite{neos_2022}.

The method of extraction is mathematically simple. As shown in Fig.~\ref{fig:ff_correlation_evolution}a, during the fuel evolution, the fission fraction of $^{238}$U is nearly constant: $f_{238} \simeq c \simeq 0.08$ . The fission fraction of $^{241}$Pu is approximately linear with respect to the fission fraction of $^{239}$Pu: $f_{241} \simeq r \cdot f_{239} \simeq 0.18 \cdot f_{239} $, because $^{241}$Pu is produced from $^{239}$Pu (see Table~\ref{table:fission_isotope}). Since all fission fractions add up to 1, the total IBD yield can be expressed as follows: 
\begin{eqnarray}\label{eq:fuel_evolution}
\sigma_f 
&\simeq& (1-f_{239}-rf_{239}-c)\sigma_{235} + f_{239}\sigma_{239} + rf_{239}\sigma_{241} + c\sigma_{238} \\ 
&=& [\sigma_{239}-\sigma_{235}+r(\sigma_{241}-\sigma_{235})] \cdot f_{239} + [\sigma_{235}+c(\sigma_{238}-\sigma_{235})], \nonumber
\end{eqnarray}
where $\sigma_i$ represents the IBD yield of each fission isotope, $r$ is the linear coefficient between $^{241}$Pu and $^{239}$Pu, and $c$ is the constant fission fraction of $^{238}$U. It is clear the IBD yield $\sigma_f$ is a linear function of $f_{239}$ to a good approximation.  
Since both $r$ and $c$ are small numbers and differences between the $\sigma_i$s are also small, the formula can be further approximated as: $\sigma_f \simeq (\sigma_{239}-\sigma_{235}) \cdot f_{239} + \sigma_{235}$, which is the case if there are only two fission isotopes. 
One can see that the isotopic IBD yield of $^{235}$U and $^{239}$Pu can be directly extracted from the intercept and slope of this linear function. 

In 2017, the Daya Bay experiment first utilized this method~\cite{DayaBay:2017jkb} to extract the individual IBD yield of $^{235}$U and $^{239}$Pu owing to their large event statistics and long-running period that covers several reactor refueling cycles. 
Fig.~\ref{fig:ff_correlation_evolution}b shows the Daya Bay data as a function of $^{239}$Pu effective fission fraction. 
Note that since there are 6 reactor cores near Daya Bay, an effective fission fraction weighted by each core's antineutrino flux is used, but the general linear relationship in Eq.~\eqref{eq:fuel_evolution} still holds. 
The data is fitted to a line (red) and compared with the prediction from the Huber-Mueller model (blue). 
The model is scaled down by $\sim$5\% in the figure to account for the difference in the total yield. 
The normalized slope, $(d\sigma_f/df_{239})/\bar\sigma_f$ represents an additional effect separated from the RAA. If the flux deficit is the same for all fission isotopes, the normalized slope would still agree with the Huber-Mueller model prediction despite the overall deficit, therefore it is an indirect test of the sterile neutrino oscillation explanation of the RAA. 
The Daya Bay result shows that the normalized slope from the Huber-Mueller model disagrees with the best fit from data at the 2.6~$\sigma$ confidence level, which indicates sterile neutrino oscillation cannot be the sole explanation of the RAA. 
Later measurements by the RENO~\cite{RENO:2018pwo} experiment in 2018 and by the updated Daya Bay results~\cite{DayaBay:2019yxq,DayaBay:2022jnn} reached similar conclusions. 
We note that the measurement of the normalized slope is still limited by event statistics, in particular by the limited data at low and high $f_{239}$ bins, where the difference between data and model manifests. 

\begin{table*}[htb!]
\begin{center}
\begin{tabular}{|c|c|c|c|c|c|}
\hline
\hline
 Year & ($10^{-43}$~cm$^2$/fission) & $^{235}$U  &  $^{239}$Pu & $^{238}$U  & $^{241}$Pu  \\ \hline
2011 & Huber-Mueller~\cite{Mueller:2011nm,Huber:2011wv} & 6.69 $\pm$ 0.15 & 4.36 $\pm$ 0.11 & 10.10 $\pm$ 1.00 & 6.04 $\pm$ 0.13  \\
2018 & SM-2018~\cite{Estienne:2019ujo}  & 6.28 & 4.42 & 10.14 & 6.23 \\
2021 & KI~\cite{Kopeikin:2021ugh}  & 6.27 $\pm$ 0.13 & 4.33 $\pm$ 0.11 & 9.34 $\pm$ 0.47 & 6.01 $\pm$ 0.13 \\
\hline
2017 & Daya Bay~\cite{DayaBay:2017jkb} & 6.17 $\pm$ 0.17  & 4.27 $\pm$ 0.26 & -- & -- \\
2018 & RENO~\cite{RENO:2018pwo} & 6.15 $\pm$ 0.19 & 4.18 $\pm$ 0.26 & -- & -- \\
2019 & Daya Bay~\cite{DayaBay:2019yxq} & 6.10 $\pm$ 0.15  & 4.32 $\pm$ 0.25 & -- & -- \\
2020 & NEOS-II~\cite{neos_2022} & 6.32 $\pm$ 0.18  & 4.66 $\pm$ 0.26 & -- & -- \\ 
2020 & STEREO~\cite{STEREO:2020fvd} & 6.34 $\pm$ 0.16  & -- & -- & -- \\ 
\hline
\hline
\end{tabular}
\end{center}
\caption{\label{table:iso_ibd_yields} Isotopic IBD yields of $^{235}$U, $^{238}$U, $^{239}$Pu, and $^{241}$Pu from model predictions in comparison with experimental measurements from Daya Bay~\cite{DayaBay:2017jkb, DayaBay:2019yxq}, RENO~\cite{RENO:2018pwo}, STEREO~\cite{STEREO:2020fvd}, and the preliminary results from NEOS-II~\cite{neos_2022}. The predictions include the Huber-Mueller model~\cite{Mueller:2011nm,Huber:2011wv}, the SM-2108 model~\cite{Estienne:2019ujo}, and the KI model~\cite{Kopeikin:2021ugh} (see Sec.~\ref{sec:beta_spectrum_ratio}). The SM-2108 model is a pure summation-based calculation, and its uncertainty is still under evaluation (see discussions in Sec.~\ref{sec:new_calculation}).} 
\end{table*}

\begin{figure}[btph]
  \begin{centering}
    \includegraphics[width=0.46\textwidth]{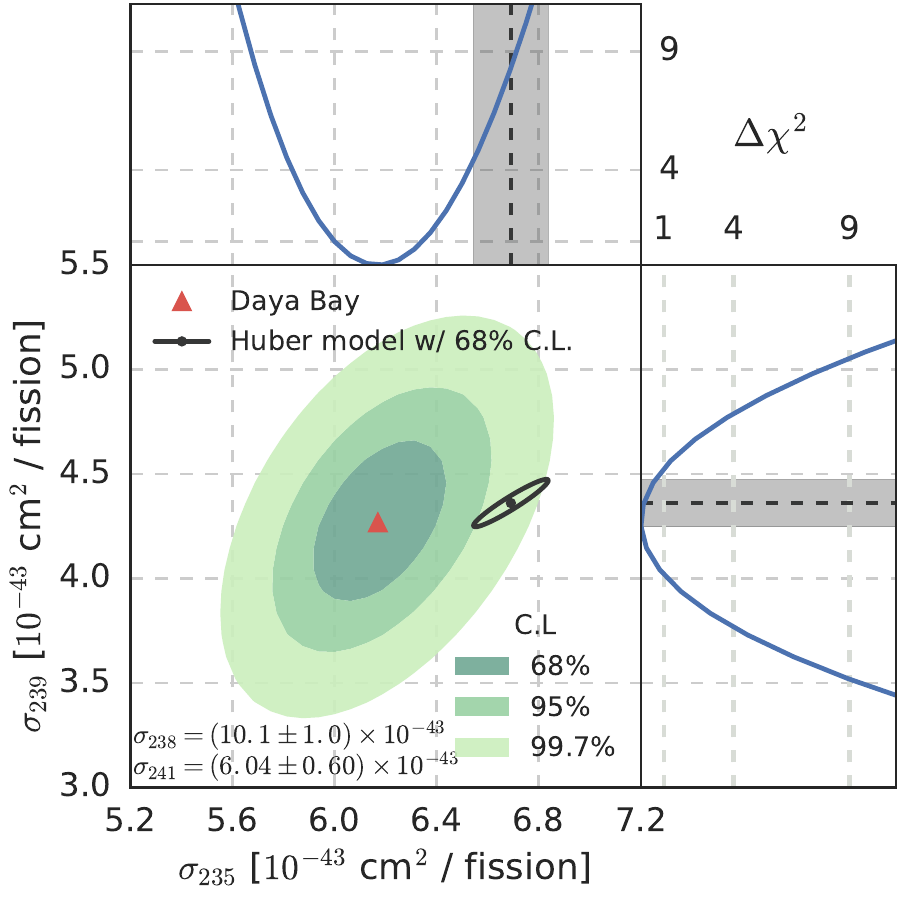}
      \includegraphics[width=0.47\textwidth]{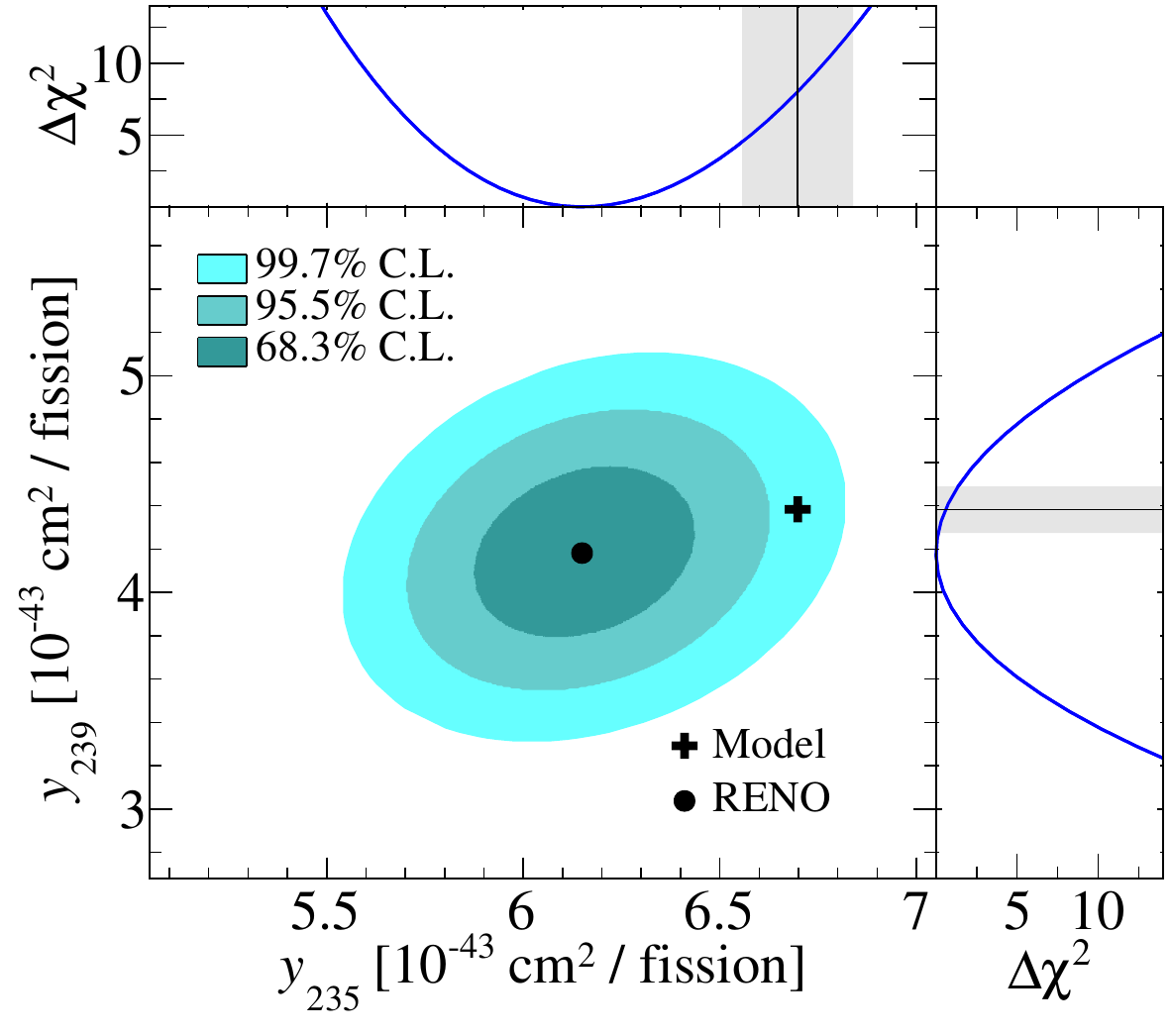}
     \put(-370,-10){a)}
      \put(-120,-10){b)}
    \par\end{centering}
    \caption{\label{fig:u235_pu239}
 (Left) Contours of measured $^{235}$U and $^{239}$Pu IBD yields per fission $\sigma_{235}$ and $\sigma_{239}$ from the Daya Bay experiment. The red triangle indicates the best-fit values of $\sigma_{235}$ and $\sigma_{239}$. Predicted values and $1\sigma$ allowed regions based on the Huber-Mueller model are shown in black. The top and side panels show one-dimensional $\Delta\chi^2$ profiles for $\sigma_{235}$ and $\sigma_{239}$, respectively. Figure taken from Ref.~\cite{DayaBay:2017jkb}.
 (Right) Similar contours of measured $^{235}$U and $^{239}$Pu IBD yields per fission from the RENO experiment in compassion with the Huber-Mueller model. Figure taken from Ref.~\cite{RENO:2018pwo}.
     }
\end{figure}

The linear fit in Fig.~\ref{fig:ff_correlation_evolution}b can be exploited further to extract the isotopic IBD yields of $^{235}$U and $^{239}$Pu using Eq.~\eqref{eq:fuel_evolution}. 
This extraction requires known values of the IBD yields of $^{241}$Pu and $^{238}$U. 
To reduce biases, model predictions of $^{241}$Pu and $^{238}$U with enlarged uncertainties (to about 10\%) are typically used in the extraction. 
Figure ~\ref{fig:u235_pu239} shows the extracted isotopic IBD yields of $^{235}$U and $^{239}$Pu from the Daya Bay and the RENO experiments. 
The experimental contours were drawn together with the theoretical predictions from the Huber-Mueller model. 
The results are summarized in Table~\ref{table:iso_ibd_yields}. 
One can see that the measured $\sigma_{235}$ is about 8\% lower than the Huber-Mueller model prediction, while the measured $\sigma_{239}$ is consistent with Huber-Mueller model within the 6\% experimental uncertainty. 
Table~\ref{table:iso_ibd_yields} also includes the preliminary measurements from the NEOS-II experiment using fuel evolution and the recent results from the STEREO experiments using HEU reactor consisting of nearly 100\% $^{235}$U fission, which are consistent with this conclusion. 
The recent fuel evolution results from the DANSS experiment~\cite{DANSS:fuel} are not included in the table, since only the ratio between the IBD yields of $^{235}$U and $^{238}$U ($1.555\pm0.052$) was reported.
These somewhat surprising results suggest that $^{235}$U may be the primary contributor to the reactor antineutrino anomaly. In other words, in the simplest scenario, a readjustment of $^{235}$U isotopic flux would resolve the RAA. 
Since $^{235}$U and $^{239}$Pu share many common fission products and therefore their beta-decay branches, it is difficult to reconcile this conclusion theoretically. 
Instead, it may hint at problems in the original ILL measurements of the electron spectra from $^{235}$U and $^{239}$Pu fission~\cite{VonFeilitzsch:1982jw, Schreckenbach:1985ep, Hahn:1989zr}, that one or both measurements may have unidentified systematic issues. 
More complex scenarios arise if one allows a combination of readjustments of individual isotopic flux and sterile neutrino oscillations. There were many literatures on this topic to study the compatibility between the sterile neutrino hypothesis and global reactor antineutrino data including fuel evolution~\cite{Giunti:2017nww,Giunti:2017yid,Gebre:2017vmm,Giunti:2019qlt,Berryman:2020agd}, and in general, the sterile neutrino oscillation hypothesis cannot be ruled out at a significant level.
This is mainly because the experimental uncertainty of $\sigma_{239}$ is still large and statistics-limited. Future experimental updates with larger data sets will help clarify the picture.
Table~\ref{table:iso_ibd_yields} further includes predictions from newer reactor antineutrino flux models such as KI~\cite{Kopeikin:2021ugh} and SM-2018~\cite{Estienne:2019ujo}. These new model predictions agree better with the results from Daya Bay and RENO, and they will be discussed in details in Sec.~\ref{sec:beta_spectrum_ratio} and \ref{sec:new_calculation}.

\begin{figure}[h]
  \begin{centering}
    \includegraphics[width=0.7\textwidth]{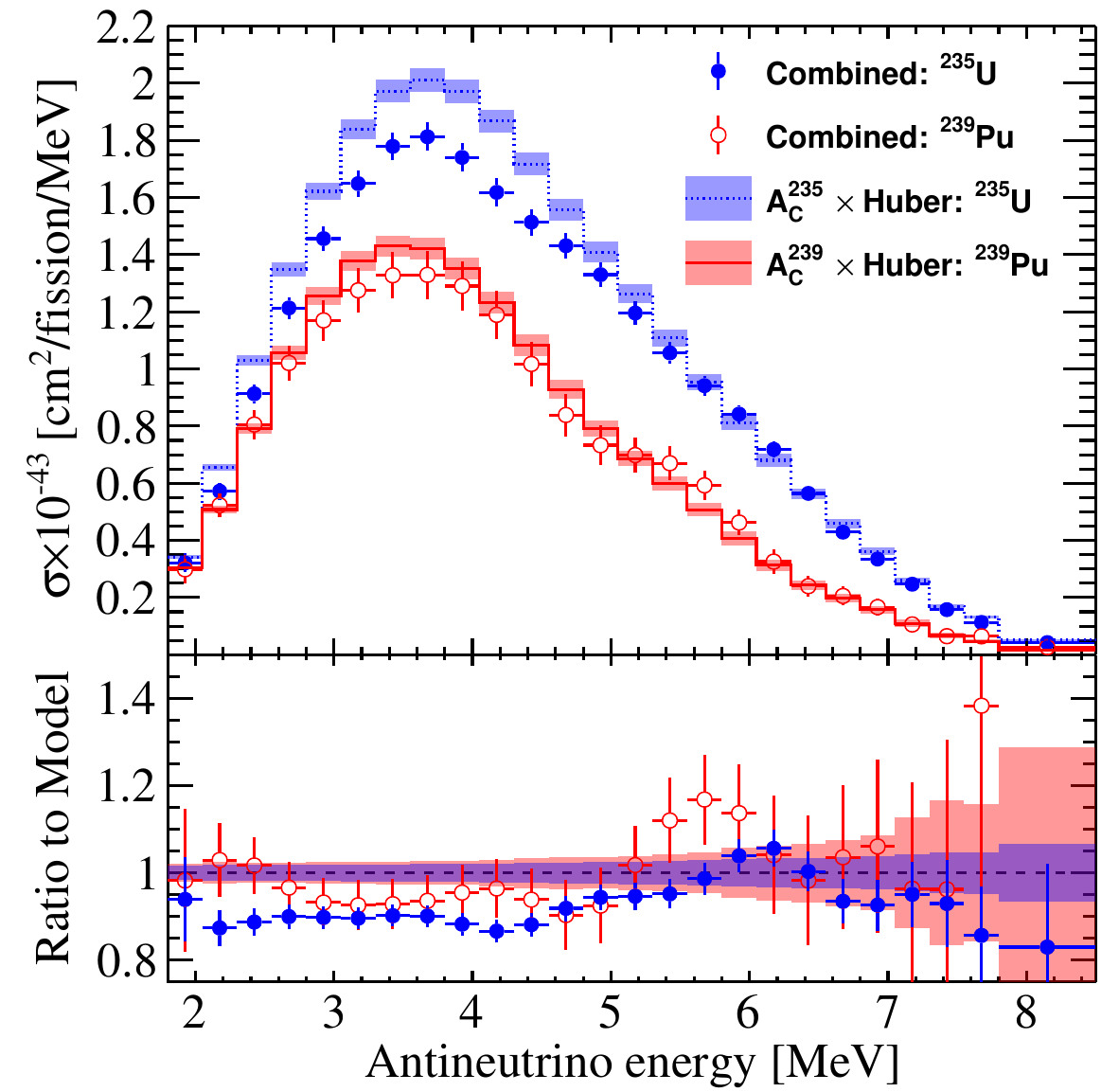}
    \par\end{centering}
    \caption{\label{fig:spec_dyb_pro}
    $^{235}$U and $^{239}$Pu antineutrino spectra unfolded from
the jointly deconvolved Daya Bay and PROSPECT measurements.
The ratio of the measurements to the Huber model prediction is shown in the bottom panel.  Figure taken from Ref.~\cite{DayaBay:2021owf}.
      }
\end{figure}

Finally, the same concept of extracting individual $^{235}$U and $^{239}$Pu IBD yields using fuel evolution can be applied to individual energy bins to extract the individual reactor antineutrino energy spectra of $^{235}$U and $^{239}$Pu. 
This obviously requires more statistics and was first performed by the Daya Bay experiment in 2019~\cite{DayaBay:2019yxq}. In 2022, the results were improved by combining the Daya Bay measurement with the PROSPECT results~\cite{DayaBay:2021owf} by the two collaborations. This is possible because the PROSPECT experiment, using a pure $^{235}$U fission reactor at HFIR, provides an additional constraint on the shape of the $^{235}$U spectrum, while the normalization of the PROSPECT spectrum is not available. 
The results are shown in Fig.~\ref{fig:spec_dyb_pro} and compared with the Huber-Mueller model predictions. We see that both the normalization and the shape do not agree with the model prediction for either isotope. In general, the deficit concentrates more at lower energy below about 5 MeV antineutrino energy. 
After normalizing to the total integral, the shape difference manifests as a ``bump'' structure at around 6 MeV (or 5 MeV in prompt energy), which is the same feature as discussed in Sec.~\ref{sec:nu_energy_spectra}. 
This ``bump'' structure is visible in both the extracted $^{235}$U and $^{239}$Pu spectra with larger statistical uncertainties for $^{239}$Pu. This hints that the biases in the model prediction for different fission isotopes could have similar origins, such as inaccurate shape factors from non-unique forbidden decays.

%% file: Beta_spectrum_ratio.tex
\subsection{Measurement of $\beta$-spectrum ratio between $^{235}$U  and $^{239}$Pu} \label{sec:beta_spectrum_ratio}

%%%% review the history
As reviewed in Sec.~\ref{sec:sub:flux_calculation}, the measurements of the cumulative $\beta^-$ spectra at ILL with the BILL
spectrometer~\cite{VonFeilitzsch:1982jw,Schreckenbach:1985ep,Hahn:1989zr} play an important role in reducing 
the uncertainties in predicting the antineutrino flux and spectra. If experimental biases exist
on the measured $\beta^-$ spectra, they would be transferred to the antineutrino spectra when using the conversion
method. The observation 
of the 4--6 MeV excess in the prompt energy spectrum, discussed in Sec.~\ref{sec:nu_energy_spectra}, 
and the observed fission-isotope dependent deficit of IBD yields
with respect to the Huber-Mueller model, discussed in Sec.~\ref{sec:evolution}, are consistent with this 
possibility. A new measurement of $\beta$-spectrum ratio between $^{235}$U spectrum and $^{239}$Pu spectrum
was recently performed at Kurchatov Institute (KI)~\cite{Kopeikin:2021rnb} to study this issue. 
Compared to the prediction of spectra,
the prediction of the ratios of the spectra (e.g. ratio between the $^{235}$U and $^{239}$Pu $\beta$ spectra: 
$\rho^5_\beta/\rho^9_\beta$; or $\nu$ spectra $\rho^5_\nu/\rho^9_\nu$) 
is expected to be less sensitive to the inputs in the calculations (e.g. nuclear 
databases)~\cite{Hayes:2016qnu,Borovoi:1982jd,Kopeikin:2012zz}. 
Specifically, the ratio of $\rho^5_\beta/\rho^9_\beta$ is estimated to agree 
with the ratio $\rho^5_\nu/\rho^9_\nu$ within 1.5--2\%~\cite{Vogel:1980bk,Kopeikin:2012zz}. 

 \begin{figure}[ht!]
  \begin{centering}
    \includegraphics[width=0.42\textwidth]{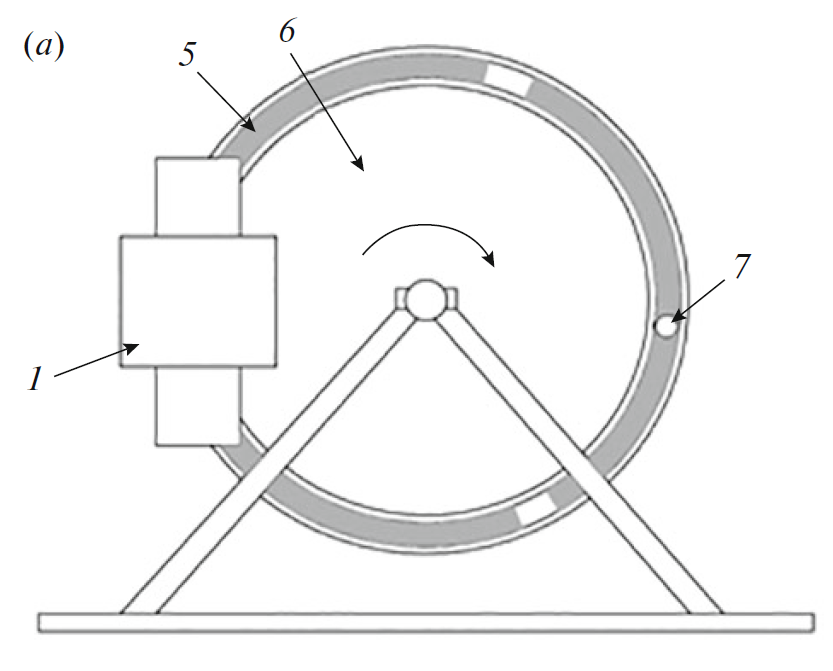}
    \includegraphics[width=0.54\textwidth]{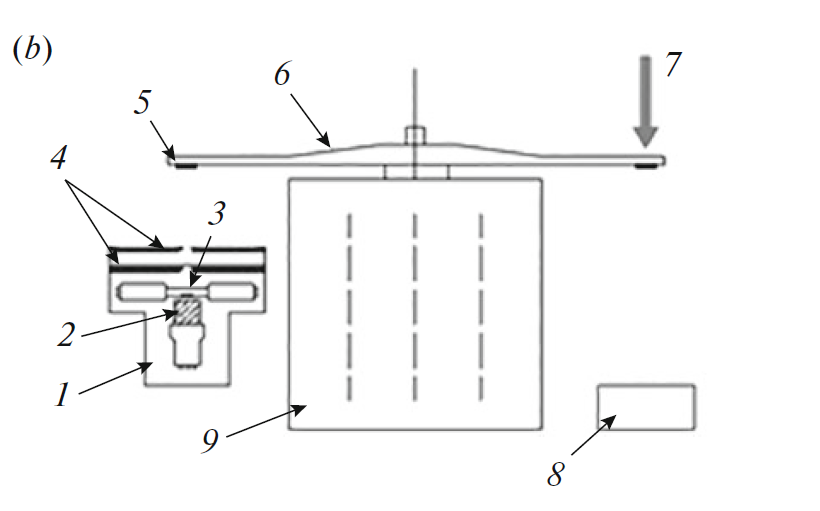}
    \par\end{centering}
    \caption{\label{fig:KI_setup} 
    Layout of the experimental setup in a beam of thermal neutrons at Kurchatov Institute (KI), 
    taken from Ref.~\cite{Kopeikin:2021rnb}:  (left) back view, (right) top view. 
    The numbers in the figure labels 1: $\beta$ spectrometer, 2: E (energy) detector,
    3: time-of-flight $\Delta$E detector, 4: diaphragms, 5: targets, 6: rotating disk used
to hold the targets, 7: neutron beam, 8: neutron trap, and 9: combined passive shielding.  }
\end{figure}

%%%% review the setup
Figure~\ref{fig:KI_setup} shows schematically the experimental setup that is deployed at the neutron-beam outlet of the IR-8 research reactor of Kurchatov Institute to measure the ratio between $^{235}$U and $^{239}$Pu $\beta$-spectra: $\rho^5_\beta/\rho^9_\beta$. 
Targets from $^{235}$U and $^{239}$Pu metallic foils (16 each) are arranged along the rim
of a rotating (at a speed of ten revolutions per second) disk, occupying 2/3 of the circle
in equal shares. The remaining 1/3 of the circle allows measuring the background.  
Targets are exposed to a neutron beam on one side of the disk, and a spectrometer records $\beta$ particles on the other side.  
A passive shielding from heavy and light materials
is placed between the neutron beam and spectrometer. 
This setup allows for a simultaneous measurement of the $\beta$-spectra of $^{235}$U, $^{239}$Pu, and background in the same neutron
beam. At the same time, the passive shielding separates the $\beta$ detection from
the target activation, which significantly reduces the background induced by beam neutrons 
as well as $\gamma$-rays in fission. 
The $\beta$ spectrometer consists of time-of-flight ($\Delta$E) and energy (E) detector layers from organic scintillators, which are separated optically. A coincidence between the 
two detectors suppresses the background of 1-MeV $\gamma$-rays by a factor of about 200.
The total energy of the two detectors yields a 12\% (ratio of the half-width to the peak position) 
energy resolution for 1-MeV electron. 
%While the $\beta$ spectra of $^{235}$U and $^{239}$Pu
%decrease fast with an increased energy, the ratio of the spectra $\rho^5_\beta/\rho^9_\beta$
%changes only by a factor of about two as the $\beta$ energy grows from 2 MeV to 7 MeV,
%which substantially relaxed the requirements on the linearity and stability of the energy scale. 
The relative measurements also naturally reduce many other systematic effects. 

%%% review the analysis ...
Measurements were performed in series of duration about 2$\times10^4$ s each with 
a total time of data acquisition of $2.3\times10^6$ s. 
%In the energy region of $E>2$ MeV,
%an average of 4, 5.5, 0.27 $\beta$ particles are recorded from $^{235}$U, $^{239}$Pu,
%and background event. 
The signal-to-background ratio is 15--20 in the region of $E>2$ MeV,
and becomes around unity at $E\approx$ 7.7 MeV. In each measurement, the spectral
distribution of $\rho^5_\beta/\rho^9_\beta$ reaches to a nearly stationary level 15 mins
after the beginning of the neutron exposure, which leads to a correction below 1.5\% at 2 MeV and decreases quickly at higher energies. 
For this relative measurement, the 
primary factor that distort the $\beta$-spectra of 
$^{235}$U  and $^{239}$Pu is the scattering and energy loss of electrons in the $^{235}$U and 
$^{239}$Pu foils and in the packing of the targets. The corrections to the $\beta$-spectra of $^{235}$U and 
$^{239}$Pu were determined by measuring the $\beta$-spectra from $^{207}$Bi, $^{56}$Mn, $^{144}$Ce-$^{144}$Pr, 
$^{42}$K, $^{38}$Cl, and $^{252}$Cf sources with different (e.g. thin vs. thick) geometry 
configurations. 
%The ratio of spectrum between the thick and thin source geometries
%$\eta{E}$ at a given energy E are nearly identical for all isotopes listed above
The correction ($\eta$) has a sizeable value at low energies with $\eta$ being 1.22, 1.10, and 1.04 at 2~MeV, 3~MeV, and 4~MeV, respectively.

The measurement of counting rate for $\beta$ particles in the spectrometer, $n_\beta$, 
is expressed as $n_\beta = \sigma \cdot F \cdot N \cdot \epsilon \cdot  \rho_\beta$,
% \begin{equation}
%     n_\beta = \sigma \times F \times N \times \epsilon \times \rho_\beta,
% \end{equation}
with $\sigma$ being the cross section for the neutron-induced nuclei fission, 
$F$ being the neutron flux density, $N$ being the number of target nuclei ($^{235}$U or 
$^{239}$Pu), $\epsilon$ being the detection efficiency of the beta particles, 
and $\rho_\beta$ being the number of $\beta$ particles emitted per fission. 
Therefore, we have 
\begin{equation}
\frac{\rho_\beta^5}{\rho_\beta^9} = \frac{\sigma^9}{\sigma^5} \cdot \frac{N^9}{N^5} \cdot \frac{n_\beta^5}{n_\beta^9},
\end{equation}
where the ratio $n_\beta^5/n_\beta^9$ is directly measured. 
The ratio $N^9/N^5$ is calculated based on the 
target mass of $^{235}$U and $^{239}$Pu. 
In order to determine the ratio 
$\sigma^9/\sigma^5$, the composition of the neutron beam is calibrated with a measurement of the Cadmium  
ratio upon the activation of a thin gold foil in the neutron beam. The beam is shown to be dominated by the thermal 
neutrons, with a small admixture ($\sim$6\%) of epithermal neutrons. The $^{235}$U ($^{239}$Pu) fission cross section 
is determined to be 553 (778) barns, after taking into account the energy-dependence of cross section and the temperature
of the neutron moderator in the research reactor. 
A reasonable variation in the amount of epithermal neutrons has 
negligible effect on the determination of $\sigma^9/\sigma^5$. 

\begin{figure}[ht!]
  \begin{centering}
   \includegraphics[width=0.7\textwidth]{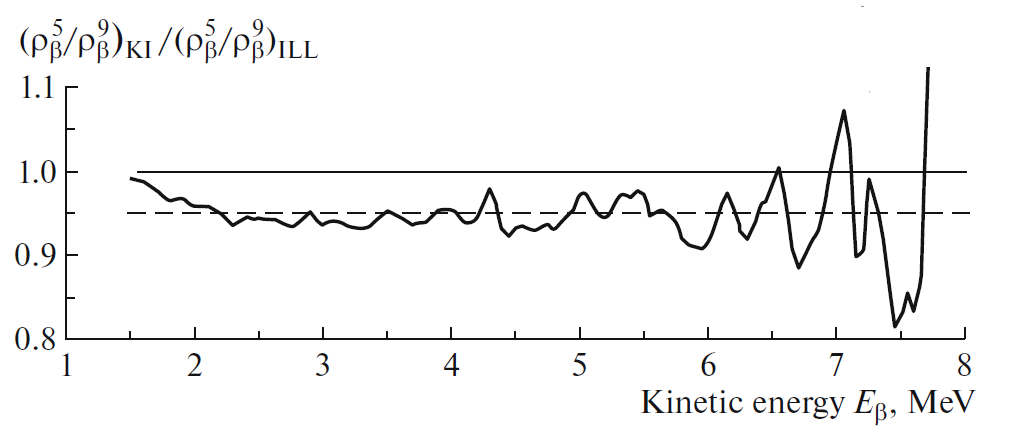}
    \par\end{centering}
    \caption{\label{fig:KI_result} The double ratio of the cumulative spectra of beta particles from 
    $^{235}$U and $^{239}$Pu fission products according to measurements at Kurchatov Institute (KI)~\cite{Kopeikin:2021rnb}
    to that of Institute Laue–Langevin 
    (ILL)~\cite{VonFeilitzsch:1982jw,Schreckenbach:1985ep,Haag:2014kia}. The double ratio is consistent with the dashed line at 0.95, and the uncertainty of the double ratio is about 1\% for electron energy below 5.5 MeV, and gradually increases to about 10\% at 7.5 MeV.
    Figure taken from Ref.~\cite{Kopeikin:2021rnb}.
%    (Top) ILL $^{235}$U/$^{239}$Pu ratios of beta-particle spectra, $\left(\rho_\beta^5/\rho_\beta^9\right)_{ILL}$ 
%(closed circles connected by the dashed curve) are compared with the KI $^{235}$U/$^{239}$Pu ratios of 
%beta-particle spectra, $\left(\rho_\beta^5/\rho_\beta^9\right)_{KI}$ (open circles connected by the 
%solid curve). (Bottom) 
%Ratio between $\left(\rho_\beta^5/\rho_\beta^9\right)_{KI}$ and 
%$\left(\rho_\beta^5/\rho_\beta^9\right)_{ILL}$ is consistent with the dashed line at 0.95. 
}
\end{figure}

%%%% review the result ...
The ratio of the beta-particle spectra $\rho_\beta^5/\rho_\beta^9$ from the KI measurement 
is compared with the previous results from ILL in Fig.~\ref{fig:KI_result}.
%are shown in Fig.~\ref{fig:KI_result}a together with . 
%The error bar represents the statistical uncertainties: $<$1\% below 5 MeV, but increasing to 30\% 
%to about 8 MeV. The general trend of the results are similar with a larger $\rho_\beta^5/\rho_\beta^9$ value
%with increasing energy followed by a sharp fall in the region around 7.5 MeV. 
%At the same time, as shown in Fig.~\ref{fig:KI_result}b, 
The double ratio between $\left(\rho_\beta^5/\rho_\beta^9\right)_{KI}$ and $\left(\rho_\beta^5/\rho_\beta^9\right)_{ILL}$
is 5\% lower than the unity over nearly the whole energy range under study, with an experimental uncertainty of about 1\% below 5.5~MeV. 
Since the ratio of $\rho^5_\beta/\rho^9_\beta$ is estimated to agree 
with the ratio $\rho^5_\nu/\rho^9_\nu$ within 1.5--2\%~\cite{Vogel:1980bk,Kopeikin:2012zz},
this result suggests a reduced $\rho^5_\nu/\rho^9_\nu$ ratio. 
This expectation is consistent with the measurements of individual $^{235}$U and $^{239}$Pu IBD yields in Table~\ref{table:iso_ibd_yields} as discussed in Sec.~\ref{sec:evolution}.

Taking into account this new result in correcting the Huber-Mueller model, the predicted IBD yield for $^{235}$U is reduced from $6.69\times10^{-43}$~cm$^2$/fission to $6.27\times10^{-43}$~cm$^2$/fission~\cite{Kopeikin:2021ugh}, which becomes much 
more consistent with the measurements: 
($6.10\pm0.15)\times10^{-43}$~cm$^2$/fission from Daya Bay~\cite{DayaBay:2019yxq} 
and ($6.15\pm0.19)\times10^{-43}$~cm$^2$/fission from RENO~\cite{RENO:2018pwo}.
Furthermore, the predicted total IBD yield~\cite{Kopeikin:2021rnb} 
is also reduced from $6.22\times10^{-43}$~cm$^2$/fission to $6.02\times10^{-43}$~cm$^2$/fission 
for an average fission fractions from a recent Daya Bay measurement~\cite{DayaBay:2019yxq} 
of 0.564, 0.304, 0.076, 0.056 for $^{235}$U, $^{239}$Pu,  $^{238}$U, and $^{241}$Pu, respectively.
In comparison, the measurement from Daya Bay is $(5.94\pm0.09)\times10^{-43}$~cm$^2$/fission. 
We see that the updated prediction taking into account new measurement of $\beta$-spectrum ratio between $^{235}$U and $^{239}$Pu~\cite{Kopeikin:2021ugh,Kopeikin:2021rnb} is consistent with the reactor antineutrino flux measurement within uncertainties.

%% file: New_calculation.tex
\section{Advancement of reactor antineutrino flux calculations}\label{sec:new_calculation} 
	
%\subsection{Evaluation of uncertainties of the conversion method}\label{sec:conversion-uncertainty}
%In view of the differences between the predictions provided by the conversion method and the measurements, a review of the conversion method and its systematic uncertainties is required.
%The converted spectra are based on a unique measurement performed at ILL with the BILL spectrometer~\cite{MAMPE1978127} using thin actinide foils exposed to a well known thermal neutron flux. This device was exceptional since it allowed a measurement between 1.8~MeV and 8~MeV of spectra by 50~keV channels with an uncertainty dominated by the normalization uncertainty of 
%3\% at 90\% C.L. except for the highest energy channels for which statistical uncertainties dominate~\cite{Schreckenbach:1981wlm,VonFeilitzsch:1982jw, Schreckenbach:1985ep,Hahn:1989zr}. Calibration of the spectrometer was performed with conversion electron sources or $(n,e^-)$ reactions on $^{207}$Pb, $^{197}$Au, $^{113}$Cd and $^{116}$In targets, providing calibration points up to 7.37~MeV.
%The irradiation time of the targets was 12 hours to two days.
%Two measurements of the energy spectrum of the electrons coming from the thermal fission of $^{235}$U have been performed with two different normalizations because of the re-evaluation of one of the calibration cross sections between the two measurement periods. It is the second measurement realized for a 12 hours irradiation that is used by the neutrino experiments of the reactors.
%% the above has been merged into earlier sections ... (XQ)

\subsection{New development of the summation method}
\label{sec:sub:new-dev-summation}

As introduced in Sec.~\ref{sec:sub:flux_calculation} and illustrated in Eq.~\ref{eq:nuspec_summation}, the summation method is based on the use of nuclear data 
assembled as a sum of all the contributions of the individual fission product's $\beta$-decay branches, weighted by the 
concentrations of the fission products in the reactor fuel (or fission target). Two types of data are 
therefore involved in the calculation: fission yield data and fission product decay property data. 
This method was originally developed by King and Perkins~\cite{King:1958zz}, followed by Refs.~\cite{Avignonne,Davis:1979gg,Vogel:1980bk} using the available data at that time. A similar calculation but supplemented by a microscopic calculation for the unknown decay schemes of neutron-rich fission products was performed in Ref.~\cite{Klapdor:1982zz}.
%The $\beta$/$\bar{\nu}$ spectra by fission of a fissionable isotope $S_k(E)$ can be decomposed into the sum of all $\beta$/$\bar{\nu}$ spectra of the individual fission products $S_{f_p}$ weighted by their activity $A_{fp}$ :
%\begin{equation}
%S_k(E)=\sum_{fp=1}^{N_{fp}}{A_{fp}\times S_{fp}(E)}
%\label{Sk}
%\end{equation} 
%Finally, the $\beta$/$\bar{\nu}$ spectrum of a fission product is the sum over the number of branches b of each $\beta$ (or $\bar{\nu}$) spectrum $S_{fp}^{b}$ (in the equation~\ref{Sfp}) from the parent nucleus to the son nucleus weighted by its branching ratio $BR_{fp}^b$ such that:
%\begin{equation}
%S_{fp}(E)=\sum_{b=1}^{N_{b}}{BR_{fp}^{b}\times S_{fp}^{b}(Z_{fp},A_{fp},E_{0 fp}^{b},E)}
%\label{Sfp}
%\end{equation}
In 1989, the measurement of 111 fission product electron spectra from Ref.~\cite{Tengblad:1989db} was used to 
perform a new summation calculation. But the agreement obtained with integral spectra measured at 
ILL~\cite{Hahn:1989zr} was only 15--20\% accurate, showing that a lot of data were still missing at that time.

In the frame of the developments performed for the Double Chooz experiment~\cite{DoubleChooz:2011ymz}, it was undertaken to re-investigate the summation method by using relevant nuclear databases.
Indeed, the summation method is the only predictive method in the absence of integrated spectra of measured electrons. 
The existing aggregate spectra needed for the conversion method are few and are measured under irradiation conditions different from those of nuclear reactors. 
% The energy distribution of the neutrons that generate the fission is different: the neutron spectrum of the ILL is purely thermal, which is not the case of a pressurized water power reactor. 
% The total spectra were measured for a very limited irradiation time, 
It is necessary to correct these spectra for off-equilibrium effects (e.g.~Table.~\ref{tab:off_eq2}), neutron capture on the fission products, and eventually to apply a correction to take into account a different energy distribution of the incident neutrons from that of 
ILL, in particular for future reactor concepts. 
Moreover, until the measurement of the spectrum of total electrons 
from the fast fission of $^{238}$U performed at Garching in 2014~\cite{Haag:2013raa}, the conversion method 
could not be applied to $^{238}$U. All these reasons have 
motivated a first re-evaluation of the summation spectra published in Ref.~\cite{Mueller:2011nm}.

\begin{figure}[hbt!]
  \begin{centering}
    \includegraphics[width=0.40\textwidth]{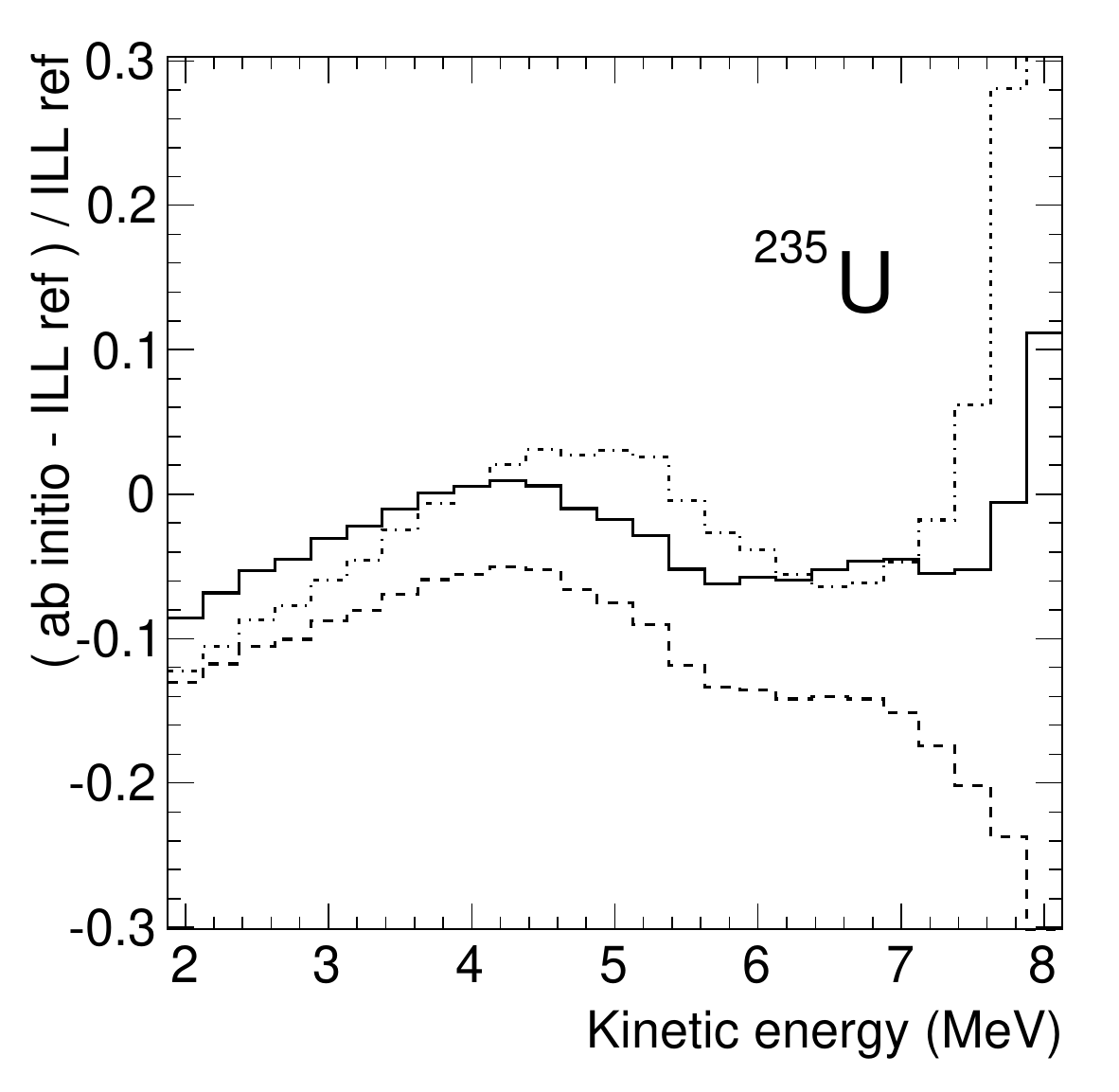}
    \includegraphics[width=0.53\textwidth]{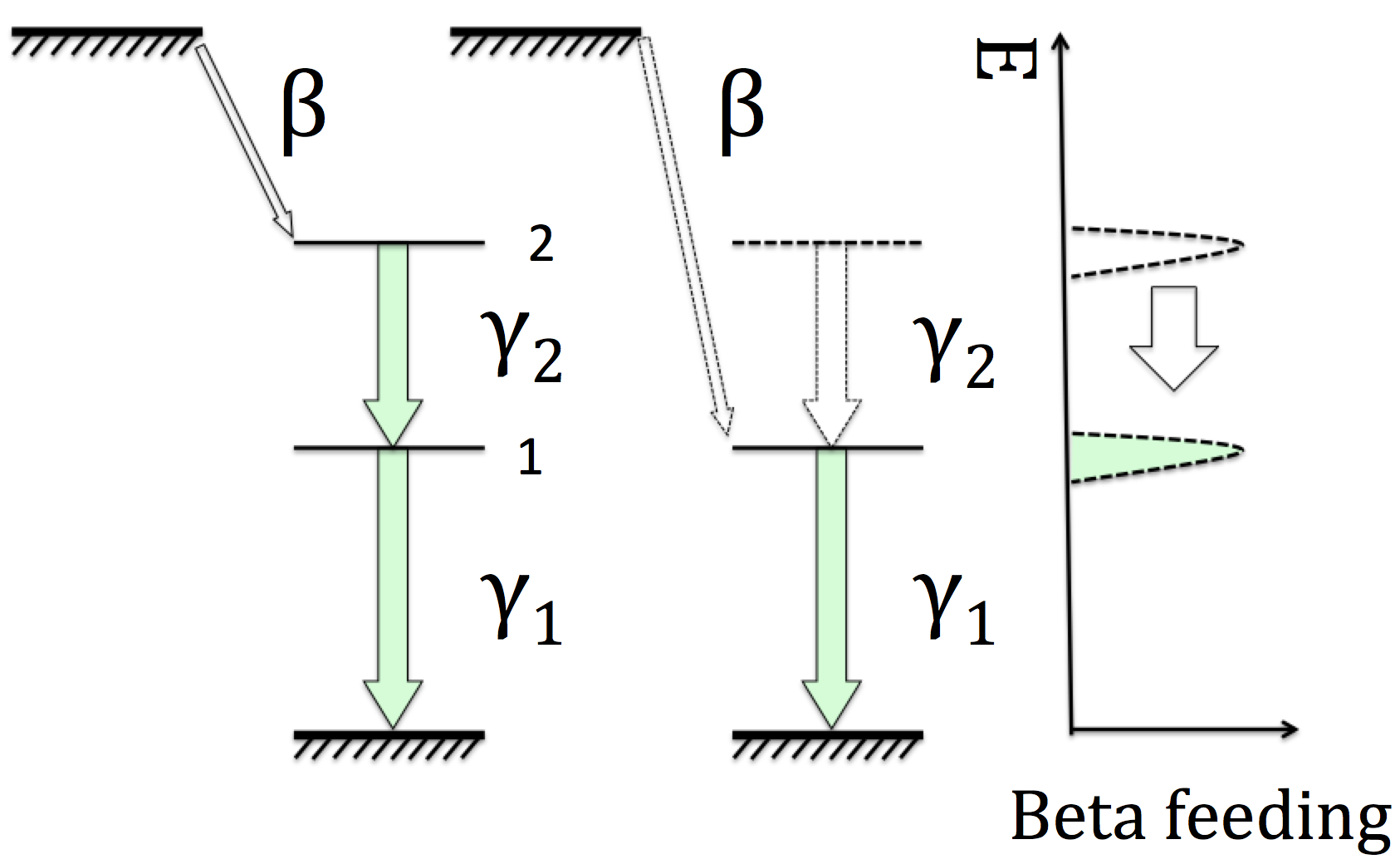} 
    \put(-370,-10){a)}
    \put(-120,-10){b)}
    \par\end{centering}
\caption{\label{fig:pande_effect} 
(left) Residues of the $^{235}$U antineutrino spectra computed as the difference of the summation calculations from Ref.~\cite{Mueller:2011nm} minus reference data from Ref.~\cite{Hahn:1989zr} divided by reference data. The different curves are the results of different choices of nuclear beta decay data: dashed--dotted curve: ENSDF data only; dashed curve: some ENSDF data replaced by pandemonium corrected data; solid curve: unmeasured $\beta$ emitters are added on top of previous curve, using the gross-theory calculations of the JENDL nuclear database and few remaining exotic nuclei described by a toy model. Figure taken from Ref.~\cite{Mueller:2011nm}. (Right) Simplified picture of a beta decay where only one excited state is populated and it de-excites by the emission of a gamma cascade. The left-hand panel represents the case. The central panel presents the Pandemonium effect, in this example represented by missing the gamma transition $\gamma_2$. The right-hand panel represents the displacement of the beta decay intensity because of the non-detection of the transition $\gamma_2$. Figure taken from Ref.~\cite{Algora:2020mhh}. 
}
\end{figure} 

% \begin{figure}
% \centering
% \resizebox{0.6\textwidth}{!}{ \includegraphics{./figs/NewCalculation/Figure3FromMueller2011.pdf}}
% \caption{\textcolor{red}{combine fig 5.1 and 5.2, find colored version of 5.1}. Taken from Ref.~\cite{Mueller:2011nm}. Residues of the $^{235}U$ antineutrino spectra computed as the difference of the summation calculations from Ref.~\cite{Mueller:2011nm} minus reference data from Ref.~\cite{Hahn:1989zr} divided by reference data. The different curves are the results of different choices of nuclear beta decay data: Blue dashed-dotted curve: ENSDF data only; red dashed curve: some ENSDF data replaced by pandemonium corrected data; solid black curve: unmeasured $\beta$ emitters are added on top of previous curve, using the gross-theory calculations of the JENDL nuclear database and few remaining exotic nuclei described by a toy model~\cite{Mueller:2011nm}. 
% }
% \label{Figure_Mueller}       % Give a unique label
% \end{figure}

Several important conclusions could be drawn regarding the summation calculations for antineutrinos from 
Ref.~\cite{Mueller:2011nm}. The nuclear databases do not contain enough data to provide accurate decay 
properties for all fission products for which fission yields are available. The evaluated databases have 
thus to be completed by other data or models for the most exotic nuclei. The relative ratio between the 
aggregate spectra of electrons and those obtained by the summation method~\cite{Mueller:2011nm} shows a 
pattern reflecting the impact of the Pandemonium effect~\cite{Hardy:1977suw}, with an 
overestimation of the high energy part of the spectrum by the nuclear data, as shown in Fig.~\ref{fig:pande_effect}a. 
The Pandemonium effect is a systematic error that affects 
some nuclei's integration data, as illustrated in Fig.~\ref{fig:pande_effect}b. 
It comes from the use of high resolution Germanium detectors. These detectors see their detection efficiency drop rapidly 
with the energy of the photons. If the levels populated during the beta decay in the daughter nucleus are 
located at high excitation energy, they can be decayed either by a cascade of low energy photons or by 
very energetic photons. In both cases, Germanium detectors have a non-negligible risk of not detecting 
these decay photons. This induces an overestimation of the beta decay branches towards the low energy 
levels of the daughter nucleus, and thus an overestimation of the energy of the electrons and 
an underestimation of the average energy of the emitted photons. The TAGS (Total Absorption 
Gamma-ray Spectroscopy) method allows removing the Pandemonium effect by using high efficiency detectors 
but with worse resolution, therefore it is complementary to high resolution spectroscopy. The principle 
of the TAGS technique is to detect (ideally) all the decay photons to be able to directly reconstruct 
the energy of the populated level by the $\beta$-decay, while the different decay $\gamma$-rays must be 
associated to each other to find the populated level in the case of Germanium detectors. The TAGS technique 
also allows the reconstruction of the beta strength distribution, which can be used to 
constrain the microscopic theoretical models~\cite{TAGS}. In 2011, the level of agreement in shape obtained 
between the summation spectra and the converted integral spectra was of the order of 10\%. Lists of nuclei 
to be measured with the TAGS technique in priority were first established by the Nantes group, which prompted a first TAGS campaign at the Jyv\"askyl\"a facility in 2009 in collaboration with the IFIC team of Valencia.

In order to establish accurate summation spectra, it is necessary to include the maximum amount of data free 
of the Pandemonium effect in the calculations. Therefore, it is necessary to analyze the data in the evaluated 
databases to identify those that might suffer from this bias and to collect the existing data that are free of it. 
The existing Pandemonium-free data in 2011 are mainly the Gamma-ray Total Absorption Spectroscopy data from 
Ref.~\cite{Greenwood} and the electron spectra from Ref.~\cite{Tengblad:1989db}. These two data sets were 
included in an optimized summation calculation that was published in 2012~\cite{Fallot:2012jv}, in which 
additional TAGS data from 7 isotopes, $^{105}$Mo, $^{102,104-107}$Tc and $^{101}$Nb were included. Among the 7 isotopes, 5 have had a large impact on the calculations of the decay heat after thermal fission of 
$^{239}$Pu~\cite{Algora:2010zz}. The impact of the new data for these 5 nuclei was also very important for the energy spectra of antineutrinos, reaching 8\% in plutonium isotopes at 6~MeV. 
% as can be seen in the left
% panel of Fig.~\ref{fig:Ratios}, which illustrates the typical shape of the correction of the Pandemonium effect 
% in the antineutrino spectra, with an increase of the spectrum at low energy (often below 2 MeV) and a decrease 
% in the high energy range, depending on the Q-values of the involved nuclei. 
It appeared after Ref.~\cite{Fallot:2012jv} 
that the summation calculations still overestimated the aggregate spectra at high energies, highlighting the 
important contributions of nuclei suffering from the Pandemonium effect in the databases.
% (see left panel of Fig.~\ref{fig:Ratios}) 
The situation is thus similar to that encountered in the summation calculations 
for the decay heat of nuclear reactors~\cite{NEA_IAEA_DecayHeat1}. These findings then reinforced the need 
for new targeted TAGS measurement campaigns, and experimental efforts were being set up in nuclear physics 
internationally.

% \begin{figure}[hbt!]
% \centering
% \includegraphics[scale=0.5]{./figs/NewCalculation/Contributors235USonzogni.png}
% \caption{\label{fig:ContributorsSonzo} Taken from Ref.~\cite{Sonzogni:2015aoa}. Calculated electron spectra 
% (solid blue [gray] line) following the thermal fission of $^{235}$U compared with the high-resolution data from 
% ILL~\cite{ILL-3}. The thin gray lines indicate the individual $\beta$ spectrum from each fission fragment and 
% thick lines highlight the 20 most important individual contributors at 5.5 MeV.}
% \end{figure} 

In Ref.~\cite{Sonzogni:2015aoa}, the authors underline the fact that in the energy region of interest around 5~MeV as indicated by the reactor antineutrino spectral anomaly (discussed in Sec.~\ref{sec:nu_energy_spectra}), the spectrum is dominated by less than 20 nuclei,
% shown in Fig.~\ref{fig:ContributorsSonzo}
reinforcing the motivations for new experimental campaigns. They also underline the fact that 65--70\% of the cross-section 
averaged antineutrino spectrum originates from the light fission fragments, among which odd-Z, odd-N nuclei dominate.
In Ref.~\cite{Sonzogni:2016yac}, a careful study of the fission yield databases has been performed. It appeared that 
the choice of databases for fission yields has an important impact on the obtained antineutrino spectra because of 
errors identified in the ENDF/B-VII.1~\cite{ENDF} database, for which corrections have been proposed. Once corrected, the 
ENDF/B-VII.1 fission yields provide spectral shapes in agreement with those obtained with the JEFF-3.1~\cite{JEFF} yields, and the measurement priority lists extracted from Ref.~\cite{Sonzogni:2016yac} are in 
agreement with those from Ref.~\cite{IGISOL:2015ifm}. New TAGS experimental campaigns were launched in parallel 
in the United States and Europe beginning in 2014--2015~\cite{Sonzogni:2015aoa,Rasco:2017dvu}.
Tables of fission products with important contributions to the reactor antineutrino spectra were published in 
Refs.~\cite{Sonzogni:2015aoa, IGISOL:2015ifm,IAEA2015}. Table~\ref{tab:Table_neutrino} is reproduced and
updated with more recent results from Ref.~\cite{Algora:2020mhh} with the indication of the measurements 
made by the TAGS collaborations worldwide with asterisks. More than half of the nuclei for which a 
priority 1 was assigned have been measured since 2009 with the TAGS technique. We will review the new TAGS
data in the Sec.~\ref{sec:sub:tags}. 

%{\bf PUT NEW UPDATED TABLE HERE}
\begin{table}[ht]
\caption{List of nuclides identified by the IAEA TAGS Consultants that should be measured using the total absorption technique to improve the 
predictions of the reactor antineutrino spectra. These nuclides are of relevance for conventional reactors based on $^{235}$U and $^{239}$Pu nuclear fuels. The list contains 34 nuclides
\cite{IAEA2015}. Relevance (Rel.)~stands for the priority of the measurement, with 1 being the highest priority. Metastable or isomeric state is indicated by letter m. Isotopes marked with asterisks show the measurements already performed by the TAGS collaborations.
%Nuclides marked with $\dagger$ are also relevant for the $^{233}$U/$^{232}$Th fuel, see additional cases in Table \ref{tab:3}.
 Table reproduced from Ref.~\cite{Algora:2020mhh}.}
\label{tab:Table_neutrino}       % Give a unique label
\centering 
\begin{tabular}{l c l c l c}
Isotope & Rel. & Isotope & Rel. & Isotope & Rel.\\\hline
36-Kr-91  	& 2  & 39-Y-97m          & 1 & 53-I-138$^*$     & 2\\
37-Rb-88      & 1 & 39-Y-98m  & 1 & 54-Xe-139$^*$       & 1\\
37-Rb-90$^*$  & 1 & 39-Y-99$^*$  & 1  & 54-Xe-141   & 2\\
37-Rb-92$^*$            & 1 & 40-Zr-101 & 1 & 55-Cs-139 & 1\\
37-Rb-93$^*$            & 1 & 41-Nb-98$^*$  & 1 & 55-Cs-140$^*$      & 1\\
37-Rb-94$^*$                         & 2 & 41-Nb-100$^*$                    & 1 & 55-Cs-141    & 2\\
38-Sr-95$^*$     & 1 & 41-Nb-101$^*$  & 1 &55-Cs-142$^*$     & 1\\
38-Sr-96                           & 1 & 41-Nb-102$^*$  &1  & 57-La-146                        & 2\\
38-Sr-97                                & 2 & 41-Nb-104m  & 2 &                  & \\
39-Y-94              & 1 & 52-Te-135                     & 1 &                      & \\
39-Y-95$^*$             & 1 & 53-I-136                     & 2 &                       & \\
39-Y-96$^*$            & 1 & 53-I-136m                    & 1\\
39-Y-97    & 2 &  53-I-137$^*$         & 1\\
\noalign{\smallskip}\hline
\end{tabular}
% Or use
% \vspace*{5cm}  % with the correct table height
\end{table}

Overall, there are two summation models, SM-2018~\cite{Estienne:2019ujo,Fallot:2012jv} and Sonzogni {\it et~al.}~\cite{Sonzogni:2017wxy,Sonzogni:2016yac}, that use the JEFF fission yields and different sets 
of decay data. These two models tend to agree on the inclusion of Pandemonium-free data in priority. 
In Ref.~\cite{Fallot:2012jv} that were later updated in SM-2018~\cite{Estienne:2019ujo},
the ingredients are detailed. The first priority is given to 
TAGS data and then to Rudstam's data~\cite{Rudstam:1990vnk} in order to minimize the Pandemonium effect. 
Then decay data from the JEFF-3.3 evaluated database~\cite{Plompen:2020due} are taken into account, 
supplemented by ENDF/B-VIII~\cite{Brown:2018jhj} and Gross Theory from JENDL~\cite{GrossTheory1,GrossTheory2}. A particularity is that in this model, all the fission products present in the fission yields database have decay data. For the 
few ones for which no decay data exist, a toy model is used with 3 decay branches with equal probability at the end-point energy of each third 
of the Q-value. The Sonzogni model~\cite{Sonzogni:2017wxy} uses the ENDF/B-VIII decay database but includes 
TAGS data as highest priority and some of the data measured by Ref.~\cite{Rudstam:1990vnk} as well. The decay 
data provided to the most exotic nuclei are the ones predicted by the model used in the evaluated database 
ENDF/B-VIII. The lists of nuclei with important contributions to the antineutrino energy spectra from the two summation models mainly agree and 
were compared in 2015 at the occasion of a TAGS Consultant meeting organized by the IAEA Nuclear Data 
Section~\cite{IAEA2015}. 

An important area of the summation models is the computation of the 
associated uncertainties. It is a very tough task since the covariance matrices associated to nuclear 
data are scarce or even non-existent. Several teams have worked on the determination of the uncertainties 
associated to summation spectra (see presentations from Sonzogni and Fallot at the 2019 IAEA meeting~\cite{IAEA2019}, and 
Ref.~\cite{Perisse:2023efm}). The first mandatory ingredient is the covariance matrix associated to 
the set of fission yields used in the model. Some of the sets of evaluated fission yields now have 
covariance matrices computed by different groups, but how they compare to each other is still 
unstudied. The second mandatory ingredient is the covariance matrices associated to the evaluated decay data. 
These matrices do not exist to date. It is also very difficult to associate a systematic error to beta feedings 
which are suspected to suffer from the Pandemonium effect, since the presence and intensity of the systematic 
effect can only be known after a TAGS measurement has been performed. More efforts in this area are 
needed. 

\subsection{New data from TAGS experiments}
\label{sec:sub:tags}

A first TAGS experimental campaign was proposed 
% by Ref.~\cite{ProposalJyv} 
in Jyv\"askyl\"a~\cite{IGISOL} as early as in 2009 motivated by antineutrino spectra. 
During this first set of measurements, the Rocinante 
detector, which is composed of 12 BaF$_2$ crystals covering a solid angle of almost 4$\pi$ around the center 
of the device, was used. The detectors are distributed according to a cylindrical geometry with a length 
and external diameter of 25~cm and a central longitudinal hole of 5~cm diameter. The total detection efficiency varies from 88\% for a 1~MeV photon to 78\% at 10~MeV. The advantage of this detector is its low sensitivity to neutrons 
(one order of magnitude less than that of NaI) and the possibility to use the rapidity of the response of 
BaF$_2$ crystals for gamma/neutron discrimination. The disadvantage lies in the energy resolution 
of these crystals, which is much worse than NaI crystals. The light response of each crystal is read independently by photomultiplier tubes (PMTs), which gives access to the 
information on the multiplicity of the cascades of gamma photons per event. During the 2009 campaign, a silicon 
electron detector, with an efficiency close to 30\%, was placed in the center of the detector to realize the 
beta-gamma coincidences and thus eliminate most of the background noise.

Fission products of interest are produced by the fission of a fine uranium target subjected to a proton beam 
with the IGISOL-IV ion guide at Jyv\"asky\"a. The ions produced are then slowed down and ionized in a hydrogen 
gas. The flow of the gas carries the ions away from the target to the acceleration and guidance system. The 
isotopes are then separated in mass with a magnetic dipole of resolution power M/$\Delta\rm{M}\sim500$~\cite{IGISOL}. 
The ions are then directed to the JYFLTRAP system at the end of the IGISOL line. JYFLTRAP is an ion trap to slow down, collect and isobarically purify the radioactive beam. It is composed of two main elements: 
a radiofrequency quadrupole, followed by two Penning traps used to purify the beam and to separate the ions of 
interest. The first Penning trap allows to reach a mass resolution of M/$\Delta\rm{M}\sim10^5$, while the second one 
allows to reach M/$\Delta\rm{M}\sim10^8$.

%% I do not understand the following statement (XQ) ... 
%The solution of the inverse problem is realized by following the expectation-maximization method~\cite{NIMTain}.

In 2009, the $^{91-94}$Rb and $^{86-88}$Br nuclei were measured with the Rocinante total absorption spectrometer~\cite{Valencia:2016rlr}
placed after the two Penning traps of the IGISOL~\cite{IGISOL} beamline. As shown in Table.~\ref{tab:Table_neutrino},
$^{92,93}$Rb were among the top priority nuclei with important contributions to the antineutrino spectrum.
$^{92}$Rb alone accounts for 16\% of the spectrum emitted by a pressurized water reactor between 5 and 8~MeV. 
Its contribution to the $^{235}$U and $^{239}$Pu spectra is 32\% and 25.7\% in 6--7~MeV 
and 34\% and 33\% in 7--8~MeV, respectively. In 2009, the intensity of the branch at the ground state (GS) of the 
daughter nucleus was fixed at 56\% in the ENSDF~\cite{ENSDF} structure database but was re-evaluated 
at 95.2\% in 2014. In 2015, the $^{92}$Rb TAGS measurement~\cite{IGISOL:2015ifm} obtained a GS to GS feeding of $(87\pm2.5)\%$, which had a large impact on the spectra of 
antineutrinos.
%from our analysis of the TAGS data, performed in the thesis of Zakari Issoufou that I co-supervised. This result  
% The case of $^{92}$Rb is particular because it is an ``anti-Pandemonium'' case, since the beta feeding to the ground state 
% was underestimated in the previous evaluations. 
% In the analysis of this core, the sensitivity of the reconstructed 
% spectrum (and thus of the $chi^2$ obtained between the data and the reconstruction) to the value of the decay branch 
% towards the fundamental level of the daughter nucleus is very large, because of the great weight of the electrons 
% in the Rocinante spectrometer. 
%The uncertainties have been obtained by varying the input parameters of the analysis 
%as the calibration parameters, the thickness of the electron detector (a silicon in this case), 
%the density of levels in the daughter nucleus, the normalization of the background, etc...
Before this measurement, the $^{92}$Rb beta decay data used in the summation calculations were those of 
Tengblad {\it et al.}~\cite{Tengblad:1989db}. After replacing these data with the TAGS measurement, the impact on the antineutrino flux is 4.5\% for $^{235}$U, 
3.5\% for $^{239}$Pu, 2\% for $^{241}$Pu, and 1.5\% for $ ^{238}$U. The impact is similar on the summation model 
in Ref.~\cite{Sonzogni:2015aoa} but much more important for Ref.~\cite{Dwyer:2014eka} in which none of the data free of Pandemonium effect has been introduced. In this case, the 
observed impact on the antineutrino flux exceeds 25\% in $^{235}$U, which shows that the quality of the summation model depends mainly 
on the selection of the data used. This election requires a critical look at the data of nuclear structure present 
in the databases, where physical inconsistencies indicate the presence of possible systematic biases 
like the Pandemonium effect.

While not in the top 10 of the main contributing nuclei to the antineutrino spectrum, the measurement of $^{86-88}$Br 
and $^{91, 94}$Rb was originally motivated by the study of gamma emission above the neutron emission threshold and was 
carried out in the second experiment. 
%$^{86-88}$Br and $^{91}$Rb were not on the priority list of~\cite{IAEA2015}. 
Although their TAGS measurement confirmed that the Pandemonium effect affected the existing data,
$^{86, 87}$Br and $^{91}$Rb did not show a large impact on the antineutrino 
spectra~\cite{Valencia:2016rlr,Tain:2015qln,Rice:2017kfj}. 
 
In contrast, measurements of $^{94}$Rb, of priority 2 in Ref.~\cite{IAEA2015}, and $^{88}$Br were found to 
produce a large impact on the spectra, highlighted by two different summation calculations~\cite{Valencia:2016rlr}. 
In the first calculation, the new TAGS data replace the high resolution spectroscopy data and so the impact obtained 
is of the typical Pandemonium effect correction, i.e. a decrease of the high energy part of the total antineutrino 
spectrum. The effect reaches 4\% in $^{235}$U and $^{239}$Pu in the case of $^{94}$Rb and only 2--3\% in the case of 
$^{88}$Br in the interval from $8$ to 9~MeV. This range explains why $^{88}$Br was not part of the published priority 
list that was established for a contribution to the spectrum of a PWR greater than 1\% between $3$ and 8~MeV.
In the second calculation, the TAGS data replace the measurements of Tengblad {\it et al.}, which in principle do not 
suffer from the Pandemonium effect. The replacement of $^{87}$Br shows little effect, that of $^{94}$Rb reaches 3\% at 
8~MeV but that of $^{88}$Br reaches 7\% between $8$ and 9~MeV with a compensation of the effect of the last two nuclei 
below 8~MeV.

\begin{figure}
\centering
%compaGreenW2017.pdf -> Figure_16.eps
\resizebox{0.44\textwidth}{!}{\includegraphics{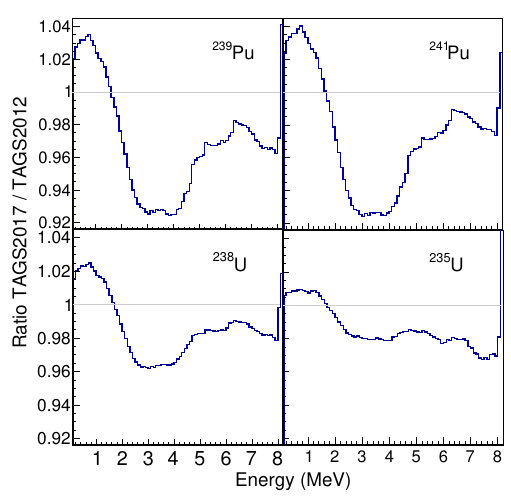}}
\resizebox{0.48\textwidth}{!}{ \includegraphics{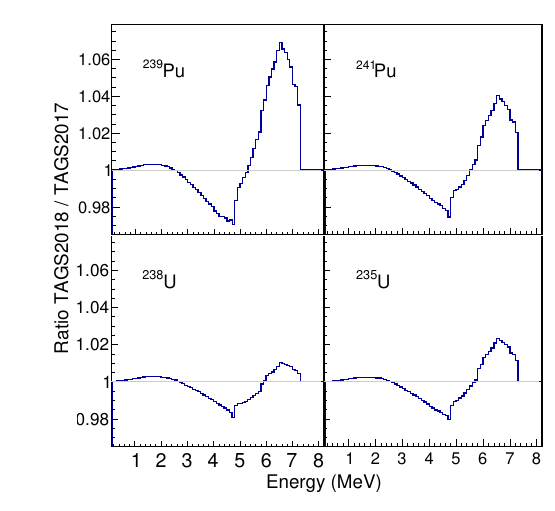}}
\caption{(Left) Accumulated impact of the beta intensities of the $^{86,87,88}$Br 
and $^{91,92,94}$Rb~\cite{Valencia:2016rlr,Tain:2015qln,IGISOL:2015ifm} decays measured with the total absorption 
spectrometer {\it Rocinante} on the antineutrino spectra with respect to that published in Ref.~\cite{Fallot:2012jv} 
for the thermal fissions of $^{235}$U, $^{239}$Pu and $^{241}$Pu, and the fast fission of 
$^{238}$U.
% ~\cite{Magali_prl}. 
(Right) Accumulated impact of the beta intensities of $^{100,100m,102,102m}$Nb measured with the DTAS detector~\cite{Guadilla:2019aiq,Guadilla:2019zwz} 
on the antineutrino spectra. 
Figures taken from Ref.~\cite{Algora:2020mhh}.
}
%The figures represent the relative impact of the $^{86,87,88}$Br and $^{91,92,94}$Rb decays.}
\label{rocinante_cumulated_impact}       % Give a unique label
\end{figure}

The cumulative impact of the beta intensities measured with the Rocinante detector at Jyv\"askyl\"a on the energy spectra 
of antineutrinos generated by the thermal fission of $^{235}$U, $^{239}$Pu and $^{241}$Pu, and the fast fission of $^{238}$U
is calculated with the summation model of Ref.~\cite{Estienne:2019ujo} and presented on the 
left panel of Fig.~\ref{rocinante_cumulated_impact}
compared to the spectrum constructed with the most recent evaluated decay databases JEFF-3.3~\cite{Plompen:2020due} and 
ENDF/B-VIII.0~\cite{Brown:2018jhj} for the same nuclei and containing only the TAGS data of Ref.~\cite{Greenwood}.
The decrease of the spectra of the two plutonium isotopes above 1.5~MeV reaches 8\%. The impact on the 
spectra of the two uranium isotopes is about 2\% and 3.8\% between $3$ and 4~MeV for $^{235}$U and $^{238}$U, respectively.

In 2014, a second experimental campaign was performed at the Jyv\"askyl\"a  facility almost exclusively dedicated 
to nuclei of importance for antineutrinos and reactor decay heat, using the DTAS detector~\cite{Guadilla:2018lcw} 
coupled to a plastic detector to perform $\beta$-$\gamma$ coincidences. 
% The DTAS detector has been installed after the exit of the precision room. 
The DTAS is composed of 18 NaI crystals of dimensions 15~cm$\times$15~cm$\times$25~cm 
each, coupled with a beta detector, a High Purity Germanium detector (HPGe) and a magnetic band to transport 
the ions. The detector is surrounded by a 5-cm thick lead shielding. The beta detector is a plastic detector of 2~mm 
thickness placed at the end of the beam tube. The role of the HPGe is to identify the possible presence of contaminants.
Twenty-three isotopes have been measured with, among them, many isomers that require the separation power of the Penning's 
traps of Jyv\"askyl\"a. An illustration of this experimental challenge is provided by the case of Niobium isomers 
$^{100, 100m}$Nb and $^{102, 102m}$Nb. Niobium is a refractory element and the isomers of $^{100}$Nb ($^{102}$Nb) are separated only by 313~keV 
(94~keV) with lifetimes close to each other: 1.5~s and 2.99~s (4.3~s and 1.3~s). $^{100}$Nb and $^{102}$Nb 
were assigned a priority 1 in Table~\ref{tab:Table_neutrino}. $^{100}$Nb is among the largest contributors to the 
antineutrino flux in the anomaly region in the shape of the spectra, similar to $^{92}$Rb, $^{96}$Y and $^{142}$Cs.

%\begin{figure}
%\centering
%compa20172018.pdf -> Figure_17.eps
%\resizebox{0.48\textwidth}{!}{ \includegraphics{./figs/NewCalculation/ImpactDTASReviewPaper.pdf}}
%\caption{Accumulated impact of the beta intensities measured with the DTAS detector on the antineutrino spectra with respect to that presented in Figure \ref{rocinante_cumulated_impact} (relative ratios) for the thermal fissions of $^{235}$U, $^{239}$Pu and $^{241}$Pu, and the fast fission of $ ^{238}$U \cite{Magali_priv}. The figure represents the relative impact of the $^{100,100m,102,102m}$Nb decays \cite{Guadilla:2019aiq,Guadilla:2019zwz}. }
%\label{DTAS_cumulated_impact}       % Give a unique label
%\end{figure}

The results showed that the high resolution measurements of $^{100, 100m}$Nb and $^{102}$Nb were affected by 
the Pandemonium effect, while the intensity distribution for $^{102m}$Nb was determined for the first
time~\cite{Guadilla:2019aiq,Guadilla:2019zwz}. The impact of these measurements on the summation calculations 
were computed in Ref.~\cite{Algora:2020mhh} and can be seen in the right panel of Fig.~\ref{rocinante_cumulated_impact}. 
The result is a decrease of the spectra between $3$ and 6~MeV and a sharp increase at 6.5~MeV in the region of the 
shape anomaly. In the calculation, the TAGS data have replaced the high resolution spectroscopy data 
from~JEFF-3.3~\cite{Plompen:2020due} and ENDF/B-VIII.0~\cite{Brown:2018jhj}. Finally, the difference between the experimentally measured antineutrino 
spectra and the summation calculation including these data becomes smaller, although the shape anomaly (see Sec.~\ref{sec:nu_energy_spectra}) has not completed disappeared~\cite{Guadilla:2019aiq,Guadilla:2019zwz}.

\begin{figure}[ht]
  \centering
\includegraphics[width=0.6\textwidth]{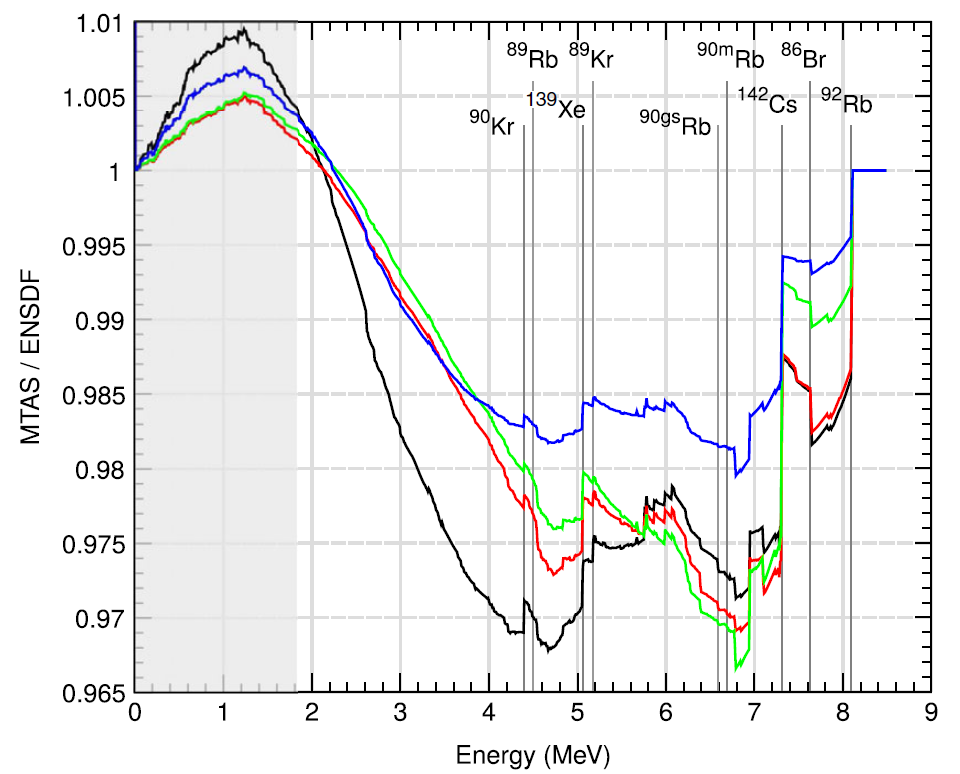}
\caption{MTAS measurement and data contained in the ENSDF database, designated 
for the most important nuclear fuel types: $^{235}$U (black), 238 U (blue), 239 Pu (red), and 241 Pu (green). The shaded 
area denotes antineutrino energies below the inverse $\beta$-decay threshold. Vertical lines indicate the Q$_{\beta}$ values of 
the measured nuclei. Figure taken from \cite{PhysRevLett.119.052503}. }
  \label{Fig:Fijal}
\end{figure}

Besides the efforts in Europe, TAGS campaigns were also carried out in the US with the Modular Total Absorption 
Spectrometer (MTAS) constructed at the Holifield Radioactive Ion Beam Facility (HRIBF)~\cite{Wolinska-Cichocka:2014ohc}. 
This spectrometer consists in 18 hexagonal modules of NaI of 21-inch length and 7-inch side to side, arranged in a honeycomb-like
structure. A 19th detector is placed in the center of the assembly with a 2.5-inch hole drilled through. The detector was 
coupled to the Online Test Facility at the Tandem accelerator of the HRIBF at ORNL. A 40 MeV proton beam was 
impinging on a $^{238}$U-carbon target placed inside a plasma source, and the fission products were mass selected by a 
magnetic separation and implanted into the Mylar tape upstream of the detector center. Radioactive samples were 
periodically transported into the center of MTAS and positioned between two silicon strip detectors by the 
tape transport system. 
In 2016, the MTAS collaboration reported the new TAGS measurements for the three main contributors 
to the reactor antineutrino spectrum above 4 MeV, $^{92}$Rb, $^{96}$Y and $^{142}$Cs~\cite{Rasco:2016leq}. $^{92}$Rb and
$^{96}$Y exhibit very large ground-state (GS) to ground-state beta branches of first-forbidden non-unique type. The GS 
to GS branch measured for $^{92}$Rb amounts to $(91 \pm 3)\%$ is consistent with the result obtained by the previous 
TAGS measurement by Ref.~\cite{IGISOL:2015ifm}. The GS-feeding measured for $^{96}$Y confirms the previous high-resolution
spectroscopy value of 95.5\% but with a larger uncertainty of $\pm$2\% (instead of 0.5\%). Indeed, the analysis of this 
latter case is complex because of an important $0^+$-$0^+$ E$_0$ transition that was not taken into account in the 
analysis of the MTAS data. The beta-feeding intensity for $^{142}$Cs was also measured, with a ($44\pm2$)\% GS-feeding as 
compared with 56\% in ENSDF~\cite{ENSDF}. 
This reduction of the GS-feeding, typical from the correction for the 
Pandemonium effect, is expected to reduce the predicted antineutrino flux. 
In 2017, the MTAS collaboration reported
in Ref.~\cite{PhysRevLett.119.052503} results of a $\beta$-decay study of fission products $^{86}$Br, $^{89}$Kr, $^{89}$Rb,
$^{90gs}$Rb, $^{90m}$Rb, $^{90}$Kr, $^{92}$Rb, $^{139}$Xe, and $^{142}$Cs, which are shown in Fig.~\ref{Fig:Fijal}. 
These TAGS measurements decreased the emitted antineutrino flux by the 4 main contributors to
fissions in PWRs by about 2\% in average with respect to summation calculations performed with the ENDF/B-VII.1 decay database. 

%MTAS publications 2022 to be added

\subsection{Results from the new summation method}

\begin{figure}[ht]
  \centering
\includegraphics[width=0.47\textwidth]{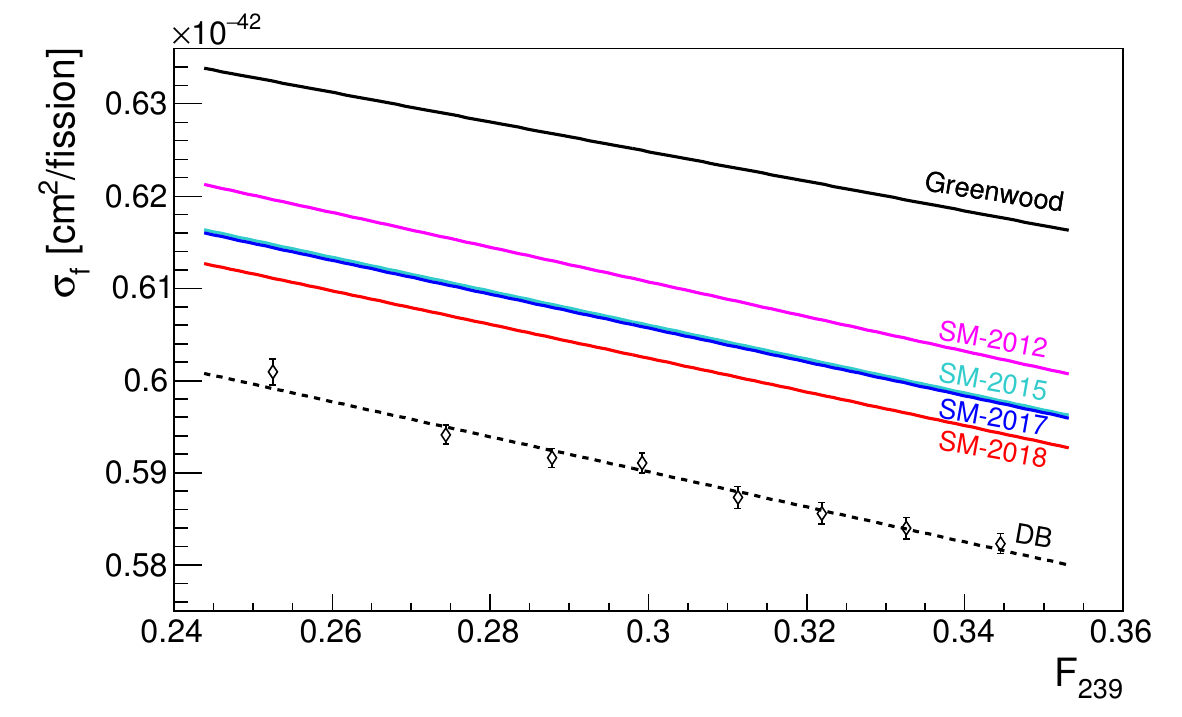}
\includegraphics[width=0.45\textwidth]{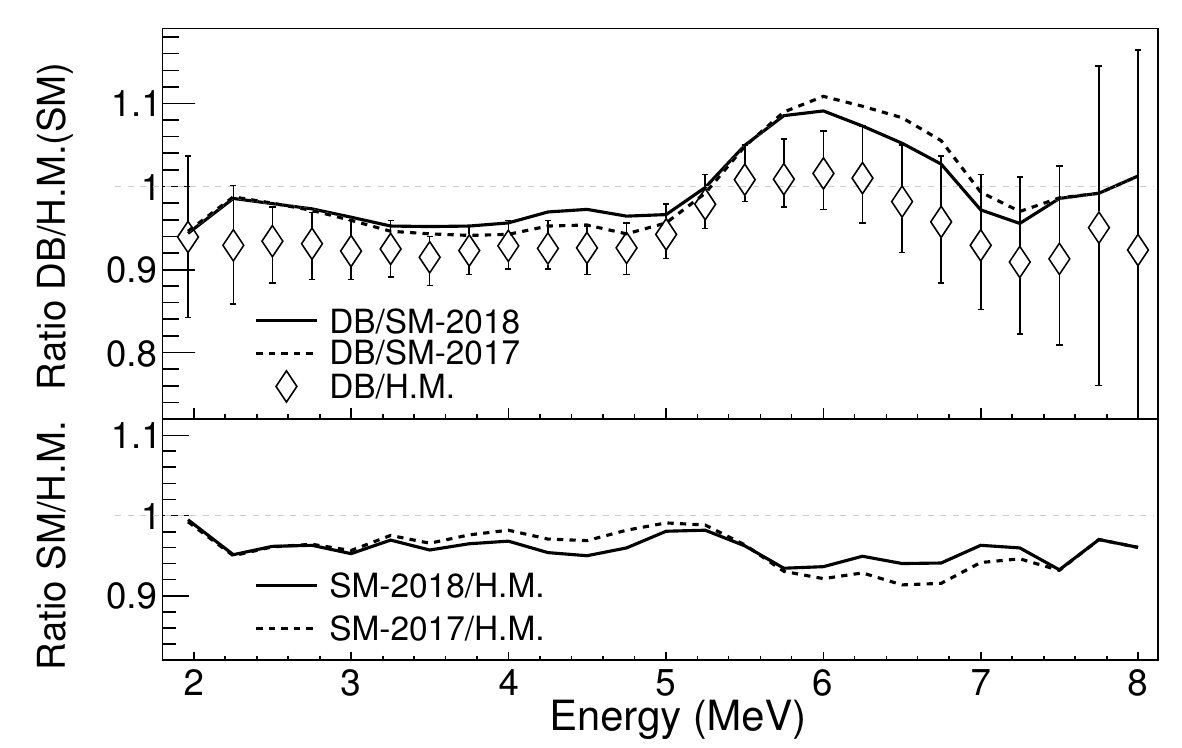}
\caption{ 
(Left) IBD yields as a function of the $^{239}$Pu fission fractions 
for different versions of the summation model are shown. Data (open diamonds) are extracted from Ref.~\cite{DayaBay:2017jkb}.
(Right) Upper Panel: Ratio of the Daya Bay antineutrino energy spectrum to that of the Huber-Mueller model as in
Ref.~\cite{DayaBay:2017jkb} (open diamonds), to that of the SM-2017 (dashed line) and the SM-2018 (continuous line) models.
Lower Panel: Ratio of the SM-2017 (dashed line) and SM-2018 (continuous line) antineutrino energy spectra 
to that of the Huber-Mueller model~\cite{Mueller:2011nm, Huber:2011wv}. Figure taken from Ref.~\cite{Estienne:2019ujo}.
}
\label{Fig:Global_impact_Flux_DB}
\end{figure}

The progression of the new summation method (SM) is summarized in Ref.~\cite{Estienne:2019ujo} as more and more TAGS results become available. The left panel of Fig.~\ref{Fig:Global_impact_Flux_DB} shows the detected antineutrino flux (i.e.~IBD yield) 
as a function of the $^{239}$Pu fission fractions obtained with the summation method updated according to the TAGS 
results that were have obtained along the way since 2009 by the Nantes-Valencia TAGS collaboration. It is remarkable 
that the inclusion of more TAGS data systematically decreases the predicted antineutrino flux to reach a difference 
with Daya Bay of only 1.9\%. This systematic decrease is the consequence of the correction of the Pandemonium effect 
from the new data of the measured nuclei, and it is expected that the remaining discrepancy with the neutrino data will 
decrease more with future TAGS measurements.
% , which do not leave much room for the reactor anomaly.
%More details can be found in~\cite{Estienne:2019ujo} where 
In addition, the individual IBD rates associated with $^{235}$U, $^{239}$Pu, $^{241}$Pu, and $^{238}$U obtained with 
the summation model are also consistent to Daya Bay results (see Table~\ref{table:iso_ibd_yields}). 
This contrasts with the Huber-Mueller model, for which a large difference is observed in the case of $^{235}$U while 
the other three cases are in very good agreement with the experiment. As discussed in Sec.~\ref{sec:evolution},
the variation of the IBD yields with the $^{239}$Pu content is also an indication of the origin of the anomaly, since it
reflects if the deficit is shared among the fission isotopes or not. The new SM model is in good agreement 
with ${d\sigma_f}/{dF_{239}}$= -1.82$\times$10$^{-43}$cm$^2$ compared with the value measured by Daya Bay
in~\cite{DayaBay:2017jkb} of ($-1.86\pm0.18$)$\times$~10$^{-43}$cm$^2$.

% The success of the summation models in the prediction of the detected antineutrino flux as a function of the 
% fuel composition is an important step toward an independent confirmation of the recent results of the reactor 
% neutrino experiments with respect to the reactor antineutrino anomaly, which now favours a nuclear physics 
% explanation. 
% In 2019 and 2023, technical meetings organized by the IAEA-NDS was devoted to the reactor antineutrino spectra, gathering neutrino physicists, reactor physicists and nuclear physicists~\cite{IAEA2019, IAEA2023}. In 2019, all Communities acknowledged the huge experimental effort with TAGS, bringing the summation method to another level~\cite{IAEA2019}. 
Despite the success in prediction of the integrated antineutrino flux, the situation regarding the prediction of the reactor 
antineutrino energy spectrum is still not satisfactory. As can be seen in the right panel of
Fig.~\ref{Fig:Global_impact_Flux_DB}, the spectrum anomaly is not explained by the summation model, which 
tends to agree well in shape with the Huber-Mueller model. This situation has triggered new experimental efforts on 
the nuclear physics side with new TAGS campaigns but also the development of new devices able to measure 
the electron shapes associated to first-forbidden transitions~\cite{IAEA2023}, since the high energy part of 
the antineutrino energy spectrum (above 4.5 MeV) is expected to be affected by the Pandemonium effect 
in nuclear databases.  A combination of the future measurements with new microscopic theoretical calculations 
of the forbidden shape factors should help understand the reactor antineutrino spectrum above 4.5 MeV.
%The contribution of the Pandemonium effect is very difficult to estimate, which makes the 
%quantitative computation of the uncertainties associated to the beta decay data tricky. 

The summation method also allows revealing the fine structures in the antineutrino energy spectrum due to 
specific fission products~\cite{Sonzogni:2017voo}. Specific patterns could be evidenced in the reactor 
antineutrino measurements due to sets of end-points of similar energies if sufficient energy resolution 
is attained by future reactor antineutrino experiments. If achieved, this would open an era of ``antineutrino spectroscopy''
able to uncover the presence of specific fission products in a reactor core. This will be relevant for the 
reactor monitoring with antineutrino detection~\cite{NuTools} and constitutes a very good benchmark for the 
study of nuclear data.

%\begin{figure}[ht]
%  \centering
%\includegraphics[width=0.45\textwidth]{./figs/NewCalculation/figure1-eps-converted-to.pdf}
%\caption{Figure 1 from \cite{Estienne:2019ujo} Upper Panel: Ratio of the Daya Bay antineutrino energy spectrum to that of the H-M model as in~\cite{DayaBay:2017jkb} (open diamonds), to that of the SM-2017 (dashed line) and the SM-2018 (continuous line) models (see text). Lower Panel: Ratio of the SM-2017 (dashed line) and SM-2018 (continuous line) antineutrino energy spectra to that of the H-M model~\cite{Mueller:2011nm, Huber:2011wv}.}
%  \label{Fig:Spectrum_DB}
%\end{figure}

We should note that the current summation model has not included the recent published TAGS 
measurements~\cite{Guadilla:2022zls, Rasco:2022pdy, Shuai:2022sud,Guadilla:2019gws}.
A forthcoming summation calculation is expected to include all the TAGS measurements published until 
now to examine the level of agreement with the most recently published reactor antineutrino spectra (e.g.~STEREO~\cite{STEREO:2022nzk} and combined measurements from Daya Bay and PROSPECT~\cite{DayaBay:2021owf}). In Ref.~\cite{Guadilla:2020pjj},
%V.Guadilla,J.L.Ta ́ın,A.Algora,etal.,Determination of $\beta$-decay ground state feeding of nuclei of importance for reactor applications, Phys. Rev. C102, 064304, 1-12 (2020)
 an independent method to determine the ground state to ground state feeding in the beta decay of fission 
 products has been published, complementary to the usual TAGS analysis from Ref.~\cite{TAS_algorithms}. 
 The combination of these analysis methods will help estimate the uncertainties associated with the 
 TAGS measurement of the beta feedings.

%\begin{figure}[hbt!]
%\centering
%\includegraphics[scale=0.5]{./figs/NewCalculation/FineStructureSonzogni.png}
%\caption{\label{fig:FineStructure} Figure 4 from \cite{Sonzogni:2017voo}. Ratio of two consecutive IBD spectrum points from 
%the Daya Bay experiment (symbols), Huber-Mueller model (dashed black line) and summation method (full red line).}
%\end{figure} 

%\begin{figure}[ht]
%  \centering
%\includegraphics[width=0.45\textwidth]{./figs/NewCalculation/DayaBayFigure7-11MeV-2022.png}
%\caption{Figure  from \cite{DayaBay:2022eyy} 
%PRL 129, 041801 (2022)}
%  \label{Fig:DB2022}
%\end{figure}

%% file: Summary.tex
\section{Conclusions and outlook}\label{sec:summary} 

It has been more than 10 years since the birth of the Reactor Antineutrino Anomaly (RAA). An anomaly is something 
that is different from what is ordinary or expected.  When that happens in physics, possible resolutions must be 
considered: are there issues in the experimental data, theoretical predictions, or both? In the rare cases where 
both experiments and theories are correct, discoveries of new physics are made, as are the cases of the solar 
neutrino problem~\cite{Cleveland:1994er,Cleveland:1998nv,Bahcall:2000nu,SAGE:1999nng,GALLEX:1998kcz,Kamiokande:1996qmi,Super-Kamiokande:1998qwk,Super-Kamiokande:2002ujc} and the atmospheric neutrino anomaly~\cite{Super-Kamiokande:1998kpq,  Kamiokande:1998ykk,Clark:1997cb} leading to the discovery of non-zero neutrino mass and mixing. In this article, 
we have reviewed the history, recent progress, and new development from both the experimental and the theoretical 
sides relevant to the RAA, from which the following conclusions can be drawn.

First, the data from reactor antineutrino flux measurements are consistent. As reviewed in Sec.~\ref{sec:model-exp-compare}, there were over 25 
measurements from about 15 independent experiments that measured the integrated reactor antineutrino flux. 
The experiments were conducted over 40 years from the 1980s to the 2020s. They use different detector technologies such as $^3$He counters and Gd or $^6$Li-loaded liquid scintillators. They use different types of nuclear reactors with various fuel
compositions, including both LEU and HEU reactors. The detectors were placed at various distances to the reactor cores, 
from a few meters up to a few kilometers. Despite the experimental challenges such as the efficiency determination and background reduction (see Sec.~\ref{sec:sub:detection}), all measurements are consistent with each other within 
their uncertainties. The combined experimental uncertainty is less than 0.5\%, much smaller than the theoretical 
uncertainty. The measurements of the integrated reactor neutrino flux are consistently smaller than the Huber-Muller 
model prediction by about 6\%.

Second, there exist significant disagreements in the reactor antineutrino energy spectrum between experiment data and 
model prediction beyond the RAA. While the RAA only focuses on the integrated reactor neutrino flux, the recent 
new measurements on the reactor neutrino energy spectrum also disagree with the Huber-Muller model prediction even 
after removing the 6\% normalization deficit (see Sec.~\ref{sec:nu_energy_spectra}). In particular, the ``5-MeV bump’’ 
was observed by $\sim$10 experiments. This shape discrepancy cannot be explained by a sterile neutrino oscillation 
hypothesis, which was one of the popular explanations of the RAA. It indicates that there must be a defect in the 
Huber-Muller model, either in the original ILL $\beta$-spectra data, or in the conversion procedure itself.

Third, the flux deficit may be different for different fission isotopes. While a direct comparison between LEU and 
HEU measurements does not show such a difference (see Fig.~\ref{fig:dyb_global_deficit_fission}), the extraction of 
individual $^{235}$U and $^{239}$Pu reactor antineutrino flux using LEU reactor fuel evolution suggest that the deficit 
is primarily due to $^{235}$U (see Table.~\ref{table:iso_ibd_yields}), although the measurement uncertainty on the 
$^{239}$Pu flux is still large and statistics limited. This isotopic dependence also cannot be explained by a sterile 
neutrino oscillation hypothesis, and again indicates issues inside the Huber-Muller model. The recent measurement of the 
$\beta$-spectrum ratio between $^{235}$U and $^{239}$Pu at the Kurchatov Institute (see Sec.~\ref{sec:beta_spectrum_ratio})
suggests that there could be a bias in the original ILL measurement of $^{235}$U, which would explain the observed isotopic dependence of the flux deficit. 
It is desirable to have independent experiments to repeat or improve the ILL measurement of cumulative $\beta$-spectra of fission isotopes from the 1980s in order to validate the foundation of the conversion method.

Fourth, there is a tremendous amount of advancement in the evaluation of reactor antineutrino modeling since the birth 
of the RAA. The uncertainty involved in the conversion method been understood more thoroughly 
(see Sec.~\ref{sec:possible_raa_explanation}). The original $\sim$2.5\% uncertainty in the Huber-Muller model is shown 
to be largely underestimated, and this uncertainty is likely to be no less than $\sim$5\%.  The summation method, 
on the other hand, has been improved to a great extent both in evaluating the nuclear databases and in improving the 
$\beta$-decay data through the campaign of TAGS experiments (see Sec.~\ref{sec:sub:new-dev-summation} and 
~\ref{sec:sub:tags}). The newly evaluated summation models agree with experimental data much better in the integrated reactor antineutrino flux from individual fission isotopes of $^{235}$U and $^{239}$Pu.

Finally, while an eV-mass-scale sterile neutrino remains a fascinating possibility, lots of progress have been 
made in clarifying the existing experimental anomalies as reviewed in Sec.~\ref{sec:possible_raa_explanation}. (e.g. MicroBooNE providing insights to the MiniBooNE 
anomalies as shown in Fig.~\ref{fig:global_ue}). The additional reactor measurements described in Sec.~\ref{sec:additional_mea} also suggest that the sterile neutrino hypothesis is unlikely to be relevant in resolving the RAA because of the large uncertainties involved in the model prediction, the 
shape discrepancies, and the isotopic dependence. For these reasons we recommend not including the RAA-allowed 
region in the electron-neutrino disappearance channel (e.g.~Fig.~\ref{fig:global_ee}, which did not include RAA), 
or enlarging the Huber-Mueller model uncertainty to at least 5\% when producing such contours.

Looking forward, to fully understand the origin of RAA and bring the reactor antineutrino flux and energy spectra in agreement with experimental data would require more effort from both the theoretical and experimental sides. 
The results from Sec.~\ref{sec:sub:new-dev-summation} show that the RAA is now disfavored by 
the summation models based on updated nuclear data, but the shape anomaly is to date still not understood. 
Above 4.5 MeV, numerous pandemonium candidates contribute to the reactor energy spectrum. More TAGS measurements 
on these shorter lived nuclei are thus necessary. Another possible explanation comes from the contribution of forbidden
transitions, for which shape factors are not well known
either theoretically or experimentally. New experiments aiming at measuring these shape factors are being setup, 
as was presented at the second meeting at IAEA in 2023~\cite{IAEA2023}. In the high energy 
part of the spectrum, isomeric states make important contributions as well, and their contribution is not well quantified yet.
Fission yields for the fission products contributing to high energy deserve more experimental constraints as well.
Finally, new high-resolution reactor antineutrino experiments such as JUNO-TAO~\cite{JUNO-TAO} could constitute an important benchmark for nuclear data, evidencing the contribution of individual components of the fission products to the reactor antineutrino spectrum.